\documentclass[aps,prx,amssymb,superscriptaddress,twocolumn]{revtex4-2}

\usepackage{amsmath,amssymb,braket,fancybox}

\usepackage{graphicx}
\usepackage{comment,color}

\newcommand{\av}[1]{\overline{#1}}
\newcommand{\mc}[1]{\mathcal{#1}}
\newcommand{\mr}[1]{\mathrm{#1}}
\newcommand{\mbb}[1]{\mathbb{#1}}
\newcommand{\mbf}[1]{\mathbf{#1}}

\newcommand{\lrs}[1]{\left( #1 \right)}
\newcommand{\lrm}[1]{{\left\{ #1 \right\}}}
\newcommand{\lrl}[1]{\left[ #1 \right]}
\newcommand{\lrv}[1]{\left| #1 \right|}
\newcommand{\braketL}[1]{\left\langle #1 \right\rangle}

\newcommand{\fracd}[2]{\frac{{d} #1 }{{d} #2 }}

\newcommand{\fracpd}[2]{\frac{\partial #1 }{\partial #2 }}

\newcommand{\aln}[1]{
\begin{align}
#1
\end{align}
}

\newcommand{\ra}{\rightarrow}

\newcommand{\Tr}{\mr{Tr}}

\newcommand{\npartial}[1]{\fracd{\braket{#1}}{t}}

\begin{document}
\title{
Quantum velocity limits for multiple observables: conservation laws, correlations, and macroscopic systems}
\date{\today}
\author{Ryusuke Hamazaki}
\affiliation{
Nonequilibrium Quantum Statistical Mechanics RIKEN Hakubi Research Team, RIKEN Cluster for Pioneering Research (CPR), RIKEN iTHEMS, Wako, Saitama 351-0198, Japan
}

\begin{abstract}
How multiple observables mutually influence their dynamics has been a crucial issue in statistical mechanics.
We here introduce a new concept, ``quantum velocity limits," to establish a quantitative and rigorous theory for non-equilibrium quantum dynamics for multiple observables.
Quantum velocity limits are universal inequalities for a vector that describes velocities of multiple observables.
They elucidate that the speed of an observable of our interest can be tighter bounded when we have knowledge of other observables, such as experimentally accessible ones or conserved quantities, compared with conventional speed limits for a single observable.
 {Moreover,  quantum velocity limits are conceptually distinct from the conventional speed limits because we need to introduce the velocity vector and solve an optimization problem for multiple variables to obtain them}.
We first derive an information-theoretical velocity limit
in terms of the generalized correlation matrix of the observables and the quantum Fisher information. 
The velocity limit has various novel consequences:
(i) {conservation law} in the system, a fundamental ingredient of quantum dynamics, can improve the velocity  limits through the correlation between the observables and conserved quantities;
(ii) speed of an observable can be bounded by a nontrivial {lower bound} from the information on another observable, while most of the previous speed limits provide only upper bounds;
(iii) there exists a notable non-equilibrium {tradeoff relation}, stating that speeds of uncorrelated observables, e.g.,
anti-commuting observables, cannot be simultaneously large;
(iv) 
%velocity limits for any  observables on a local subsystem in locally interacting many-body systems remain convergent even in the thermodynamic limit, unlike the naive application of the conventional speed limits.
 {velocity limits for local  observables  in locally interacting many-body systems are described by the fluctuation of a local Hamiltonian, which is convergent even in the thermodynamic limit, with a nontrivial finite-size correction.}
Moreover, we discover another distinct velocity limit for multiple observables on the basis of the local conservation law of probability current, which
becomes advantageous for macroscopic transitions of multiple quantities.
Our newly found velocity limits ubiquitously apply not only to unitary quantum dynamics but to classical  and quantum stochastic dynamics, offering a key step towards universal theory of far-from-equilibrium dynamics for multiple observables.

\end{abstract}

\maketitle

\section{Introduction}
Mutual influence of multiple observables has played a pivotal role in statistical mechanics. As a classic example, correlations between heat and electric currents are widely recognized as the thermoelectric effect~\cite{callen1998thermodynamics}.
As another famous example, a special type of observables, i.e., conserved quantities, lead to anomalous quantum transport properties for other observables, which is understood through the  Mazur-Suzuki bound~\cite{mazur1969non,suzuki1971ergodicity,PhysRevB.55.11029}.
Investigating such an interplay of multiple observables has now become an active area of research in various contexts, from the generalized Gibbs ensemble describing stationary states of isolated systems with many conserved quantities~\cite{PhysRevLett.98.050405}
to the thermodynamic uncertainty relations~\cite{PhysRevLett.114.158101,PhysRevLett.116.120601,horowitz2020thermodynamic} with multiple observables~\cite{dechant2018multidimensional,PhysRevX.11.041061} for the stationary dynamics in classical stochastic systems.
Besides the fundamental interest, establishing a theory of non-equilibrium dynamics for multiple observables results in practical advantages; one can understand the behavior of an observable from the knowledge of other observables, which are easy to evaluate theoretically or experimentally. 
However, previous studies mainly focused on systems near equilibrium or stationary states.
Therefore, despite its importance, universal theory that governs far-from-equilibrium (or stationary) quantum dynamics for multiple observables  has remained largely unexplored.

\begin{figure*}
\begin{center}
\includegraphics[width=\linewidth]{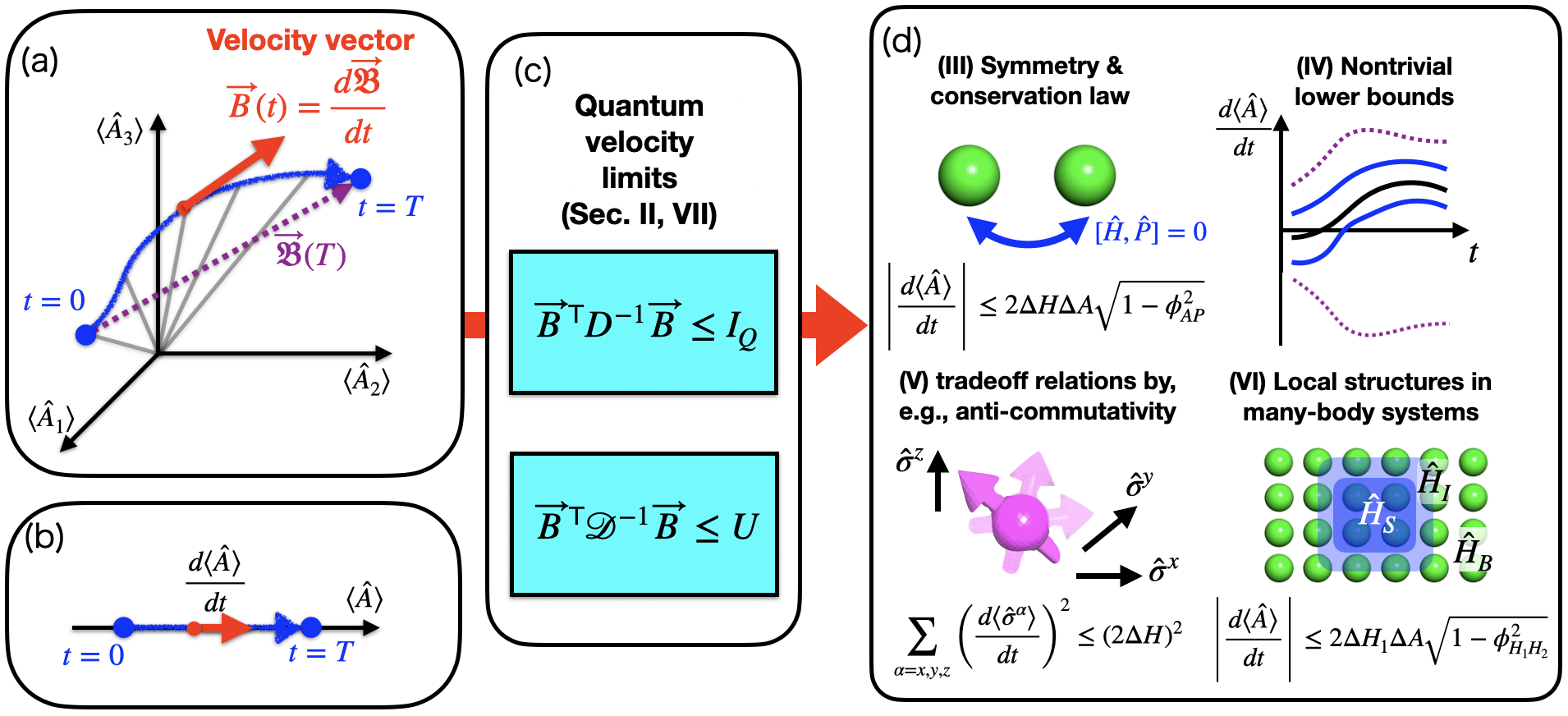}
\end{center}
\caption{
Schematic illustrations of our achievements.
(a) We consider the time evolution of a set of multiple observables simultaneously. From the trajectory of the expectation values of the observables $\vec{\mathfrak{B}}$, we introduce a velocity vector $\vec{B}$. 
This treatment enables us to understand  mutual influence of the observables, unlike (b) the treatment of the conventional quantum speed limit for a single observable.
(c) The velocity vector satisfies two distinct quantum velocity limits (QVLs), the one based on quantum information theory (Sec. II) and the one following from the continuity equation of probability (Sec. VII).
(d) Our QVLs lead to many novel applications that the conventional speed limits cannot address. In Sec. III, we elucidate how symmetry and conservation law $\hat{P}$ of the dynamics improve our ability to evaluate the rate of the transitions.
In Sec. IV, we show that the QVLs can result in nontrivial lower bounds on the speed.
This is because our bounds (blue) on the actual speed (black) are asymmetric, unlike conventional speed limits (dotted purple).
In Sec. V, we discover novel non-equilibrium tradeoff relations, some of which are unique to quantum systems.
For example, we show that anti-commuting observables cannot be simultaneously fast.
In Sec. VI, We also elucidate how locally interacting structures in quantum many-body systems are crucial in evaluating the speed of local observables 
 { $\hat{A}$ on a subsystem $S$. For example, let us consider a  local Hamiltonian given by $\hat{H}=\hat{H}_{S}+\hat{H}_{I}+\hat{H}_B$, where $\hat{H}_S$, $\hat{H}_I$, and $\hat{H}_B$  respectively represent the Hamiltonians for the subsystem,  the interaction, and the bath.
Then, we can generally take $\hat{H}_1=\hat{H}_{S}+\hat{H}_{I}$ and $\hat{H}_2=\hat{H}_{B}$ in the figure. Moreover, when $[\hat{A},\hat{H}_I]=0$, we can even take $\hat{H}_1=\hat{H}_{S}$ and $\hat{H}_2=\hat{H}_{B}+\hat{H}_{I}$.}
Inequalities presented in (d) are representative examples of our findings for the case of unitary dynamics, where $\phi_{AB}=\frac{\mr{cov}(\hat{A},\hat{B})}{\Delta A\Delta B}$.
However, we discuss more generalized versions of these inequalities and other fruitful applications in the main text.
}
\label{schem}
\end{figure*}

Recently, universal and rigorous theories on non-equilibrium state transitions have been developed in the context of quantum speed limits (QSLs).
The first seminal work was  put forward by Mandelstam and Tamm in 1945~\cite{mandelstam1945energy},
who derived that the time for an initial state to evolve into an orthogonal state under the unitary time evolution is lower bounded using the inverse of the energy fluctuation. 
Since then, such QSLs have been generalized and refined by numerous studies~\cite{fleming1973unitarity,bhattacharyya1983quantum,PhysRevLett.65.1697,PhysRevLett.70.3365,margolus1998maximum,PhysRevA.67.052109,PhysRevLett.103.160502,mondal2016quantum,PhysRevA.93.052331,PhysRevLett.120.060409,PhysRevA.103.062204,PhysRevLett.129.140403,PhysRevA.103.022210,deffner2017quantum,gong2022bounds} with experimental verifications~\cite{bason2012high,PhysRevX.11.011035,ness2021observing}.
Indeed, speed limits are generalized to open quantum systems~\cite{PhysRevLett.110.050402,PhysRevLett.111.010402,PhysRevLett.110.050403,PhysRevA.89.012307,zhang2014quantum,PhysRevA.94.052125,funo2019speed},  classical systems~\cite{PhysRevLett.120.070402,PhysRevLett.120.070401,PhysRevLett.121.030605,PhysRevX.10.021056,nicholson2020time,PhysRevLett.121.070601,PhysRevLett.125.120604}, and even  nonlinear population dynamics~\cite{adachi2022universal,garcia2022diversity,hoshino2022geometric}.
Moreover, refined bounds are obtained in light of the geometry of states~\cite{uhlmann1992energy,PhysRevA.82.022107,deffner2013energy,PhysRevA.91.042330,PhysRevX.6.021031,PhysRevLett.121.030605}, (quantum) information theory~\cite{PhysRevX.10.021056,nicholson2020time,PhysRevX.12.011038}, local conservation law of probability and optimal transport theory~\cite{PhysRevE.97.062101,PhysRevLett.126.010601,PhysRevResearch.3.043093,PhysRevResearch.4.L012034,dechant2022minimum,PRXQuantum.3.020319,PhysRevX.13.011013,PhysRevResearch.5.013017,PhysRevLett.130.010402}.
Besides, speed limits also turn out to provide constraints in controlling non-equilibrium systems~\cite{PhysRevA.74.030303,PhysRevLett.103.240501,PhysRevA.82.022318,PhysRevA.85.052327,PhysRevX.9.011034,PhysRevX.11.011035,PhysRevLett.118.100601,PhysRevLett.118.100602,giovannetti2001quantum,giovannetti2004quantum,toth2014quantum}, which is crucial for practical applications represented as quantum technology.

While many previous studies discussed QSLs of a state in light of metrics of the Hilbert space, recent works start to focus on the speed of the expectation value of a physical observable  $\hat{A}$~\cite{PhysRevX.10.021056,nicholson2020time,PhysRevA.106.042436,PhysRevX.12.011038,PRXQuantum.3.020319}.
Indeed, such observable-based speed limits provide a better bound than the metric-based one for an observable of our interest, which is more directly relevant to experiments than the quantum state itself~\cite{PhysRevX.12.011038}.
Interestingly, the first observable-based QSL was already obtained in the Mandelstam-Tamm paper~\cite{mandelstam1945energy}: they derived that the instantaneous speed of $\braket{\hat{A}(t)}:=\mr{Tr}[\hat{A}\hat{\rho}(t)]$ for a quantum state $\hat{\rho}(t)$ at time $t$ is given by 
%\aln{\label{MT}
$\lrv{\fracd{\braket{\hat{A}(t)}}{t}}\leq \frac{2}{\hbar}\Delta A\Delta H$
%}
for a unitary dynamics whose Hamiltonian is $\hat{H}$,
where $\Delta A=\sqrt{\braket{(\hat{A}-\braket{\hat{A}})^2}}$ is the quantum fluctuation of $\hat{A}$.
As recently discussed in Ref.~\cite{PhysRevX.12.011038}, this QSL can be generalized and tightened for general dynamics as $\lrv{\fracd{\braket{\hat{A}(t)}}{t}}\leq \Delta A\sqrt{I_Q}$, where $I_Q$ is the quantum Fisher information.

Although observable-based QSLs have attracted growing attention, they are  discussed only for {every single} observable.
However, we may be able to  evaluate the speed of observables of our interest better when we already have some 
knowledge on the dynamical behavior of other observables. 
For example, we expect that the speed will be slowed down if we know some fundamental structures of dynamics, e.g., conserved quantities and locality of the interactions of the Hamiltonians.
Discovering a quantitative and rigorous theory that justifies such an expectation is crucial for our understanding of far-from-equilibrium quantum dynamics.

\subsection{Summary of the achievements}

In this paper, we make the first step towards the theory of  far-from-equilibrium quantum dynamics for multiple observables by introducing the concept of quantum velocity limit (QVL).
Quantum velocity limits are universal inequalities concerning the out-of-equilibrium dynamics for expectation values of multiple observables.
As a notable observation crucially different from the conventional QSLs, we notice that the dynamics of the set of observables define a vector for the \textit{velocity} instead of the (scalar) speed for a single observable.
 {Indeed, the QVLs can be obtained only after introducing such a velocity vector and solving an optimization problem for multiple variables (see Appendix~\ref{sec:multiproof}), which are first employed to understand quantum dynamics in our manuscript. }
Our fundamental QVLs, illustrated in Eqs.~\eqref{matrix_bound},~\eqref{scalar_bound}, and~\eqref{trans_bound},
become better as we increase the number of observables.
In particular, our bounds are tighter than the conventional QSLs for a single observable~\cite{PhysRevX.12.011038}.
Furthermore, QVLs have conceptually distinct consequences from the previous speed limits: they enable us to   evaluate the rate of the dynamics under the knowledge of fundamental structures of systems, e.g., conservation law and correlations of operators.
Therefore, QVLs offer practical advantages as well as being a novel and fundamental concept towards a universal understanding of far-from-equilibrium quantum dynamics for multiple observables.
Our achievements are illustrated in Fig.~\ref{schem}, which are summarized in the following.

\subsubsection{Information-theoretical quantum velocity limit}
We begin with presenting the information-theoretical QVLs in Sec.~\ref{SecII}.
The essential observations are threefold: i) defining a velocity vector $\vec{B}=\lrs{\frac{d\braket{\hat{A}_1}}{dt},\cdots,\frac{d\braket{\hat{A}_K}}{dt}}$ for a set of $K$ time-independent  observables (Fig.~\ref{schem}(a));
ii) defining a set of $M$ special observables, called invariant observables, which are a closely related concept of conserved quantities;
iii) constructing a generalized correlation matrix $D$, which is a correlation matrix of $\{\hat{A}_k\}$ after optimally removing the effect of invariant components.
Then, we find information-theoretical QVLs as  a novel matrix inequality (Eq.~\eqref{matrix_bound}) and  {an equivalent} scalar inequality~\eqref{scalar_bound}, the latter of which reads (Fig.~\ref{schem}(c))
\aln{
\vec{B}^\mathsf{T}D^{-1}\vec{B}\leq I_Q.
}
 {The essence of the proof of this inequality is to consider the Cauchy-Scwarz inequality for the inner product between the velocity vector and some auxiliary vector, on which we solve an optimization problem. 
Indeed, introducing velocity vectors and solving an optimization problem are crucial for deriving the QVLs and are first employed to understand quantum dynamics in our manuscript.}
For the unitary dynamics, we further have $I_Q\leq 4\Delta H^2$  (with setting $\hbar=1$).
We  argue that the QVLs are refined versions of many conventional speed limits and become improved when we increase the number of observables ($K$ and $M$).
Furthermore, our QVLs provide  hitherto unknown generalizations of the quantum Cram\'er-Rao bound, which is the fundamental inequality of the quantum information theory, by considering the effect of invariant observables.
%We also show that our discussion can be extended to the vectors $\vec{\mathfrak{B}}=(\braket{\hat{A}_1(T)}-\braket{\hat{A}_1(0)},\cdots,\braket{\hat{A}_K(T)}-\braket{\hat{A}_K(0)})$ describing the displacement of observables for a finite-time interval $T$.

Despite its conciseness, our QVLs lead to many distinct applications (Fig.~\ref{schem}(d)), which the conventional approaches of the QSLs cannot address.
Indeed, as summarized below, our work reveals the fundamental and rigorous relationship between far-from-equilibrium dynamics and crucial structures that govern it, such as symmetry and conservation laws, correlations of operators and their connection with quantum non-commutativity, and local interactions of the many-body Hamiltonian.
In addition, the QVLs can provide nontrivial \textit{lower bound} of the speed, unlike most of the conventional QSLs that only give upper bounds.
%Furthermore, our QVLs predict novel tradeoff relations for observables without correlations in the generalized sense, including anti-commuting observables.

While most of the concrete examples in this manuscript are for unitary quantum dynamics, we stress that one can readily apply our QVLs to quantum stochastic systems.
Furthermore, the velocity limits also apply to classical stochastic systems and even nonlinear population dynamics~\cite{adachi2022universal}.
Therefore, velocity limits introduced in this paper offer a universal framework for understanding non-equilibrium dynamics concerning multiple observables.

\subsubsection{Symmetry and conservation law}
Our bound elucidates a fundamental  and novel relation between the speed of observables and the conservation law of a system for the first time (Sec.~\ref{SecIII}).
We show that invariant observables, which are related to conserved quantities of dynamics,  tighten the velocity and speed limits (see \eqref{inv_speed} for the speed limit).
Our result rigorously and quantitatively demonstrates how  conservation laws can slow down the dynamics, which is consistent with our naive expectations.

For the unitary dynamics,
a symmetry operator  $\hat{P}$ satisfying $[\hat{H},\hat{P}]=0$ becomes an invariant observable, and the following universal bound is obtained:
\aln{
\left|\frac{d\langle\hat{A}\rangle}{dt}\right|
\leq \Delta A\sqrt{I_Q}\sqrt{1-\phi_{AP}^2}
\leq 2\Delta H\Delta A\sqrt{1-\phi_{AP}^2},
}
where $\phi_{AB}=\frac{\mr{cov}(\hat{A},\hat{B})}{\Delta A\Delta B}$ with $\mr{cov}(\hat{A},\hat{B})=\frac{\braket{\hat{A}\hat{B}}+\braket{\hat{B}\hat{A}}}{2}-\braket{\hat{A}}\braket{\hat{B}}$ being the symmetrized covariance.
Thus, the correlation between the observable and the symmetry reduces the bound by the factor $\sqrt{1-\phi_{AP}^2}$ compared with the previously discussed bound~\cite{PhysRevX.12.011038}.
Note that we can generalize the inequality to a more complicated situation with multiple conserved quantities.

Notably, the Hamiltonian itself always becomes an invariant observable for 
the unitary dynamics,  {which does not hold in, e.g., stochastic systems}.
In particular, taking $\hat{P}=\hat{H}$ is the above inequality, we always have the speed limit stronger than the previous ones~\cite{mandelstam1945energy,PhysRevX.12.011038} for \textit{any} unitary quantum dynamics.
Importantly, the newly found bound is not only quantitative but also qualitative improvement in that it enables us to achieve the equality condition for much broader situations.
Indeed, we show that our bound satisfies the equality condition for any pure initial state and Hamiltonian in a single spin-1/2 system, unlike the previously known  bounds~\cite{mandelstam1945energy,PhysRevX.12.011038}.

\begin{comment}
Similarly, employing the conservation of the purity, which again always holds for the unitary dynamics, we discover a distinct speed limit:
\aln{
\left|\frac{d\langle\hat{A}\rangle}{dt}\right|
\leq {\Delta}' A\sqrt{I_Q}
\leq 2{\Delta}' A\Delta H,
}
where ${\Delta}' A^2=\sum_\mu\rho_\mu\Delta A_\mu^2\leq {\Delta} A^2$ 
with
$\Delta A_\mu^2=\braket{\rho_\mu|\hat{A}^2|\rho_\mu}-\braket{\rho_\mu|\hat{A}|\rho_\mu}^2$
for the density matrix diagonalized as $\hat{\rho}=\sum_\mu\rho_\mu\ket{\rho_\mu}\bra{\rho_\mu}$.
For certain mixed states, such as the ones where $\ket{\rho_\mu}$ coincide with the eigenvectors of $\hat{A}$, this inequality becomes much tighter than the conventional speed limits.
\end{comment}

\subsubsection{Nontrivial lower bounds}
As shown in Sec.~\ref{SecIV}, the QVL for two observables leads to a unique asymmetric lower and upper bound of the velocity of one observable $\hat{A}$ from the knowledge of the other one $\hat{B}$, which can lead to the nontrivial lower bound for the speed.
This is in stark contrast with many conventional speed limits, which only provide the upper bound.
More concretely, our bound takes the form (see \eqref{uplo} for the general expression), e.g.,
\aln{
v_{B}\phi_{AB}-f\sqrt{1-\phi_{AB}^2}\leq v_A\leq v_{B}\phi_{AB}+f\sqrt{1-\phi_{AB}^2},
}
where $v_X=\frac{1}{\Delta X}{\frac{d\braket{\hat{X}}}{dt}}$ and $f=\sqrt{I_Q-\lrs{\frac{1}{\Delta B}{\frac{d\braket{\hat{B}}}{dt}}}^2}$.
Thus, when $v_{B}\phi_{AB}>f\sqrt{1-\phi_{AB}^2}$, our lower bound indicates the nontrivial lower bound for the speed, $\lrv{\frac{d\braket{\hat{A}}}{dt}}$.

Our  bound relies on the knowledge of the velocity of the other reference observable $\frac{d\braket{\hat{B}}}{dt}$ and the correlation between the two observables $\phi_{AB}$.
Notably, our inequality indicates that we can precisely determine the velocity of the observable of interest  as $v_A\simeq v_B\phi_{AB}$, if we know that the QSL for the reference observable $\lrv{\frac{d\braket{\hat{B}}}{dt}}\leq \sqrt{I_Q}\Delta B$ is tight.
We demonstrate that this situation actually occurs in a single spin-1/2 system.

\subsubsection{Non-equilibrium tradeoff relations}
Our QVL indicates a new non-equilibrium tradeoff relation for uncorrelated observables; that is, the speeds of uncorrelated observables cannot be  simultaneously fast (see \eqref{uncor} in Sec.~\ref{SecV}).
More concretely, we show the additivity principle that the sum of the squares of the normalized speeds of the observables is upper bounded by the quantum Fisher information.

As a remarkable example, we discover that  the tradeoff relation is caused by the anti-commutativity, a nontrivial quantum property for certain operators.
In particular, if we take a set of observables satisfying the anti-commutation relation 
$\hat{A}_1,\cdots, \hat{A}_K\:(\hat{A}_i\hat{A}_j+\hat{A}_j\hat{A}_i=2\delta_{ij})$, we obtain a ubiquitous bound
%When we take, e.g., a general spin 1/2 system, we can consider a set of anti-commuting Pauli strings $\hat{\Sigma}_1,\cdots, \hat{\Sigma}_K\:(\hat{\Sigma}_i\hat{\Sigma}_j+\hat{\Sigma}_j\hat{\Sigma}_i=2\delta_{ij})$ to find
\aln{
\sum_{k=1}^K\lrv{\fracd{\braket{\hat{A}_k}}{t}}^2\leq I_Q
}
for \textit{any} quantum states and dynamics.
As physically important examples, we discuss the cases for a set of anti-commuting Pauli strings and the Majorana fermions for the operators $\{\hat{A}_k\}$.

Our result implies hitherto unknown tradeoff relations due to anti-commutativity, reminiscent of the famous uncertainty relations in quantum mechanics.
While the standard uncertainty relation states that quantum fluctuations of two non-commuting observables cannot be simultaneously small, our tradeoff relation states that 
speeds of multiple anti-commuting observables cannot be large simultaneously.
Therefore, this tradeoff relation  
demonstrates that nontrivial commutativity properties of observables can even affect their dynamics, as well as their fluctuations.
 {We stress that no counterpart of such relations exists in classical systems.}

\subsubsection{Locally interacting many-body systems}
We also discover inequalities that dramatically improve the evaluation of the velocities (or speeds) of local observables in quantum many-body systems (see \eqref{mbbounds} in Sec. \ref{SecVI}).
Many conventional speed limits, such as the Mandelstam-Tamm bound~\cite{mandelstam1945energy} and Margolous-Levitin bound~\cite{margolus1998maximum}, typically become meaningless in large quantum many-body systems~\cite{PhysRevX.9.011034,PhysRevX.11.011035,PRXQuantum.3.020319}.
In contrast, using our method, 
we show that the speed of observables $\hat{A}$ in a local subsystem under unitary dynamics is bounded using an energy fluctuation only in the subsystem, e.g.,
\aln{\label{intro6}
\lrv{\frac{d\braket{\hat{A}}}{dt}}\leq 2\Delta A\Delta H_{SI}\sqrt{1-\phi_{H_{SI}H_B}^2}.
}
 {Here,  we have decomposed the total Hamiltonian as  
$\hat{H}=\hat{H}_{S}+\hat{H}_{I}+\hat{H}_B$ and defined $\hat{H}_{SI}=\hat{H}_{S}+\hat{H}_{I}$, where $\hat{H}_S$, $\hat{H}_I$, and $\hat{H}_B$  respectively represent the Hamiltonians for the subsystem,  the interaction, and the bath.
Note that we can obtain novel inequalities  other than \eqref{intro6}, if there is a nontrivial commutativity structure between the observable and the Hamiltonian.
For example, when $[\hat{A},\hat{H}_I]=0$, we can replace $\Delta{H}_{SI}$ and $\phi_{H_{SI}H_B}$ in \eqref{intro6} with 
$\Delta{H}_S$ and $\phi_{H_{S}H_{IB}}$, respectively, where $\hat{H}_{IB}=\hat{H}_I+\hat{H}_B$. }

Importantly, our bound elucidates the fundamental relation between local structures of the system and the speed of an observable.
Indeed, the right-hand side does not increase with the total system size since $\Delta H_{SI}$ is convergent for locally interacting systems, in contrast to the total energy fluctuation $\Delta H$.
Our results provide a rigorous and valuable bound for arbitrary observables in the local subsystem, unlike many previous approaches~\cite{PhysRevX.9.011034,PhysRevX.11.011035,PRXQuantum.3.020319,PhysRevLett.130.010402}.
 {We note that while Ref.~\cite{PhysRevResearch.3.023074} considered a metric-based speed limit for a local subsystem based on $\Delta H_{SI}$, our results based on observables lead to many nontrivial features that are not directly obtained by the approach in~\cite{PhysRevResearch.3.023074} (see Sec.~\ref{distinct}).}

Furthermore, as another notable consequence of considering the QVL, our bound in \eqref{intro6} indicates that the correlation $\phi_{H_{SI}H_B}$ between the Hamiltonian acting on the subsystem and the rest nontrivially improves the bound.
We argue that this factor  becomes especially crucial when the size of the bath is small, which is relevant for  experiments in artificial quantum systems, e.g., trapped ions.
 {
Moreover, this correction ensures that the bound in \eqref{intro6} becomes always better than the conventional bound $2\Delta A\Delta H$, as detailed in Sec.~\ref{distinct}.
}

\subsubsection{Bound based on the conservation law of probability}

In addition to the above QVL based on the quantum Fisher information,
we also derive distinct speed limits using the local conservation law of probability (see \eqref{trans_bound} in Sec.~\ref{SecVII}).
This is advantageous for macroscopic transitions of multiple observables, improving the recently found bound for a single observable~\cite{PRXQuantum.3.020319}.
This velocity limit
relies on the (generalized) correlation matrix of the \textit{gradient} of observables  
and the local probability current.
We argue that this velocity limit leads to distinct consequences that are not obtained by the velocity limit based on quantum Fisher information.
As a remarkable example of a single-particle transport, we demonstrate the nontrivial tradeoff relation between the speeds of  the position and the even-odd probability density of the particle.

As exemplified by the discovery of the two distinct types of the QVLs,
our method provides a general framework to 
derive a wide variety of velocity limits as  generalizations of different types of speed limits.
Indeed, speed limits whose proof relies on the Cauchy-Schwarz inequality, such as that for classically chaotic systems~\cite{PhysRevResearch.5.L012016}, will be extended to velocity limits with our general procedure.

\subsection{Organization of this paper}

The rest of this paper is organized as follows.
In Sec. \ref{SecII},  we show our information-theoretical velocity limit for multiple observables on the basis of the quantum Fisher information and a generalized correlation matrix.
We also illustrate that our bound is regarded as a generalization of many previously obtained inequalities.
In Sec. \ref{SecIII}, we discuss how invariant observables of the system tighten the speed limit by showing several important applications.
In Sec. \ref{SecIV}, we derive the asymmetric upper and lower bound of the speed of an observable and explain its meaning.
In Sec. \ref{SecV}, we demonstrate the non-equilibrium tradeoff relation among the speeds of uncorrelated observables, especially anti-commutating observables.
In Sec. \ref{SecVI}, we argue that our velocity limit can be applied to obtain useful inequalities in quantum many-body systems.
In Sec. \ref{SecVII}, we discuss the different type of  velocity limit based on the local conservation law of probability.
After a formulation for the single-observable case, which slightly generalizes the treatment in Refs.~\cite{PRXQuantum.3.020319}, we discuss the multiple-observable case and its application.
In Sec. \ref{SecVIII}, we conclude the paper with future outlook.

\section{Information-theoretical Velocity limits for multiple observables}\label{SecII}

\subsection{Setup}
We consider general quantum dynamics, where a density matrix $\hat{\rho}(t)$ at time $t$ follows an equation of motion $d\hat{\rho}(t)/dt =\mathcal{L}[\hat{\rho}(t)]$ with some (generally time-dependent) super-operator $\mc{L}$.
For unitary dynamics, we have $\mc{L}[\hat{\rho}]=-i[\hat{H}(t),\hat{\rho}]$, where  $\hbar$ is set to unity in the following.
For quantum stochastic dynamics, we can consider the Liouvillian of, e.g., the Gorini-Kossakowski-Sudarshan-Lindblad equation~\cite{gorini1976completely,lindblad1976generators} as $\mc{L}$.
We can also treat classical stochastic systems with our setup by focusing only on diagonal elements of $\hat{\rho}(t)$ and their transitions.

Now, the dynamics can be rewritten as~\cite{liu2019quantum}
\aln{\label{sldeq}
{\fracd{{\hat{\rho}}}{t}} =\frac{1}{2}\{\hat{\rho},\hat{L}\},
}
where 
$
\hat{L}
%=2\int_0^\infty e^{-\hat{\rho}s}\mathcal{L}[\hat{\rho}(t)]e^{-\hat{\rho}s}ds
$
is the symmetric logarithm derivative (SLD) and $\{\hat{A},\hat{B}\}=\hat{A}\hat{B}+\hat{B}\hat{A}$ is the anti-commutator.

We focus on a set of linearly independent observables 
$\hat{A}_k\;(k=1,\cdots, K)$ and define the velocity vector $\vec{B}$ as
\aln{
\vec{B}(\{\hat{A}_k\})=\lrs{{\fracd{\braket{\hat{A}_1}}{t}},\cdots,{\fracd{\braket{\hat{A}_K}}{t}}}^\mathsf{T}.
}
For simplicity, we assume that these observables are independent of time, $d\hat{A}_k/dt=0$, although the generalization to the time-dependent case is straightforwardly done in a manner similar to Ref.~\cite{nicholson2020time,PhysRevX.12.011038}.
In this case, Eq.~\eqref{sldeq} leads to
\aln{
\fracd{\braket{\hat{A}_k}}{t}=\braket{\hat{A}_k,\hat{L}},
}
where 
$
\braket{\hat{A},\hat{B}}=\frac{1}{2}\Tr\lrl{\hat{\rho}(t)\{\hat{A},\hat{B}\}}
$
is the symmetrized correlation function.

Besides $\{\hat{A}_k\}$, we also identify a set of (generally time-dependent) observables $\hat{\Lambda}_\mu\:(1\leq \mu\leq M)$ that satisfy
\aln{\label{invariant}
\npartial{\hat{\Lambda}_\mu}-\braketL{\fracd{\hat{\Lambda}_\mu}{t}}=\braket{\hat{\Lambda}_\mu,\hat{L}}=0
}
for all $\mu$. 
We call $\hat{\Lambda}_\mu$ as invariant observables.
For example, time-independent operators that conserve during time evolutions are taken as 
 $\hat{\Lambda}_\mu$.
Without loss of generality, we can apply the Gram-Schmidt orthonormalization and assume  $\braket{\hat{\Lambda}_\nu,\hat{\Lambda}_\mu}=\delta_{\nu\mu}$.
We also assume that $\hat{A}_1,\cdots,\hat{A}_K, \hat{\Lambda}_1,\cdots, \hat{\Lambda}_M$ are linearly independent.

As shown below, by distinguishing the invariant observables $\{\hat{\Lambda}_\mu\}$ from the other observables $\{\hat{A}_k\}$, we can obtain the concise inequality of $\vec{{B}}$ where the role of invariant observables is evident.

\subsection{Quantum velocity limit}
Under the above setup, we show the following QVL in the form of the matrix inequality as our main result:
\aln{\label{matrix_bound}
\vec{B}\vec{B}^\mathsf{T}\preceq I_Q D,
}
where $A\preceq B$ means that the operator $B-A$ is positive semi-definite.
Here, we define the SLD quantum Fisher information
$
I_Q=\braket{\hat{L},\hat{L}}
$
and
a $K\times K$ generalized correlation matrix $D=D(\{A_k\};\{\Lambda_\mu\})$, whose matrix elements are given by
\aln{\label{gcor}
&D(\{A_k\};\{\Lambda_\mu\})_{kl}\nonumber\\
&=\braketL{\hat{A}_k-\sum_{\mu=1}^M\braket{\hat{A}_k,\hat{\Lambda}_\mu}\hat{\Lambda}_\mu,\hat{A}_l-\sum_{\mu=1}^M\braket{\hat{A}_l,\hat{\Lambda}_\mu}\hat{\Lambda}_\mu} \nonumber\\
&=\braket{\hat{A}_k,\hat{A}_l}
-\sum_{\mu=1}^M\braket{\hat{A}_k,\hat{\Lambda}_\mu}\braket{\hat{\Lambda}_\mu,\hat{A}_l}.
}
Note that $D$ is generally positive semi-definite.
In the following, we assume that $D$ is also positive definite and has an inverse~\footnote{If $D$ has $\bar{K}$ zero eigenvalues,
we can construct a projection operator $\hat{\mc{P}}$ to the subspace with no zero eigenvalues. From the matrix inequality in Eq.~\eqref{matrix_bound}, we have $\hat{\mc{P}}\vec{B}\vec{B}^\mathsf{T}\hat{\mc{P}}\preceq \hat{\mc{P}}D\hat{\mc{P}}$. When we regard $\hat{\mc{P}}D\hat{\mc{P}}$ as a $(K-\bar{K})\times(K-\bar{K})$ matrix, it is positive-definite and has an inverse. Then, Eq.~\eqref{scalar_bound} holds true by replacing $D$ with $\hat{\mc{P}}D\hat{\mc{P}}$}.
%As discussed in Appendix~\ref{}, this assumption typically holds; for example, this is shown to be the case when $d\braket{\hat{A}_k}/dt\neq 0$ for all $k$.

With this assumption, Eq.~\eqref{matrix_bound} leads to the  {equivalent} QVL in the form of the scalar inequality:
\aln{\label{scalar_bound}
\mathcal{K}(\{A_k\};\{\Lambda_\mu\}):=\vec{B}^\mathsf{T}D^{-1}\vec{B}\leq I_Q.
}
Importantly, for unitary quantum dynamics, we have $I_Q\leq 4\Delta H^2$, where the equality condition is achieved for,  e.g., pure states.
Then we have the bound based on the energy fluctuation of the system,
$
\vec{B}^\mathsf{T}D^{-1}\vec{B}\leq I_Q\leq 4\Delta H^2.
$

The proofs of Eqs.~\eqref{matrix_bound} and \eqref{scalar_bound} are given in Appendix~\ref{sec:multiproof};
 {the essence is to consider the Cauchy-Scwarz inequality for the inner product between the velocity vector and some auxiliary vector, on which we solve an optimization problem. }
There, we also discuss that Eqs.~\eqref{matrix_bound} and \eqref{scalar_bound} are optimal under the knowledge of invariant observables $\{\hat{\Lambda}_\mu\}$ in the following sense:
if we consider a matrix $D^f$ whose matrix elements read
\aln{
D^f_{kl}=\braketL{\hat{A}_k-\sum_{\mu=1}^Mf_{k\mu}\hat{\Lambda}_\mu,\hat{A}_l-\sum_{\mu=1}^Mf_{l\mu}\hat{\Lambda}_\mu}
}
for a set of real variables $\{f_{k\mu}\}$,
we have
\aln{
D\preceq D^f \text{   \quad and   \quad} \vec{B}^\mathsf{T}(D^f)^{-1}\vec{B} \leq \vec{B}^\mathsf{T}D^{-1}\vec{B},
}
where the equality condition is given by $f_{k\mu}=\braket{\hat{A}_k,\hat{\Lambda}_\mu}$
for all $k$ and $\mu$.
 {Furthermore, we show that generalized inequalities for multiple observables and multiple parameters (i.e., not just a single parameter of the time $t$) are obtained by a suitable optimization technique.}

We stress that our velocity limits are for the velocity vector of the expectation values of multiple observables and should not be confused with the standard speed limits for the state vector $\vec{p}$ and the density matrix $\hat{\rho}$~\cite{deffner2017quantum,gong2022bounds}.
As discussed throughout the manuscript, our velocity limits enable us to better evaluate the dynamics of an observable from the knowledge of other observables, unlike the previous speed limits.

 {
We also remark on how we may be able to experimentally access the invariant observables and correlations of observables, which play an important role in our QVLs.
For invariant observables, we can show  that invariant observables are time-independent for many cases of our interest (see Sec.~\ref{invsaisyo}).
We can also argue that, even when invariant observables change in time, evaluating them is experimentally easier than evaluating the entire density matrices, which usually requires the full quantum tomography.
For correlations of observables, we can demonstrate some cases where their evaluation  can become simplified and experimentally friendly (see Secs. \ref{HSIO} and \ref{ACPS}).
Furthermore, even in a situation where direct evaluation of the speed of an observable is easier than that of the correlation, our inequalities are meaningful.
Indeed, our inequalities, in turn, offer a way to evaluate correlations that are not easy to measure directly from the speed of an observable (see Sec. \ref{HSIO}).
}

\subsection{Connections with previous literature}
Let us discuss connections and distinctions with previous literature.
First of all, most of the previous information-theoretical speed limits consider the case of $M=1$ with $\hat{\Lambda}_1=\hat{\mbb{I}}$.
In contrast, as seen in the next subsection, our bounds become tighter if we include more invariant observables, if they exist.

Even when we consider the case of $M=1$ with $\hat{\Lambda}_1=\hat{\mbb{I}}$, our result has novel consequences.
In this case, $D$ reduces to the covariance matrix $C$, whose matrix elements are given by
\aln{
C_{kl}=\braket{\hat{A}_k,\hat{A}_l}
-\braket{\hat{A}_k}\braket{\hat{A}_l}.
}
Then, our bound can be regarded as the multi-dimensional quantum 
Cram\'er-Rao inequality generalized to arbitrary observables (i.e., not restricted to unbiased estimators~\cite{liu2019quantum}).
The application of the general multi-dimensional quantum 
Cram\'er-Rao inequality to dynamics has  {never} been discussed previously.

Note that the classical multi-dimensional Cram\'er-Rao bound for vector-valued observables
has recently been used to understand classical stochastic dynamics~\cite{dechant2018multidimensional,PhysRevX.10.021056,PhysRevX.11.041061}. 
However, our QVL \eqref{scalar_bound} is more general in that it can be used even in quantum systems, where non-commutativity comes into play.
If we assume that the off-diagonal matrix elements of $\hat{\rho}$ and $\hat{A}_k$ do not appear during dynamics, Eq.~\eqref{scalar_bound} reduces to  
the classical multi-dimensional Cram\'er-Rao bound for vector-valued observables,
$
\vec{B}^\mathsf{T}C_C^{-1}\vec{B}\leq I_C,
$
where $C_C=\braket{{A}_k{A}_l}
-\braket{{A}_k}\braket{{A}_l}$ is the classical covariance matrix and $I_C$ is the classical Fisher information.
As detailed later, by considering Eq.~\eqref{scalar_bound} in quantum dynamics, we obtain many conceptually different consequences from the previous literature.
 {For example, inequality~\eqref{inv_schrodinger} is ensured by the fact that the dynamics of the generator, i.e., the Hamiltonian, itself becomes a conserved quantity for isolated systems, which is not the case for stochastic systems considered in Refs.~\cite{dechant2018multidimensional,PhysRevX.10.021056,PhysRevX.11.041061}.
As another example, there is no direct counterpart in classical systems for the tradeoff relations for anti-commuting observables discussed in Sec.~\ref{SecV}.
We also note that the quantum generalization of the classical version of the velocity limit is not trivial in that the correlation function of operators and the Fisher information are not unique due to the quantum non-commutativity. 
Our manuscript focuses on the quantum velocity limit based on the symmetrized correlation function and the SLD quantum Fisher information.
}
\begin{comment}
 {We also note that the quantum generalization of the classical version of the velocity limit is neither unique nor trivial. Indeed, in Appendix~\ref{} we discover another type of QVL, which reduces to $
\vec{B}^\mathsf{T}C_C^{-1}\vec{B}\leq I_C,
$ in classical systems
but is distinct from \eqref{scalar_bound}. In unitary dynamics, this QVL contains the Wigner-Yanase skew information~\cite{PhysRevA.91.042330}, instead of quantum Fisher information.}
\end{comment}

To obtain the QSL for a single observable obtained previously, we again take 
$M=1$ with $\hat{\Lambda}_1=\hat{\mbb{I}}$ and $K=1$.
Then, we have
\aln{\label{single_bound}
\lrv{\npartial{\hat{A}}}\leq \Delta A \sqrt{I_Q},
}
which is equivalent to the QSL obtained in Ref.~\cite{PhysRevX.12.011038} when $\hat{A}$ is independent of time.
For unitary dynamics, we have $I_Q\leq 4\Delta H^2$, and thus
\aln{\label{MTpure}
\lrv{\npartial{\hat{A}}}\leq\mc{B}_\mr{MT}:= 2\Delta A \Delta H,
}
which is the Mandelstam-Tamm bound for an observable $\hat{A}$.
As mentioned in the next subsection, the QVL \eqref{scalar_bound} for multiple observables becomes tighter than that in Eq.~\eqref{single_bound}  for a single observable by increasing the number of observables $K$ (as well as that of invariant observables $M$).

Finally, if we consider $M\geq 2$, QVLs in \eqref{matrix_bound} and \eqref{scalar_bound} can  be regarded as a hitherto unknown generalized version of the quantum Cram\'er-Rao bound, where conserved quantities (or, more generally, invariant observables) are taken into account.

\subsection{Better bounds from more observables}

Our velocity limits in Eqs.~\eqref{matrix_bound} and \eqref{scalar_bound} become tighter if we include more observables.
In particular, when $\{A_k\}\subset \{A'_k\}$,
we have 
\aln{\label{moreobs}
\mathcal{K}(\{A_k\};\{\Lambda_\mu\})\leq \mathcal{K}(\{A_k'\};\{\Lambda_\mu\})\leq I_Q.
}
Likewise, when $\{\Lambda_\mu\}\subset \{\Lambda_\mu'\}$, we have a matrix inequality \aln{
D(\{A_k\};\{\Lambda_\mu'\})\preceq D(\{A_k\};\{\Lambda_\mu\})
}
and thus 
\aln{\label{moreinv}
\mathcal{K}(\{A_k\};\{\Lambda_\mu\})\leq \mathcal{K}(\{A_k\};\{\Lambda_\mu'\})\leq I_Q.
}
We skip the proof of \eqref{moreobs} since it is essentially equivalent to Eqs. (7)-(12) in Ref.~\cite{PhysRevX.11.041061}.
Instead of the classical covariance matrix treated there, we can perform a similar discussion for the quantum generalized correlation matrix $D$.

To prove 
$D(\{A_k\};\{\Lambda_\mu'\})\preceq D(\{A_k\};\{\Lambda_\mu\})$,
it is sufficient that we show the case with $\{\Lambda_\mu'\}_{\mu=1}^{M+1}=\{\Lambda_\mu\}_{\mu=1}^M\cup \{\Lambda_{M+1}\}$.
In this case, we have
\aln{
\sum_{kk'}a_ka_{k'}&\lrl{D(\{A_k\};\{\Lambda_\mu\})_{kk'}-D(\{A_k\};\{\Lambda_\mu'\})_{kk'}}\nonumber\\
&=\lrv{\sum_ka_k\braket{\hat{A}_k,\hat{\Lambda}_{M+1}}}^2\geq 0.
}
for any nonzero vector $\{a_k\}$.
Thus, we have $D(\{A_k\};\{\Lambda_\mu'\})\preceq D(\{A_k\};\{\Lambda_\mu\})$, which is known to lead to $D(\{A_k\};\{\Lambda_\mu'\})^{-1}\succeq D(\{A_k\};\{\Lambda_\mu\})^{-1}$. Consequently, inequality~\eqref{moreinv} follows.

\begin{table*}
{
 \caption{Summary of the applications of our velocity limits described in each of the sections with the corresponding numbers of observables $K$ and $M$. Note that conventional quantum speed limits for the observable correspond to $K=M=1$.}}
 \label{summary}
 \centering
  \begin{tabular}{cccc}
   \hline\hline
Section&  Application & $K$ & $M$\\
   \hline
   III & Speed limits under invariant observables & 1 & \:\:Arbitrary\:\: \\
      IV & Asymmetric upper and lower bound & \:\:2\:\: & Arbitrary \\
            V & Tradeoff relation for uncorrelated observables &\:\: Arbitrary \:\:& \:\:Arbitrary\:\: \\
            VI & Many-body systems & \:\:Arbitrary\:\: & \:\:Arbitrary\:\: \\ \hline
              & (Conventional speed limits) &\:\: 1 \:\:& \:\:1\:\: \\
   \hline   \hline
  \end{tabular} 
\end{table*}

\subsection{Bound for finite time interval}
While we mainly consider the instantaneous speed of the expectation value of multiple observables in this paper,  we can also find the corresponding inequality for a finite time interval.
For this purpose, we focus on a time interval $t\in [0,T]$ and define the displacement vector of observables of our interest (see Fig.~\ref{schem}(a)),
\aln{
\vec{\mathfrak{B}}=\lrs{\braket{\hat{A}_1(T)}-\braket{\hat{A}_1(0)},\cdots, \braket{\hat{A}_K(T)}-\braket{\hat{A}_K(0)}}.
}
In this case, we have a matrix inequality (see Appendix~\ref{appfintime} for proof)
\aln{\label{finitetime}
\vec{\mathfrak{B}}\vec{\mathfrak{B}}^\mathsf{T}
\preceq T^2\av{DI_Q},
}
where
$
\av{X}:=\frac{1}{T}\int_0^Tdt X(t)
$
is the average over time.
Assuming that ${{D}}$ is positive definite for all $t\in[0,T]$, we find that $\av{{D}}$ is also positive definite.
Then, we have the following scalar inequality, which relates the displacement for multiple observables and the time interval:
\aln{\label{finitescalar}
\sqrt{\vec{\mathfrak{B}}^\mathsf{T}\av{\frac{D}{I_Q}}^{-1}\vec{\mathfrak{B}}}\leq T
}
For the time-independent unitary dynamics, we further have 
\aln{\label{vecfin}
\frac{\sqrt{\vec{\mathfrak{B}}^\mathsf{T}\av{{D}}^{-1}\vec{\mathfrak{B}}}}{2\Delta H}\leq T.
}
We will discuss some concrete applications of this result in Sec.~\ref{secadditive}.

For $K=M=1$ with $\hat{\Lambda}_1=\hat{\mbb{I}}$,
it reduces to
\aln{\label{scafin}
\frac{|\braket{\hat{A}(T)}-\braket{\hat{A}(0)}|}{2\av{\Delta A}\Delta H}\leq T.
}
Again, our inequality for multiple observables, such as Eq.~\eqref{vecfin}, is better than the inequality for a single observable, such as 
Eq.~\eqref{scafin}.

\section{Speed limit and invariant observables}
\label{SecIII}

In the following sections, we demonstrate that the QVL obtained in the previous section is not just a theoretical generalization but has various notable consequences.
In Table~I, we summarize applications in each of the sections with the corresponding numbers of observables, $K$ and $M$.
When $K=1$, we can call the bound as the speed limit (or QSL) instead of the velocity limit (or QVL).

\subsection{Speed limit with invariant observables and its meaning}
\label{invsaisyo}

As a first application, we discuss the improved QSL under invariant quantities, which are also related to symmetry and the conservation law of the system.
Let us focus on $K=1$ in Eq.~\eqref{scalar_bound}, which leads to
\aln{\label{inv_speed}
\lrv{\npartial{\hat{A}}}\leq \sqrt{\braket{\hat{A}^2}-\sum_{\mu=1}^M\braket{\hat{A},\hat{\Lambda}_\mu}^2}\sqrt{I_Q}.
}

\begin{figure}
\begin{center}
\includegraphics[width=\linewidth]{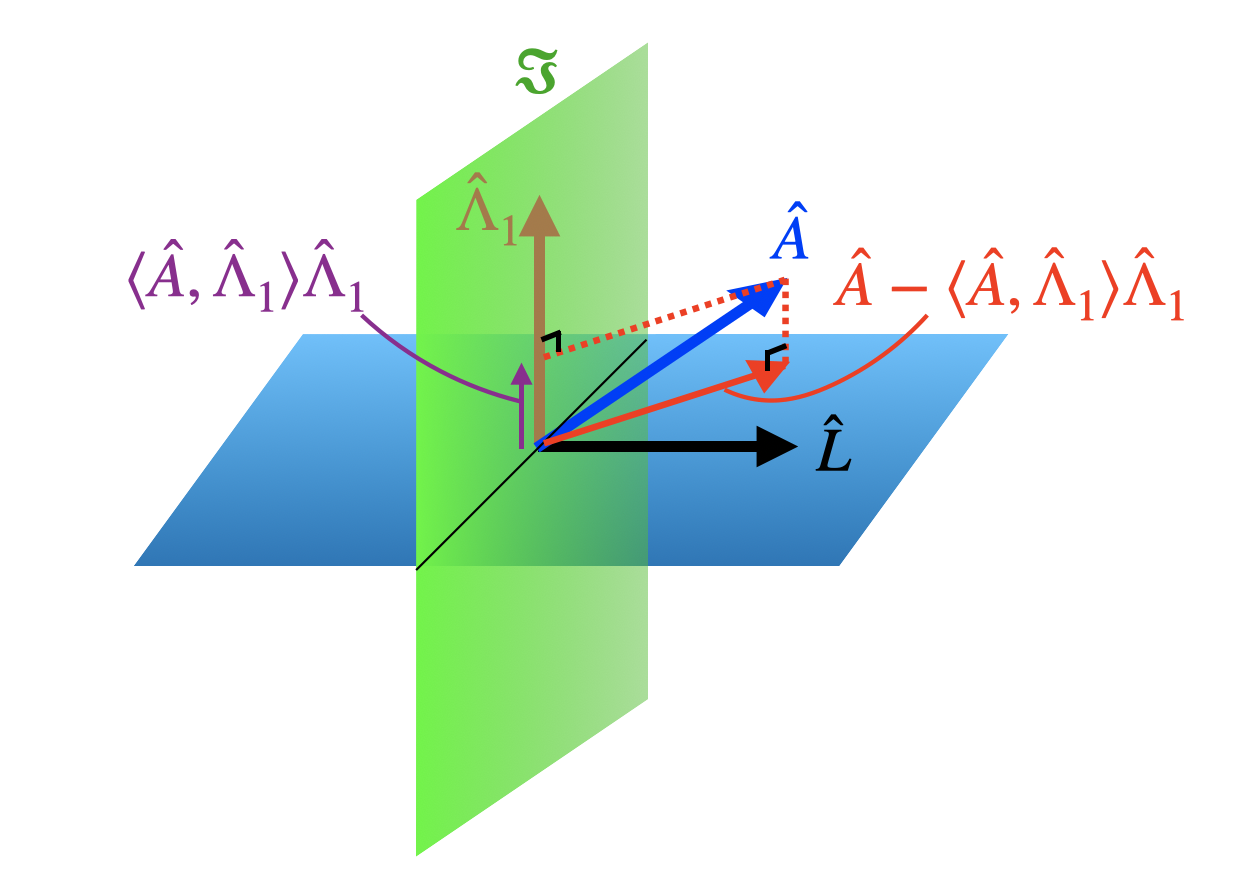}
\end{center}
\caption{
A schematic illustration that geometrically motivates Eq.~\eqref{inv_speed}.
The operators are represented by the vectors, and their inner product describes the symmetric correlation $\braket{\hat{X},\hat{Y}}=\frac{\braket{\hat{X}\hat{Y}}+\braket{\hat{Y}\hat{X}}}{2}$.
We want to optimize the upper bound of $\braket{\hat{A},\hat{L}}$ from the knowledge of $\{\hat{\Lambda}_\mu\}_{\mu=1}^M$ ($M=1$ is shown for simplicity).
This is accomplished by the orthogonal decomposition of $\hat{A}$ into the projected observable onto $\{\hat{\Lambda}_\mu\}$, $\sum_\mu \braket{\hat{A},\hat{\Lambda}_\mu}\hat{\Lambda}_\mu$, and the rest $\hat{A}-\sum_\mu \braket{\hat{A},\hat{\Lambda}_\mu}\hat{\Lambda}_\mu$.
Note that the green surface shows the set of all possible invariant observables $\mathfrak{I}:=\{\hat{X}|\braket{\hat{X},\hat{L}}=0\}$,
where $\{\hat{\Lambda}_\mu\}_{\mu=1}^M\subseteq\mathfrak{I}$ holds.
}
\label{ortho}
\end{figure}

This inequality can also be understood as follows.
Since $\braket{\hat{\Lambda}_\mu,\hat{L}}=0$, we have $d\braket{\hat{A}}/dt=\braket{\hat{A}-\sum_\mu f_\mu\hat{\Lambda}_\mu,\hat{L}}$ for any $f_\mu\in\mbb{R}$.
Using the Cauchy-Schwarz inequality and the optimization of the value of 
$
\braketL{\hat{A}-\sum_\mu f_\mu\hat{\Lambda}_\mu,\hat{A}-\sum_\mu f_\mu\hat{\Lambda}_\mu}
$
result in $f_\mu=\braket{\hat{A},\hat{\Lambda}_\mu}$ and then the above inequality.
Geometrically, the subtraction of $\sum_\mu \braket{\hat{A},\hat{\Lambda}_\mu}\hat{\Lambda}_\mu$ from $\hat{A}$ means that we can only focus on an observable projected onto the operator space that is orthogonal to $\{\hat{\Lambda}_\nu\}$. Indeed, we have
\aln{
\braketL{\hat{A}-\sum_\mu \braket{\hat{A},\hat{\Lambda}_\mu}\hat{\Lambda}_\mu,\hat{\Lambda}_\nu}=0
}
for any $\nu$ (see Fig.~\ref{ortho} for the case with $M=1$).
By this orthogonal decomposition of $\hat{A}$, we can optimize the Cauchy-Schwarz inequality under the knowledge of $\{\hat{\Lambda}_\nu\}$ and tighten the speed limit.

Interestingly, the factor $-\sum_{\mu=1}^M\braket{\hat{A},\hat{\Lambda}_\mu}^2$ in Eq.~\eqref{inv_speed} also appears in the Mazur bound~\cite{mazur1969non} if we consider time-dependent $\hat{\Lambda}_\mu$ that conserve during all times.
The Mazur bound is a bound on the long-time average of the temporal auto-correlation function near equilibrium.
Our speed limit~\eqref{inv_speed} (or, more generally, the velocity limit in Eq.~\eqref{scalar_bound}) demonstrates that overlap of an observable with conserved quantities even affects the transient dynamics far from equilibrium.
Note, however, that $\hat{\Lambda}_\mu$ in our bounds may not necessarily be a conserved quantity for all times. 
Instead, it is sufficient that $\hat{\Lambda}_\mu$, which may depend on $t$ itself, satisfies Eq.~\eqref{invariant} at each $t$.
We also note that similar orthogonal decompositions have recently been developed in different contexts, i.e., improving the bound on the estimation of 
parameters under some constraints in quantum metrology~\cite{PhysRevX.10.031023} and evaluating the effect of local conserved quantities on the eigenstate thermalization hypothesis~\cite{PhysRevLett.124.040603}.

Inequality~\eqref{inv_speed} becomes better as we increase $M$.
Correspondingly, the set of observables that satisfy the equality condition becomes larger for larger $M$.
Indeed, let us consider a set $\mc{A}^\mr{eq}_{M}$ of observables  satisfying the equality condition of \eqref{inv_speed} at a fixed time $t$.
Then, we have
\aln{\label{equality_1}
\mc{A}^\mr{eq}_{M+M'}\supseteq\lrm{\hat{A}+\sum_{\mu=M+1}^{M'}f_\mu\hat{\Lambda}_\mu|\hat{A}\in\mc{A}_M,f_\mu\in\mbb{R}}\supset\mc{A}^\mr{eq}_{M},
}
where $M'\geq 1$.
As seen from the following example, the equality condition for \eqref{inv_speed} is universally satisfied for a single spin-1/2 system.

 We remark on time dependence of the invariant observables $\{\hat{\Lambda}_\mu\}$. In general, invariant observables are chosen such that they are orthonormalized as $\braket{\hat{\Lambda}_\mu,\hat{\Lambda}_\nu}=\delta_{\mu\nu}$ for each $t$. 
However, in many important situations, $\{\hat{\Lambda}_\mu\}$ is independent of time.
 {As one of the relevant classes, let us consider 
(possibly time-dependent) unitary quantum dynamics whose Hamiltonian $\hat{H}(t')$ is commutative with time-independent operators $\hat{P}_1,\hat{P}_2,\cdots,\hat{P}_{{M}}$ (see the following sections in detail) for times $t'\in[0,t]$.
After the orthonormalization, each $\hat{\Lambda}_\mu$ is described by the linear combination of $\hat{P}_1,\hat{P}_2,\cdots,\hat{P}_{{M}}$, whose coefficients are described using the symmetrized correlations at time $t$,
\aln{\label{normalinv}
\braket{\hat{P}_{\mu},\hat{P}_{\mu'}}(t)=\frac{1}{2}\mathrm{Tr}[\hat{\rho}(t)(\hat{P}_{\mu}\hat{P}_{\mu'}+\hat{P}_{\mu'}\hat{P}_{\mu})].
}
However, since $\hat{\rho}(t)=U(t)\hat{\rho}(0)U(t)^\dag$ and $[\hat{P}_\mu,U(t)]=0$ for all $\mu$, it is evident that \eqref{normalinv} does not depend on time.}
Indeed, the orthonormalization factors are solely determined by the initial-state information in this important class of dynamics.

 {
In more general cases, invariant observables may depend on time.
However, even in that case, our inequalities can have advantages from an experimental viewpoint.
 {For this purpose, we stress that obtaining invariant observables from $\{\hat{P}_\mu\}$  only requires expectation values and correlations of $\{\hat{P}_\mu\}$, not the full information of the density matrix.}
Such expectation values are  easier to evaluate in experiments than the full density matrix, which usually requires full quantum tomography.
Thus, even though we should evaluate the normalization constant for invariant observables at all times in some cases, we do not have to know the density matrices, which is experimentally favorable.
}

\subsection{Examples for unitary dynamics}

\subsubsection{Single spin system}\label{singlespin}

\begin{figure}
\begin{center}
\includegraphics[width=\linewidth]{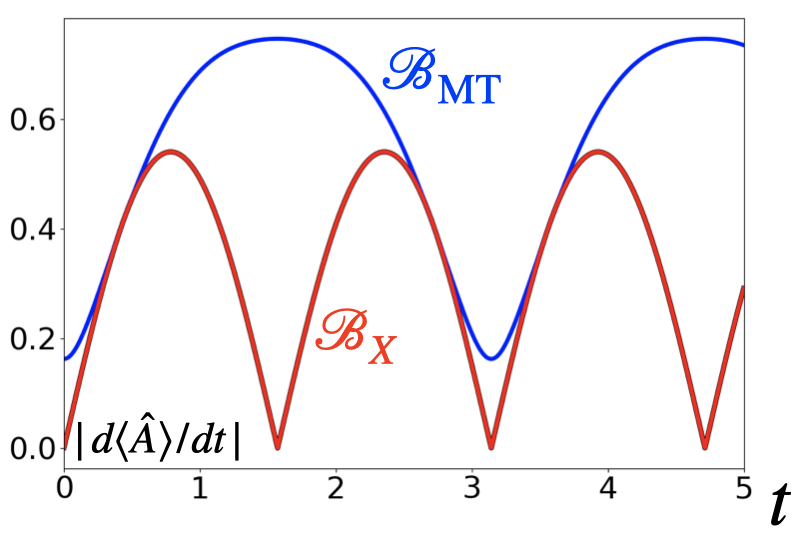}
\end{center}
\caption{
Time dependence of the speed of the expectation value of $\hat{A}=\frac{{\hat{\sigma}^x}+{\hat{\sigma}^z}}{2}$ (black) and speed limits for the single-spin system $\hat{H}=g\hat{\sigma}^x$.
We show the Mandelstam-Tamm bound $\mc{B}_\mr{MT}$ (blue) in \eqref{MTpure} and our bound based on the $\hat{\sigma}^x$-conservation $\mc{B}_{X}$ (red). 
Our bound satisfies the equality condition $\mc{B}_X=\lrv{\frac{d\braket{\hat{A}}}{dt}}$, while $\mc{B}_\mr{MT}$ does not.
We use $g=1.0$ and the initial state $\ket{\psi(0)}=\cos0.5\ket{\uparrow}+\sin0.5\ket{\downarrow}$, where $\ket{\uparrow}/\ket{\downarrow}$ is  the eigenstate of $\hat{\sigma}^z$ with the eigenvalue $+1/-1$.
}
\label{speedinv0}
\end{figure}

As a first example, let us consider a single spin system that undergoes unitary time evolution with a Hamiltonian $\hat{H}=g\hat{\sigma}^x$.
We can generally parametrize the initial state as $\ket{\psi(0)}=\cos(\theta/2)\ket{\uparrow}+e^{i\phi}\sin(\theta/2)\ket{\downarrow}$\:($0\leq \theta \leq\pi,0\leq \phi< 2\pi$), where $\ket{\uparrow}/\ket{\downarrow}$ is  the eigenstate of $\hat{\sigma}^z$ with the eigenvalue $+1/-1$. In this setting, we have $\braket{\hat{\sigma}^x(t)}=\sin\theta\cos\phi$,
$\braket{\hat{\sigma}^y(t)}=\cos 2gt \sin\theta\sin\phi-\sin 2gt\cos \theta$, 
$\braket{\hat{\sigma}^z(t)}=\sin 2gt \sin\theta\sin\phi+\cos 2gt\cos \theta$, and $\Delta H=g\sqrt{1-\sin^2\theta\cos^2\phi}$.
Then, 
$\hat{A}=c_I\hat{\mbb{I}}+c_y\hat{\sigma}^{y}+c_z\hat{\sigma}^{z}\:(c_I, c_y,c_z\in\mbb{R})$ satisfies the 
equality condition of the previous speed limit \eqref{MTpure} ($M=1$ with $\hat{\Lambda}_1=\hat{\mbb{I}}$) for arbitrary later times if $\theta=0,{\pi}$.
However, the equality condition for \eqref{MTpure} is not satisfied for a more general observable $\hat{A}=c_I\hat{\mbb{I}}+c_x\hat{\sigma}^x+c_y\hat{\sigma}^{y}+c_z\hat{\sigma}^{z}$ with $c_x\neq 0$, even when $\theta=0,{\pi}$.

In contrast, since $\hat{\sigma}^x$ is conserved, we can obtain a tighter bound using the above general method.
Indeed, our inequality 
\eqref{inv_speed} with $M=2, \hat{\Lambda}_1=\hat{\mbb{I}},$ and $\hat{\Lambda}_2=\frac{\hat{\sigma}^x-\braket{\hat{\sigma}^x}}{\sqrt{\braket{(\hat{\sigma}^x-\braket{\hat{\sigma}^x})^2}}}$ still satisfies the equality condition, in accordance with Eq.~\eqref{equality_1}.
Notably, the equality condition holds for any $\theta$ and $\phi$ in this case.

In Fig.~\ref{speedinv0}, we show one example that our bound $\mc{B}_X=2\Delta H\sqrt{\braket{\hat{A}^2}-\sum_{\mu=1}^2\braket{\hat{A},\hat{\Lambda}_\mu}^2}$ discussed above satisfies the equality condition, while the Mandelstam-Tamm bound $\mc{B}_\mathrm{MT}$ in \eqref{MTpure} does not.

 More generally, we obtain the following striking fact for a single spin-1/2 system (or a two-level system):
 For \textit{any} Hamiltonians, observables, and initial pure states, our speed limit 
\eqref{inv_speed} with $M=2, \hat{\Lambda}_1=\hat{\mbb{I}},$ and $\hat{\Lambda}_2=\frac{\hat{H}-\braket{\hat{H}}}{\sqrt{\braket{(\hat{H}-\braket{\hat{H}})^2}}}$ (corresponding to \eqref{schpure} given later) satisfies the equality condition (see Appendix~\ref{equality_anyspin1} for a proof).
Therefore, our inequalities attain the equality condition in much broader situations than the previous QSLs for an observable~\cite{mandelstam1945energy,PhysRevX.12.011038}.
Given that the attainability of the equality condition is a crucial subject on speed limits~\cite{PhysRevLett.103.160502}, our inequalities are regarded as providing qualitative improvement (not to mention quantitative improvement) in evaluating the speed of observables.

\subsubsection{Hamiltonian and symmetry as invariant observables}
\label{HSIO}

In the following, we show more complicated examples  beyond the single spin system.
As one of the primary examples of Eq.~\eqref{inv_speed}, let us consider a general system driven by a Hamiltonian $\hat{H}(t)$, i.e., $\mc{L}[\hat{\rho}(t)]=-i[\hat{H}(t),\hat{\rho}(t)]$.
In this case, we can find some invariant operators, such as the power of $\hat{H}(t)$ or  $\hat{\rho}(t)$, the projection to the eigenstates of $\hat{H}(t)$ or  $\hat{\rho}(t)$, and nontrivial symmetries commuting with  $\hat{H}(t)$.

Let us first take the powers of the shifted Hamiltonian, 
$
\hat{\mbb{I}}, \delta\hat{H}, \delta\hat{H}^2, \cdots,
$
where $\delta \hat{H}:=\hat{H}-\braket{\hat{H}}$ ($t$ is omitted for brevity).
By the Gram-Schmidt orthonormalization, we have, e.g., 
$
\hat{\Lambda}_1=\hat{\mbb{I}}, \quad \hat{\Lambda}_2= \frac{\delta\hat{H}}{\sqrt{m_2},} 
$
and
\aln{
\hat{\Lambda}_3=\frac{\delta \hat{H}^2-\frac{m_3}{m_2}\delta \hat{H}-m_2}{\sqrt{m_4-\frac{m_3^2}{m_2}-m_2^2}},
}
where $m_z=\braket{\delta\hat{H}^z}$ is the $z$th central moment of the Hamiltonian.
Applying Eq.~\eqref{inv_speed}, we find the conventional speed limit $|d\braket{\hat{A}}/dt|\leq \Delta A\sqrt{I_Q}$ for $M=1$, as we have seen in the previous section.
Furthermore, setting $M=2$ leads to
\aln{\label{inv_schrodinger}
\lrv{\npartial{\hat{A}}}&\leq \sqrt{\Delta{{A}^2}-\frac{\mathrm{cov}(\hat{A},\hat{H})^2}{\Delta H^2}}\sqrt{I_Q}\nonumber\\
&=\Delta A\sqrt{I_Q}\sqrt{1-\phi_{AH}^2},
}
where $\mathrm{cov}(\hat{X},\hat{Y})=\braket{\hat{X},\hat{Y}}-\braket{\hat{X}}\braket{\hat{Y}}$ is the symmetrized covariance, $\phi_{XY}=\frac{\mathrm{cov}(\hat{X},\hat{Y})}{\Delta X\Delta Y}$ is the quantum version of the Pearson correlation coefficient, and we have used $m_2=\Delta H^2$.
Inequality \eqref{inv_schrodinger} means that the knowledge about the correlation between the observable and the Hamiltonian improves the speed limit.
We stress that this inequality ubiquitously holds for \textit{any} unitary quantum dynamics without further assumptions.

When $\hat{\rho}$ is a pure state, inequality~\eqref{inv_schrodinger} becomes
\aln{\label{schpure}
%\lrv{\npartial{\hat{A}}}\leq\mc{B}_H:= 2\sqrt{\Delta{{A}^2}\Delta H^2-\mathrm{cov}(\hat{A},\hat{H})^2},
\lrv{\npartial{\hat{A}}}\leq\mc{B}_H:= 2\Delta A\Delta H\sqrt{1-\phi_{AH}^2}.
}
Note that this can also directly be obtained from the Schr\"odinger uncertainty relation~\cite{schrodinger1999heisenberg}, which states 
$
\lrv{\frac{[\hat{X},\hat{Y}]}{2i}}^2+|\mr{cov}({\hat{X},\hat{Y}})|^2\leq \Delta X^2\Delta Y^2
$
for two Hermitian operators $\hat{X}$ and $\hat{Y}$.
However, we stress that \eqref{inv_schrodinger} generally goes beyond this uncertainty relation since $I_Q\leq 4\Delta H^2$ for general mixed states.
Furthermore, we can obtain tighter inequality even for pure states by including higher-order invariant observables, $\hat{\Lambda}_3,\cdots$.

Another example for which our inequality is relevant is where the system possesses some additional symmetry $\hat{P}$, or conservation law, which commutes with the Hamiltonian, $[\hat{H},\hat{P}]=0$.
If we choose $M=2$ with $\hat{\Lambda}_1=\mbb{\hat{I}}$ and 
 {$\hat{\Lambda}_2=\frac{{\hat{P}}-\braket{\hat{P}}}{\Delta P}$}, 
we obtain
\aln{
\lrv{\npartial{\hat{A}}}\leq \Delta A\sqrt{I_Q}\sqrt{1-\phi_{AP}^2},
}
which becomes
\aln{\label{parpure}
\lrv{\npartial{\hat{A}}}\leq\mc{B}_P:=2\Delta A\Delta H\sqrt{1-\phi_{AP}^2}
}
for pure states.

To confirm the advantage our our inequalities in Eqs.~\eqref{schpure} and~\eqref{parpure}, we consider spin-1/2 system on $L=2$ lattice sites, whose Hamiltonian is given by 
%\aln{\label{Ham}
%\hat{H}=\sum_{j=1}^{L-%1}J\hat{\sigma}_j^z\hat{\sigma}_{j+1}^z+\sum_{j=%1}^{L}g\hat{\sigma}_j^x+h\hat{\sigma}_j^z.
%}
\aln{\label{Ham22}
\hat{H}=J\hat{\sigma}_1^z\hat{\sigma}_{2}^z+J'(\hat{\sigma}_1^x\hat{\sigma}_{2}^x+\hat{\sigma}_1^y\hat{\sigma}_{2}^y)+g(\hat{\sigma}_1^x+\hat{\sigma}_2^x)+h(\hat{\sigma}_1^z+\hat{\sigma}_2^z).
}
This Hamiltonian respects a permutation symmetry given by
\aln{
\hat{P}=\frac{1}{2}\sum_{\alpha=x,y,z}\hat{\sigma}_1^\alpha\hat{\sigma}_2^\alpha+\frac{1}{2}\hat{\mbb{I}}.
}

\begin{figure}
\begin{center}
\includegraphics[width=\linewidth]{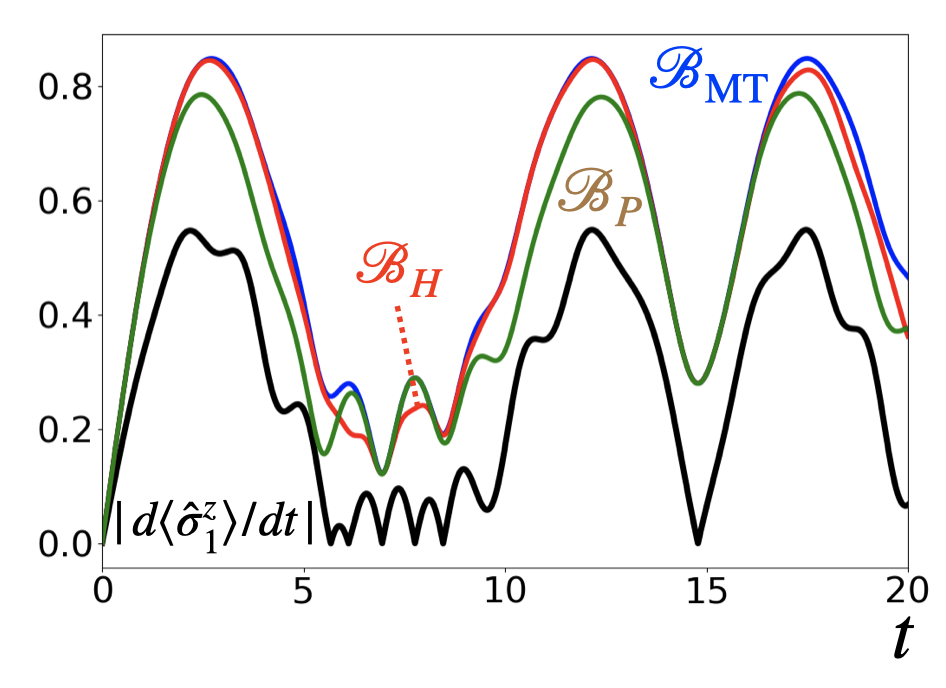}
\end{center}
\caption{
Time dependence of the speed of $\braket{\hat{\sigma}_1^z}$ (black) and speed limits for a spin system in Eq.~\eqref{Ham22}.
We show the Mandelstam-Tamm bound $\mc{B}_\mr{MT}$ (blue), the bound based on the Hamiltonian conservation $\mc{B}_H$ (red), and  the bound based on the permutation conservation $\mc{B}_P$ (green). 
Our bounds $\mc{B}_H$ and $\mc{B}_P$ provide better bounds than $\mc{B}_\mr{MT}$.
We use $J=1, J'=0, g=0.3$, $h=1$, and the initial state $\ket{\uparrow\downarrow}$, where $\ket{\uparrow}/\ket{\downarrow}$ is the eigenstate of $\hat{\sigma}^z$ with eigenvalue $+1/-1$.
}
\label{speedinv}
\end{figure}

Figure~\ref{speedinv} shows the speed of $\hat{A}=\hat{\sigma}_1^z$ and the speed limits.
Assuming an initial pure state, we compare the speed with $\mc{B}_\mr{MT}$ in Eq.~\eqref{MTpure}, $\mc{B}_H$ in Eq.~\eqref{schpure}, and $\mc{B}_P$ in Eq.~\eqref{parpure}.
We can see that $\mc{B}_H$ and $\mc{B}_P$ are better than $\mc{B}_\mr{MT}$, and that no hierarchy exists between $\mc{B}_H$ and $\mc{B}_P$.

 {
Note that the bound $\mc{B}_P$ is easy to access in experiments, even though it involves the correlation between $\hat{A}$ and $\hat{P}$. Indeed, since an explicit calculation leads to}
\aln{
\mc{B}_P=2\Delta H\sqrt{1-\braket{\hat{\sigma}_1^z}^2-\frac{(\braket{\hat{\sigma}_1^z}+\braket{\hat{\sigma}_2^z}-\braket{\hat{\sigma}_1^z}\braket{\hat{P}})^2}{\Delta P}},
}
this bound is obtained only by the measurement of the single-site expectation values at time $t$ and the conserved quantities $\Delta H, \braket{\hat{P}}$, and $\Delta P$, which  are obtained from the initial state.
This is especially relevant  {and advantageous for experiments} when we do not have enough resolution to directly measure $d\braket{\hat{\sigma}_1^z}/dt$ because of, e.g., the lack of time resolution or impossibility of taking the time derivative due to temporal noise on $\hat{\sigma}_1^z$.

 {
Even in  a situation where the direct evaluation of the speed of an observable is easier than that of the correlation, our inequalities~\eqref{schpure} and~\eqref{parpure} have advantages.
Indeed, our inequalities, in turn, offer a way to evaluate correlations that are challenging to measure directly from the speed of the observable.
More concretely,} when we can measure $|d\braket{\hat{A}}/dt|$, $\Delta A$, and the conserved quantities, we have upper bounds of the covariances, i.e., 
\aln{
|\mr{cov}(\hat{A},\hat{H})|\leq \sqrt{\Delta A^2\Delta H^2-\frac{1}{4}\lrs{\frac{d\braket{\hat{A}}}{dt}}^2},\nonumber\\
|\mr{cov}(\hat{A},\hat{P})|\leq \Delta P\sqrt{\Delta A^2-\frac{1}{4\Delta H^2}\lrs{\frac{d\braket{\hat{A}}}{dt}}^2},
}
which may be difficult to measure directly in general.
Note that these inequalities are strictly tighter than the usual Caushy-Scwarz inequality, e.g., $|\mr{cov}(\hat{A},\hat{P})|\leq \Delta P\Delta A$.

\subsubsection{Tighter bounds due to the purity conservation}
As yet another interesting example of \eqref{inv_speed}, we consider the conservation law of any power of the density matrix $\hat{\rho},\hat{\rho}^2,\hat{\rho}^3,\cdots$ for the unitary dynamics. 
This purity conservation is also regarded as the conservation of the projection to the basis of $\hat{\rho}$.
We can thus take 
\aln{
\hat{\Lambda}_\mu=\frac{\ket{\rho_\mu}\bra{\rho_\mu}}{\sqrt{\rho_\mu}}
}
for a general mixed state $\hat{\rho}$, which is diagonalized as 
$
\hat{\rho}=\sum_\mu \rho_\mu\ket{\rho_\mu}\bra{\rho_\mu}.
$
In this case, \eqref{inv_speed} leads to
\aln{\label{rhos}
\lrv{\npartial{\hat{A}}}\leq\sqrt{\sum_\mu \rho_\mu\Delta A_\mu^2}\sqrt{I_Q}
=\sqrt{\sum_{\mu\neq \nu} \rho_\mu |\braket{\rho_\mu|\hat{A}|\rho_\nu}|^2}\sqrt{I_Q},
}
where 
\aln{
\Delta A_\mu^2=\braket{\rho_\mu|\hat{A}^2|\rho_\mu}-\braket{\rho_\mu|\hat{A}|\rho_\mu}^2
}
is the fluctuation for each basis $\ket{\rho_\mu}$.
Note that this bound is tighter than the conventional bound $|d\braket{\hat{A}}/dt|\leq\Delta A\sqrt{I_Q}$, since $\lrs{\sum_\mu\rho_\mu\braket{\rho_\mu|\hat{A}|\rho_\mu}}^2\leq \sum_\mu\rho_\mu\braket{\rho_\mu|\hat{A}|\rho_\mu}^2 $ and thus
\aln{
\sum_\mu \rho_\mu\Delta A_\mu^2\leq \Delta A^2.
}
We stress that this inequality ubiquitously holds for \textit{any} unitary quantum dynamics without further assumptions, as in \eqref{inv_schrodinger}.

As an elementary example where our bound is advantageous, let us consider the state whose diagonal basis coincides with those of $\hat{A}=\sum_\mu a_\mu\ket{a_\mu}\bra{a_\mu}$, i.e., $\hat{\rho}=\sum_\mu\rho_\mu\ket{a_\mu}\bra{a_\mu}$.
Then, we have $|d\braket{\hat{A}}/dt|=|\Tr[[\hat{\rho},\hat{A}]\hat{H}]|=0$.
In this case, the right-hand side of Eq.~\eqref{rhos} vanishes, as desired (i.e., the equality condition is satisfied).
In contrast, the previous bound $\Delta A\sqrt{I_Q}$ does not vanish general.
%As a very simple example where our bound is advantageous, let us consider an infinite-temperature state for $\rho$, for which $d\braket{\hat{A}}/dt=0$.
%In this case, we can choose $\ket{\rho_\mu}$ as the eigenstate of $\hat{A}$, which leads to the vanishment of the right-hand side of Eq.~\eqref{rhos}, as desired.
%In contrast, the previous bound $\Delta A\sqrt{I_Q}$ does not vanish general.

Finally, we note that other choices of invariant observables are possible.
In Appendix~\ref{otherinv}, we show several other applications of Eq.~\eqref{inv_speed}.

\section{Asymmetric upper and lower bound}\label{SecIV}
Inequality~\eqref{scalar_bound} provides a general relation concerning multiple observables.
As an application, let us take $K=2$ observables.
In this case, we explicitly have
\aln{\label{two_obs}
\frac{D_{22}\lrv{\npartial{\hat{A}_1}}^2-2D_{12}\npartial{\hat{A}_1}\npartial{\hat{A}_2}+D_{11}\lrv{\npartial{\hat{A}_2}}^2}{D_{11}D_{22}-|D_{12}|^2}\leq I_Q,
}
where the right-hand side is upper bounded by $4\Delta H^2$ for the unitary evolution.
Note that, when we take $M=1$ and $\hat{\Lambda}_1=\mbb{\hat{I}}$, we have $D=C$, and Eq.~\eqref{two_obs} gives the quantum generalization of the result presented in Ref.~\cite{PhysRevX.10.021056}.

After straightforward calculations from this inequality, we find the  nontrivial asymmetric lower and upper bound for the velocity,
\begin{comment}
\aln{\label{two_obs}
\frac{D_{22}|\partial_t\braket{\hat{A}_1}|^2-2D_{12}\partial_t\braket{\hat{A}_1}\partial_t\braket{\hat{A}_2}+D_{11}|\partial_t\braket{\hat{A}_2}|^2}{D_{11}D_{22}-|D_{12}|^2}\leq I_Q,
}
where the right-hand side is upper bounded by $2\Delta H$ for the unitary evolution.
When we take $M=1$ and $\hat{\Lambda}_1=\mbb{\hat{I}}$, we have $D=C$ and Eq.~\eqref{two_obs} gives the quantum version of the result presented in Ref.~\cite{}.
\end{comment}
\begin{align}\label{uplo}
{\chi V_2-\sqrt{(1-\chi^2)(I_Q-V_2^2)}}
&\leq V_1\nonumber\\
&\leq {\chi V_2 +\sqrt{(1-\chi^2)(I_Q-V_2^2)}}.
\end{align}
Here, we have introduced the normalized velocity 
\aln{
V_k:=\frac{1}{\sqrt{D_{kk}}}\npartial{\hat{A}_k}
}
and the generalized Pearson correlation coefficient 
\aln{
\chi:=\frac{{D}_{12}}{\sqrt{D_{11}D_{22}}}\leq 1,
}
which reduces to $\phi_{A_1A_2}$ for $D=C$.
Note that $V_2^2\leq I_Q$ is ensured because of the single-observable speed limit for $\hat{A}_2$.
We also note that $|V_1-\chi V_2|$ is upper bounded by $\sqrt{(1-\chi^2)(4\Delta H^2-V_2^2)}\:(\leq 2\Delta H\sqrt{1-\chi^2})$ for the unitary evolution.

\begin{figure}
\begin{center}
\includegraphics[width=\linewidth]{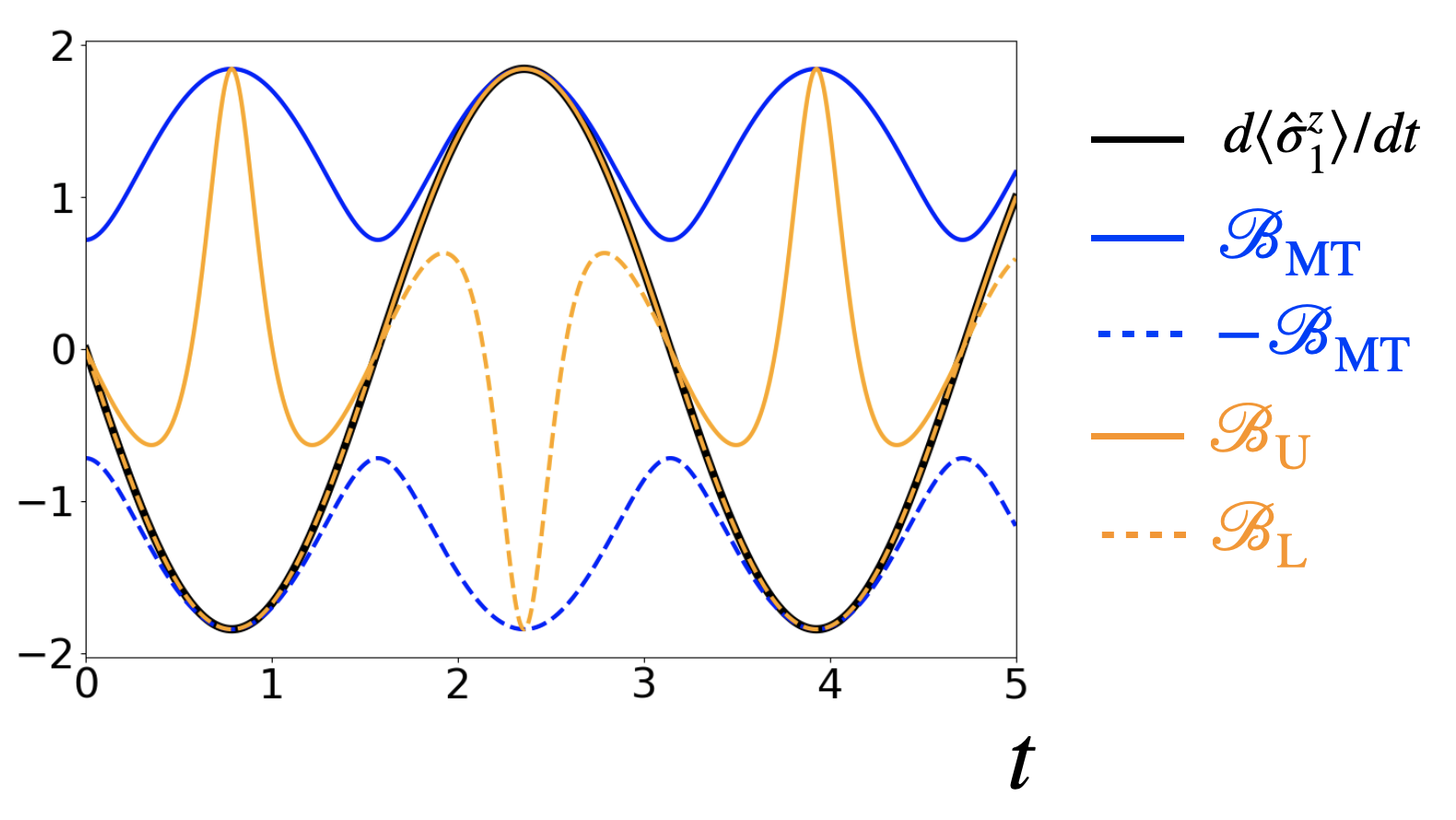}
\end{center}
\caption{
Time evolution of the velocity $d\braket{\hat{A}_1}/dt=d\braket{\hat{\sigma}_1^z}/dt$ and its lower (dashed lines) and upper (solid lines) bounds for the single-spin dynamics whose Hamiltonian is given by $\hat{H}=g\hat{\sigma}^x$.
We show $\pm\mc{B}_\mr{MT}$ (blue) and $\mc{B}_\mr{L/U}$ (orange) in Eq.~\eqref{uplo} with $\hat{A}_2=\hat{\sigma}^y$, $V_k=(d\braket{\hat{A}_k}/dt)/\Delta A_k$, and $\chi=\phi_{A_1A_2}=\mr{cov}(\hat{A}_1,\hat{A}_2)/\Delta A_1\Delta A_2$.
The bounds $\mc{B}_\mr{L}$ and $\mc{B}_\mr{U}$ can provide better bounds than $\pm \mc{B}_\mr{MT}$.
We use $g=1.0$ and the initial state $\ket{\psi(0)}=\cos0.2\ket{\uparrow}+\sin0.2\ket{\downarrow}$, where $\ket{\uparrow}/\ket{\downarrow}$ is  the eigenstate of $\hat{\sigma}^z$ with the eigenvalue $+1/-1$.
}
\label{upper0}
\end{figure}

The inequality \eqref{two_obs} provides both nontrivial upper and lower bounds for the velocity of $\hat{A}_1$ (i.e., $V_1$), given the correlation $\chi$ and the velocity $V_2$ of the other observable $\hat{A}_2$. 
Such asymmetric bounds for the velocity have seldom been obtained in previous literature.
In particular, when ${\chi V_2-\sqrt{(1-\chi^2)(I_Q-V_2^2)}}>0$, our inequality indicates the nontrivial lower bound of the speed $|V_1|$, while many previous speed limits only indicate the upper bounds.

Furthermore, when the single-observable speed limit for $\hat{A}_2$ becomes tighter (i.e., $|V_2^2- I_Q|$ is small), \eqref{two_obs} also becomes tight and $V_1$ becomes close to $\chi V_2$.
Interestingly, if we know that an observable $\hat{A}_2$ satisfies the equality condition for the single-observable speed limit, i.e., $V_2=\sqrt{I_Q}$, the speed of another arbitrary observable $\hat{A}_1$ is precisely determined as 
\aln{\label{A1A2}
\npartial{\hat{A}_1}=\frac{D_{12}}{D_{22}}\npartial{\hat{A}_2}.
}

In the following examples, we take $M=1$ and $\hat{\Lambda}_1=\hat{\mbb{I}}$, which leads to $V_k=(d\braket{\hat{A}_k}/dt)/\Delta A_k$ and $\chi=\phi_{A_1A_2}=\mr{cov}(\hat{A}_1,\hat{A}_2)/\Delta A_1\Delta A_2$.
As the first example, let us consider the single spin system as in Sec.~\ref{singlespin}.
For simplicity, we here take $\hat{A}_2=\hat{\sigma}^y$.
In this case,  when $\theta=0$ or $\pi$, Eq.~\eqref{A1A2} holds. 
For example, if we take $\hat{A}_1=\hat{\sigma}^z$, Eq.~\eqref{A1A2} reduces to 
$\frac{d\braket{\hat{\sigma}^z}}{dt}=\frac{-\braket{\hat{\sigma}^z}\braket{\hat{\sigma}^y}}{1-\braket{\hat{\sigma}^y}^2}\frac{d\braket{\hat{\sigma}^y}}{dt}$, which actually holds true.
Even when $\theta$ is not exactly $\theta=0$ or $\pi$, $I_Q-V_2^2$ becomes small when $\theta$ is close to those values, and 
inequality~\eqref{uplo} provides a good evaluation for other observables. 
Figure~\ref{upper0} demonstrates this fact: the upper and lower bounds ($\mc{B}_\mathrm{L}$ and $\mc{B}_\mathrm{U}$, respectively) on $\frac{d\braket{\hat{\sigma}^z}}{dt}$ indicated by \eqref{uplo} are
tighter than the standard single-observable speed limit for $\hat{\sigma}^z$, i.e., 
$-2\Delta H\Delta\sigma^z\leq \frac{d\braket{\hat{\sigma}^z}}{dt}\leq 2\Delta H\Delta\sigma^z$.

\begin{figure}
\begin{center}
\includegraphics[width=\linewidth]{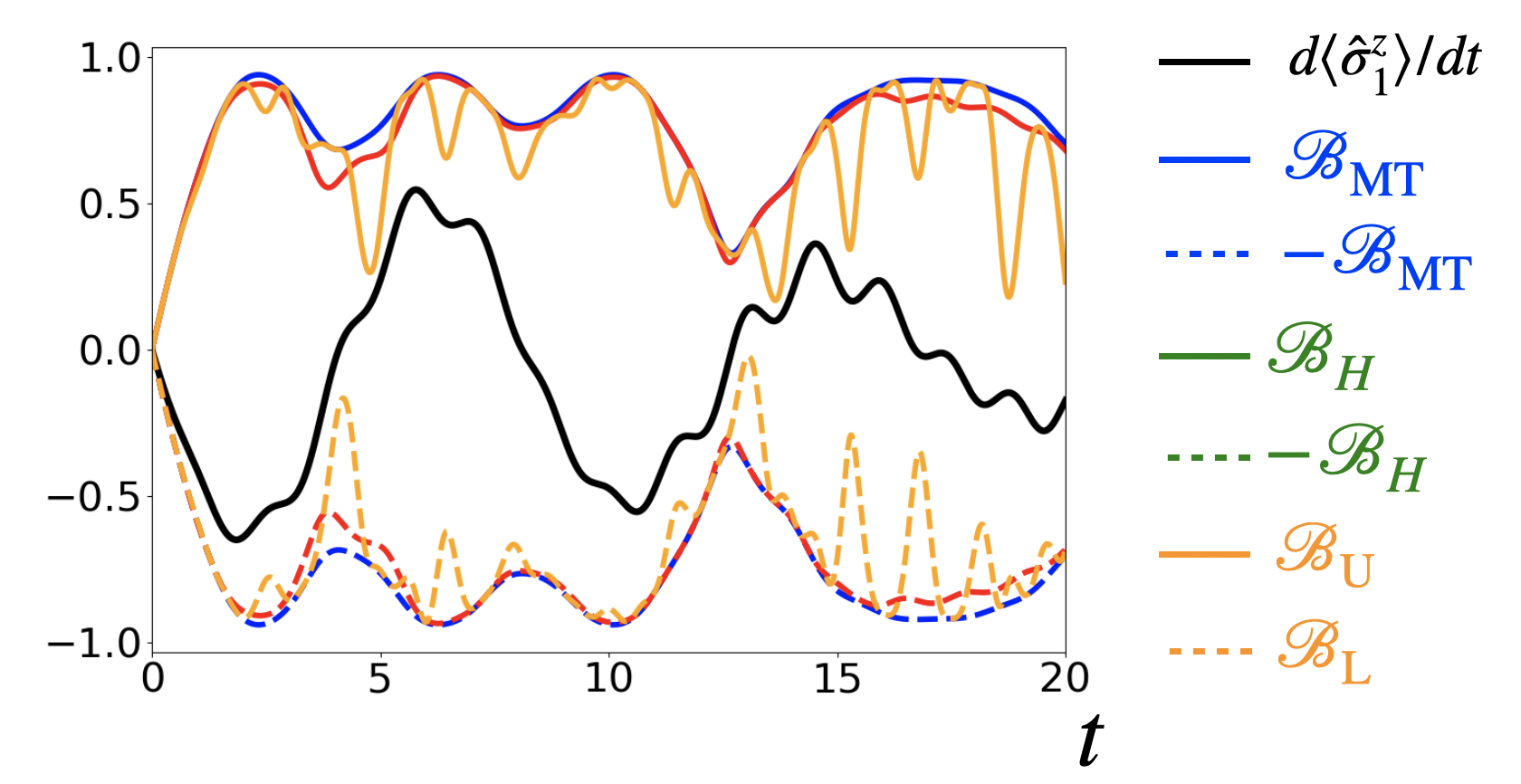}
\end{center}
\caption{
Time evolution of the velocity $d\braket{\hat{A}_1}/dt=d\braket{\hat{\sigma}_1^z}/dt$ and its lower  (dashed lines) and upper (solid lines) bounds for the dynamics whose Hamiltonian is given by Eq.~\eqref{Ham22}.
We show $\pm\mc{B}_\mr{MT}$ (blue), $\pm\mc{B}_{H}$ (red), and $\mc{B}_\mr{L/U}$ (orange) in Eq.~\eqref{uplo} with $\hat{A}_2=\hat{\sigma}_1^x$, $V_k=(d\braket{\hat{A}_k}/dt)/\Delta A_k$, and $\chi=\phi_{A_1A_2}=\mr{cov}(\hat{A}_1,\hat{A}_2)/\Delta A_1\Delta A_2$.
The bounds $\mc{B}_\mr{L}$ and $\mc{B}_\mr{U}$ can provide better bounds than the other bounds.
We use $J=1.0,J'=0.1,g=0.3, h=1.0$, and the initial state $\ket{\uparrow\downarrow}$, where $\ket{\uparrow}/\ket{\downarrow}$ is the eigenstate of $\hat{\sigma}^z$ with eigenvalue $+1/-1$.
}
\label{upper}
\end{figure}

To demonstrate our bound in a more complicated setup, we next consider the coupled two spins whose Hamiltonian is given by Eq.~\eqref{Ham22}.
%\aln{\label{Ham2}
%\hat{H}=J\hat{\sigma}_1^z\hat{\sigma}_{2}^z+J'(\hat{\sigma}_1^x\hat{\sigma}_{2}^x+\hat{\sigma}_1^y\hat{\sigma}_{2}^y)+g(\hat{\sigma}_1^x+\hat{\sigma}_2^x)+h(\hat{\sigma}_1^z+\hat{\sigma}_2^z).
%}
%We again take $M=1$ and $\hat{\Lambda}_1=\hat{\mbb{I}}$.
%, which leads to $V_k=(d\braket{\hat{A}_k}/dt)/\Delta A_k$ and $\chi=\mr{cov}(\hat{A}_1,\hat{A}_2)/\Delta A_1\Delta A_2$.
As observables, we choose $\hat{A}_1={\hat{\sigma}_1^z}$,
and $\hat{A}_2={\hat{\sigma}_1^x}$.
Figure~\ref{upper} shows the lower and upper bounds $\mc{B}_\mr{L}$ and $\mc{B}_\mr{U}$ of $d\braket{\hat{A}_1}/dt$
indicated from Eq.~\eqref{uplo}, i.e., $\mc{B}_\mr{L}\leq d\braket{\hat{A}_1}/dt\leq \mc{B}_\mr{U}$, 
the Mandeltam-Tamm bound $-\mc{B}_\mr{MT}\leq d\braket{\hat{A}_1}/dt\leq\mc{B}_\mr{MT}=2\Delta A\Delta H$,
and the bound based on the Hamiltonian conservation in Eq.~\eqref{schpure} 
$-\mc{B}_{H}\leq d\braket{\hat{A}_1}/dt\leq\mc{B}_{H}=2\Delta A\Delta H\sqrt{1-\phi_{AH}^2}$.
The bounds $\mc{B}_\mr{L}$ and $\mc{B}_\mr{U}$ can provide better bounds than the other bounds, while there is no absolute hierarchy between 
$\mc{B}_\mr{L}/\mc{B}_\mr{U}$  and $\mc{B}_{H}$.

 {Finally, we note that \eqref{two_obs} is also useful to evaluate the (generalized) correlation from the fluctuation and the speed of observables if the latter two are easier to observe directly in experiments.
For this purpose, we notice that \eqref{two_obs} can be solved in terms of $\chi$ as
\aln{
&\frac{1}{I_Q}\lrs{V_1V_2-\sqrt{(I_Q-V_1^2)(I_Q-V_2^2)}}\nonumber\\
&\leq \chi\leq \frac{1}{I_Q}\lrs{V_1V_2+\sqrt{(I_Q-V_1^2)(I_Q-V_2^2)}}.
}
For isolated systems and $M=1$ with $\hat{\Lambda}_1=\hat{\mbb{I}}$, we have
\aln{
&\frac{v_1v_2}{4\Delta H^2}-\sqrt{\lrl{1-\lrs{\frac{v_1}{2\Delta H}}^2}\lrl{1-\lrs{\frac{v_2}{2\Delta H}}^2}}\nonumber\\
&\leq\phi_{A_1A_2}\leq\frac{v_1v_2}{4\Delta H^2}+\sqrt{\lrl{1-\lrs{\frac{v_1}{2\Delta H}}^2}\lrl{1-\lrs{\frac{v_2}{2\Delta H}}^2}},
}
where $v_k=\frac{1}{\Delta A_k}\frac{d\braket{\hat{A}_k}}{dt}$.
This can be tighter than the naive bound obtained from the Cauchy-Schwarz inequality, $|\phi_{A_1A_2}|\leq 1$.
}

\section{Tradeoff relation for uncorrelated observables}\label{SecV}

\subsection{Additivity principle}
We next show a new non-equilibrium tradeoff relation between the speeds of uncorrelated observables.
We assume that $K$ observables of our interest are uncorrelated with one another in the generalized sense that 
\aln{\label{uncorr}
D_{kl}=0\:(k\neq l).
}
Then, using Eq.~\eqref{scalar_bound}, we have a stronger inequality than the one for a single observable,
\aln{\label{uncor}
\sum_{k=1}^K\frac{1}{D_{kk}}\lrv{\npartial{\hat{A}_k}}^2=\sum_{k=1}^KV_{k}^2\leq I_Q.
}
Namely, we have a simple additivity principle that the  sum of the squares of the (normalized) speeds becomes the lower bound of the quantum Fisher information.

If Eq.~\eqref{uncor} holds true during the finite time-interval $t\in [0,T]$ of our interest, 
we can discuss the additivity principle for the displacement $\vec{\mathfrak{B}}$ by applying Eq.~\eqref{finitescalar}.
Indeed, we obtain
\aln{\label{uncorfin}
\sum_k\frac{|\braket{\hat{A}_k(T)}-\braket{\hat{A}_k(0)}|^2}{\av{D_{kk}I_Q}}\leq T^2.
}

As the first application of Eq.~\eqref{uncor}, if we assume $M=1$ and $\hat{\Lambda}_1=\mbb{\hat{I}}$, we have
\aln{\label{uncor2}
\sum_{k=1}^K\frac{1}{\Delta A_k^2}\lrv{\npartial{\hat{A}_k}}^2\leq I_Q,
}
given that $C_{kl}=\mr{cov}(\hat{A}_k,\hat{A}_l)=0$ for $k\neq l$.
Defining the characteristic speed of $\hat{A}_k$ as $v_k=|V_k|= |d\braket{\hat{A}_k}/dt|/\Delta A_k$~\cite{nicholson2020time}, this can simply written as $\sum_kv_k^2\leq I_Q$, which is regarded as a quantum extension of what is obtained by Ref.~\cite{nicholson2021thermodynamic}.

\subsection{Additivity for anti-commuting observables}\label{secadditive}
The uncorrelated structure, which is necessary for the above additivity principle, naturally appears under certain situations.
As a notable case, let us
choose $M=0$ and consider observables for which $\braket{\hat{A}_k,\hat{A}_l}=0$ for $k\neq l$.
%Notable applications of Eq.~\eqref{uncor} are obtained when we choose $M=0$ and consider observables for which $\braket{\hat{A}_k,\hat{A}_l}=0$ for $k\neq l$.
This holds true for any state $\hat{\rho}$ when the observables anti-commute, i.e., $\{\hat{A}_k,\hat{A}_l\}=0$  for $k\neq l$.
In this case, we have the tradeoff inequality
\aln{\label{tradegenanti}
\sum_{k=1}^K\frac{1}{\braket{\hat{A}_k^2}}\lrv{\npartial{\hat{A}_k}}^2\leq I_Q
}
for \textit{any} state and dynamics.
We also find its finite-time version
\aln{
\sum_k\frac{|\braket{\hat{A}_k(T)}-\braket{\hat{A}_k(0)}|^2}{\av{\braket{\hat{A}_k^2}I_Q}}\leq T^2,
}
since the anti-commutation condition holds any time.

These non-equilibrium tradeoff relations, stating that two (or more than two) anti-commuting observables cannot have large speeds simultaneously, are reminiscent of the standard uncertainty relation that two non-commuting observables cannot have small variances simultaneously.
Therefore, our inequalities offer a fundamental and useful principle in non-equilibrium dynamics caused by the nontrivial commutativity property~\footnote{Strictly speaking, operators can anti-commute $\{\hat{A}_k,\hat{A}_l\}=0\:(k\neq l)$ even for the classical case, where they also commute $[\hat{A}_k,\hat{A}_l]=0$. However, such operators are limited to the trivial ones satisfying $\hat{A}_k\hat{A}_l=\hat{A}_l\hat{A}_k=0\:(k\neq l)$.
In that case, each operator  $\hat{A}_k$ should act on the different Hilbert subspace $\mc{H}_k$, where we perform the orthogonal decomposition of the total Hilbert space as  $\mc{H}=\bigoplus_k\mc{H}_k$.
In stark contrast, by allowing non-commutativity, anti-commutation relation holds for more nontrivial sets of operators, which act on the same Hilbert space without the orthogonal decomposition.}. 
Note that, applying this to unitary quantum dynamics, our result leads to
\aln{
\sum_k\frac{{\braket{i[\hat{X}_k,\hat{Y}]}^2}}{{\braket{\hat{X}_k^2}}}\leq 4\Delta Y^2,
}
for an operator $\hat{Y}$ and any set of operators satisfying the anti-commutation relation $\{\hat{X}_k,\hat{X}_l\}=2\delta_{kl}\hat{X}_k^2$.

\subsubsection{Majorana fermions}
The anti-commutation condition is profoundly connected to quantum particle statistics.
For example, let us first consider a system composed of multiple Majorana fermions labeled by $k$.
The set of such Majorana fermion operators $\{\hat{\gamma}_k\}$ satisfy
$\hat{\gamma}_k^\dag=\hat{\gamma}_k$ and $\{\hat{\gamma}_k,\hat{\gamma}_l\}=2\delta_{kl}$.
Then, we readily have
\aln{
\sum_{k=1}^K\lrv{\npartial{\hat{\gamma}_k}}^2\leq {I_Q}
}
and 
\aln{\label{Majsum}
\sum_{k=1}^K{|\braket{\hat{\gamma}_k(T)}-\braket{\hat{\gamma}_k(0)}|^2}\leq {T^2{\av{I_Q}}},
}
for \textit{any} dynamics of Majorana fermions.
If we consider the time-dependent unitary dynamics, such as the dynamics described by the Sachdev-Ye-Kitaev model~\cite{PhysRevD.94.106002}, we further have
\aln{\label{Majsum2}
\sqrt{\sum_{k=1}^K{|\braket{\hat{\gamma}_k(T)}-\braket{\hat{\gamma}_k(0)}|^2}}\leq {T\sqrt{I_Q}}\leq 2T\Delta H.
}

\begin{figure}
\begin{center}
\includegraphics[width=\linewidth]{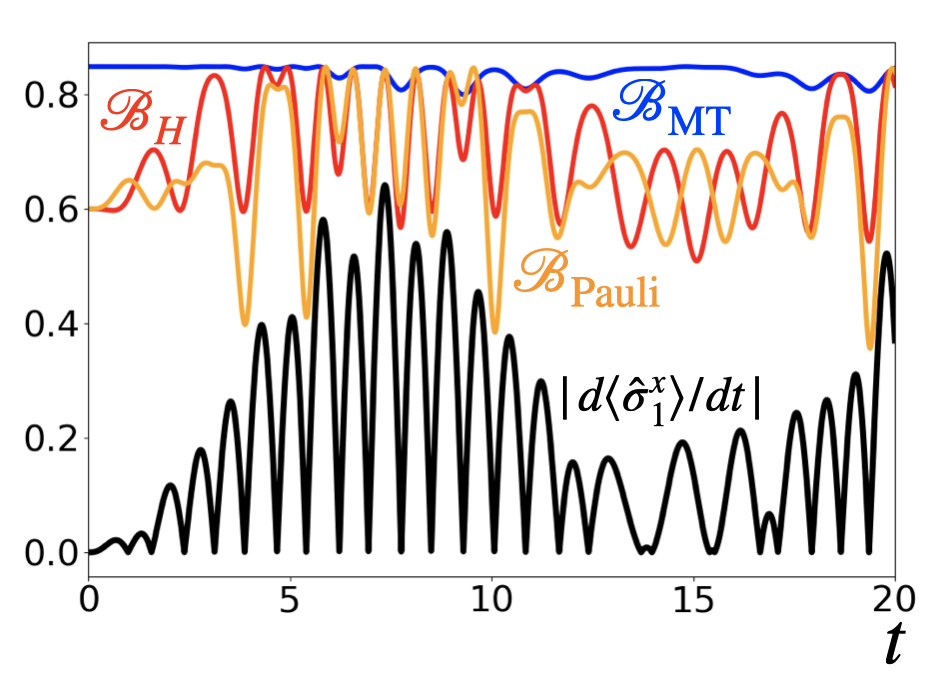}
\end{center}
\caption{
Time dependence of the speed of $\braket{\hat{\sigma}_1^x}$ (black) and speed limits for the two-spin system in Eq.~\eqref{Ham22}.
We show  $\mc{B}_\mr{MT}$ (blue),  $\mc{B}_H$ (red), and  the bound based on the anti-commuting Pauli matrices $\mc{B}_\mr{Pauli}$ (orange). 
Our bounds $\mc{B}_H$ and $\mc{B}_\mr{Pauli}$ can provide better bounds than $\mc{B}_\mr{MT}$.
We use $J=1, J'=0, g=0.3$, $h=1$, and the initial state $\ket{\uparrow\downarrow}$, where $\ket{\uparrow}/\ket{\downarrow}$ is the eigenstate of $\hat{\sigma}^z$ with eigenvalue $+1/-1$.
}
\label{speedpauli}
\end{figure}

\subsubsection{Anti-commuting Pauli strings}
\label{ACPS}
As another notable example, we next consider any spin-1/2 system and Pauli strings $\{\hat{\Sigma}_q\}$, where 
$
\hat{\Sigma}_q=\prod_l\hat{\sigma}_l^{\alpha_l}
$
with $\alpha_l=0,x,y,z$ ($\hat{\sigma}_l^0=\mbb{\hat{I}}_l$). 
Taking a set $\mc{P}_A$ ($|\mc{P}_A|=K$) of mutually anti-commuting Pauli strings, we find
\aln{\label{paulisum}
\sum_{\hat{\Sigma}_q\in\mc{P}_A}^K\lrv{\npartial{\hat{\Sigma}_q}}^2\leq I_Q
}
for \textit{arbitrary} dynamics.

As a first application of inequality~\eqref{paulisum}, let us  consider the unitary dynamics in the single spin-1/2 system (or a two-level system).
Notably, in this case, for \textit{any} Hamiltonians, observables, and initial pure states, our velocity limit 
\eqref{paulisum} with 
taking $\hat{\sigma}^x,\hat{\sigma}^y$, and $\hat{\sigma}^z$ satisfies the equality condition.
That is, 
\aln{\label{pnoeq}
\lrv{\npartial{\hat{\sigma}^x}}^2+\lrv{\npartial{\hat{\sigma}^y}}^2+\lrv{\npartial{\hat{\sigma}^z}}^2=4\Delta H^2=I_Q,
}
always holds true (see Appendix \ref{pnoeqproof}).
%, in stark contrast with the previously seen speed limits (such as the Mandelstam-Tamm bound), where the equality condition is achievable only for $\theta=0,\frac{\pi}{2}$.

To demonstrate our
 inequality for a more complicated setup, we next consider a unitary dynamics  for a pure state in the two-spin system whose Hamiltonian is given by Eq.~\eqref{Ham22}.
We especially consider $\braket{\hat{\sigma}_1^x}$ as an observable of our interest and compare the Mandelstam-Tamm bound $\mc{B}_\mr{MT}$ in Eq.~\eqref{MTpure},
$\mc{B}_H$ in Eq.~\eqref{schpure}, and a new bound obtained from Eq.~\eqref{paulisum} with $\mc{P}_A=\{\hat{\sigma}_1^x,\hat{\sigma}_1^y,\hat{\sigma}_1^z\}$,
\aln{\label{threepaulix}
\lrv{\npartial{\hat{\sigma}_1^x}}\leq \mc{B}_\mr{Pauli}=\sqrt{4\Delta H^2-\lrv{\npartial{\hat{\sigma}_1^y}}^2-\lrv{\npartial{\hat{\sigma}_1^z}}^2}
}
Figure~\ref{speedpauli} shows that $\mc{B}_\mr{Pauli}$ can provide a better bound than $\mc{B}_\mr{MT}$ and $\mc{B}_H$.

Finally, we note that as in  Eqs.~\eqref{Majsum} and \eqref{Majsum2}, we have a finite-time version of \eqref{paulisum}, i.e.,
\aln{
\sum_{\hat{\Sigma}_q\in\mc{P}_A}^K|\braket{\hat{\Sigma}_q(T)}-\braket{\hat{\Sigma}_q(0)}|^2\leq T^2\av{I_Q}
}
for general dynamics and
\aln{\label{Psum2}
\sqrt{\sum_{\hat{\Sigma}_q\in\mc{P}_A}^K |\braket{\hat{\Sigma}_q(T)}-\braket{\hat{\Sigma}_q(0)}|^2}\leq T\sqrt{I_Q}\leq 2T\Delta H
}
for time-independent unitary dynamics.

 {
Notably, the tradeoff relations for anti-commuting observables are easy to verify in experiments since they do not involve the correlation of observables, which is sometimes hard to measure. 
Furthermore, the inequalities in, e.g., \eqref{Majsum2}, \eqref{threepaulix}, and \eqref{Psum2} involve time-independent quantities, such as $\Delta H$ and $I_Q$, which make the evaluation even easier.
For example, the lower bound of the transition time $T$ for which $\{\braket{\hat{\Sigma}_q}\}$ changes from $\{\braket{\hat{\Sigma}_q(0)}\}=\{\Sigma_q^\mathrm{ini}\}$ to $\{\braket{\hat{\Sigma}_q(T)}\}=\{\Sigma_q^\mathrm{fin}\}$ is easily evaluated from \eqref{Psum2} just by measuring the initial value of $\Delta H$ (or $I_Q$).
}

\subsubsection{Comparison with the previous speed limit based on the metric}\label{commet}
Interestingly, the inequalities~\eqref{Majsum2} and \eqref{Psum2} 
can be verified only by the information of the initial and final times.
Furthermore, if we include a sufficient number of anti-commuting observables, the inequalities become better than  conventional speed limits based on the metric of the quantum state.
To see this, we remind the general inequality for a single observable  {(note that a similar inequality is discussed to apply to classical stochastic thermodynamics~\cite{hasegawa2023thermodynamic})}
\aln{
|\braket{\hat{A}(T)}-\braket{\hat{A}(0)}|^2&\leq
\frac{\Delta_A^2}{4}\|\hat{\rho}(T)-\hat{\rho}(0)\|_1^2\nonumber\\
&\leq
{\Delta_A^2}(1-F(T)),
}
where $\Delta_A=\max_{\ket{v}}\frac{{\braket{v|\hat{A}|v}}}{\braket{v|v}}-\min_{\ket{v}}\frac{\braket{{v|\hat{A}|v}}}{\braket{v|v}}$ is the spectral width of $\hat{A}$, $\|X\|_1=\mr{Tr}\sqrt{X^\dag X}$ is the trace-1 norm, and 
$
F(T)=\lrs{\mr{Tr}\lrm{\sqrt{\sqrt{\hat{\rho}(0)}\hat{\rho}(T)\sqrt{\hat{\rho}(0)}}}}^2
$
is the mixed-state fidelity.
Here, we have used the Fuchs-van de Graaf inequality~\cite{nielsen2002quantum}.
For the time-independent untiary dynamics, we have a well-known speed limit based on the metric of the state~\cite{deffner2017quantum},
$
\mr{arccos}(\sqrt{F(T)})\leq T\Delta H 
$
for $0\leq T\Delta H\leq\frac{\pi}{2}$.
We then have
\aln{
|\braket{\hat{A}(T)}-\braket{\hat{A}(0)}|^2
\leq
{\Delta_A^2}\sin^2(T\Delta H).
}
Using this inequality for $K$ times, we have
\aln{\label{metric}
\sqrt{\sum_{k=1}^K|\braket{\hat{A}_k(T)}-\braket{\hat{A}_k(0)}|^2}\leq \sqrt{\sum_{k=1}^K\Delta_{A_k}^2}|\sin(T\Delta H)|
}
for $0\leq T\Delta H\leq\frac{\pi}{2}$.

While Eq.~\eqref{metric} holds for general observables, our bounds for anti-commuting observables, such as~\eqref{Majsum2} and \eqref{Psum2}, can be tighter for large $K$.
For example, if we consider anti-commuting Pauli strings, we have $\Delta_{\Sigma_q}=2$,  and 
Eq.~\eqref{metric} becomes
\aln{\label{metsum}
\sqrt{\sum_{\hat{\Sigma}_q\in\mc{P}_A}^K |\braket{\hat{\Sigma}_q(T)}-\braket{\hat{\Sigma}_q(0)}|^2}\leq 2\sqrt{K}|\sin(T\Delta H)|.
}
Thus, inequality \eqref{Psum2} provides a better bound when
\aln{
K>\min\lrm{\lrs{\frac{T\Delta H}{\sin(T\Delta H)}}^2, \lrs{\frac{\pi}{2}}^2}.
}
In particular, if we take $K\geq 3$, \eqref{Psum2} always gives a better bound than 
\eqref{metsum} (defined for $0\leq T\Delta H\leq\frac{\pi}{2}$).
We have a similar result for the Majorana fermion case.

\subsection{Remark on the coherent and incoherent speed limits}
Before ending this section, we briefly remark on the coherent and incoherent speed limits discussed in Ref.~\cite{PhysRevX.12.011038}.
To explain these speed limits in our context, let us diagonalize the density matrix at a fixed time $t$ as $\hat{\rho}=\sum_\mu \rho_\mu\ket{\rho_\mu}\bra{\rho_\mu}$.
Then, we can decompose an observable $\hat{A}$ as $\hat{A}=\hat{A}_C+\hat{A}_I$, where $\hat{A}_C=\sum_{\mu\neq\nu}\braket{\rho_\mu|\hat{A}|\rho_\nu}\ket{\rho_\mu}\bra{\rho_\nu}$ and $\hat{A}_I=\sum_\mu\braket{\rho_\mu|\hat{A}|\rho_\mu}\ket{\rho_\mu}\bra{\rho_\mu}$.
Here, we consider that $\hat{A}_C$ and $\hat{A}_I$ are fixed and independent of time.
Then, at time $t$, we can show speed limits separately for $\hat{A}_C$ and $\hat{A}_I$.
Indeed, we have $\lrv{\npartial{\hat{A}_C}}\leq \Delta A_C\sqrt{I_{QC}}$ and $\lrv{\npartial{\hat{A}_I}}\leq \Delta A_I\sqrt{I_{QI}}$, where
$I_{QC}$ and $I_{QI}$ are coherent and incoherent parts of the quantum Fisher information, respectively (see Ref.~\cite{PhysRevX.12.011038} for their explicit expressions).
Importantly, $I_Q=I_{QC}+I_{QI}$.

Now, it is straightforward to see that 
$\mr{cov}(\hat{A}_C,\hat{A}_I)=0$.
Thus, we can use our general discussion in \eqref{uncor2} to obtain
\aln{\label{sumineq}
\frac{1}{\Delta A_C^2}\lrv{\npartial{\hat{A}_C}}^2+
\frac{1}{\Delta A_Q^2}\lrv{\npartial{\hat{A}_Q}}^2\leq I_Q,
}
which is consistent with the coherent and incoherent speed limits.
While inequality~\eqref{sumineq} has less information than the separate speed limits, 
we stress that it will be improved if we can find additional uncorrelated observables and include them in the left-hand side of \eqref{uncor2}.

\section{Application to quantum many-body systems}\label{SecVI}

\begin{figure*}
\begin{center}
\includegraphics[width=\linewidth]{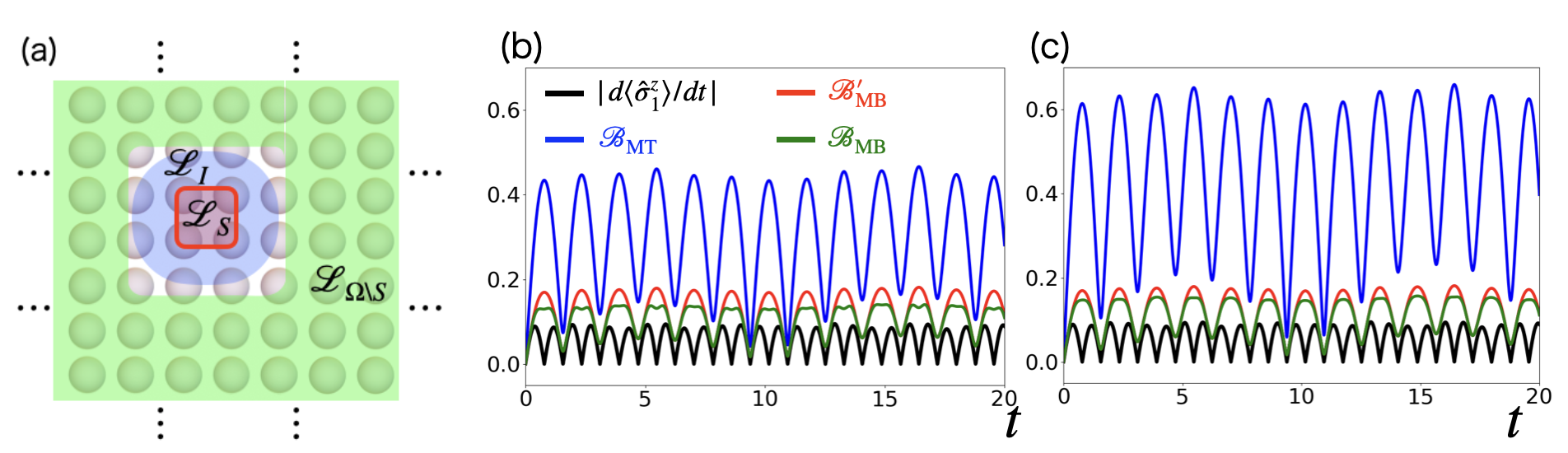}
\end{center}
\caption{
(a) Schematic illustration of a many-body system, whose dynamics is decomposed as $\mc{L}=\mc{L}_S+\mc{L}_I+\mc{L}_{\Omega\backslash S}$,
where $\mc{L}_S\:(\mc{L}_{\Omega\backslash S})$ acts nontrivially only  on $S\:(\Omega\backslash S)$ and $\mc{L}_I$ represents the interaction between $S$ and $\Omega\backslash S$.
(b,c) Time dependence of the speed of observables and speed limits for a spin system in Eq.~\eqref{Ham} with (b) $L=6$ and (c) $L=12$.
We show the speed of $\braket{\hat{\sigma}_1^z}$ (black), the Mandelstam-Tamm bound $\mc{B}_\mr{MT}$ (blue), the bounds for the many-body system $\mc{B}_\mr{MB}'$ (red) and  $\mc{B}_\mr{MB}$ (green). 
Our bounds $\mc{B}_\mr{MB}'$ and $\mc{B}_\mr{MB}$ provide good bounds even when $L$ increases, in contrast with $\mc{B}_\mr{MT}$.
We also discover that $\mc{B}_\mr{MB}$ becomes better than $\mc{B}_\mr{MB}'$ especially for relatively small $L$, indicating that the correlation factor can capture finite-size corrections for the speed limit.
We use $J=1, g=0.3$, $h=1$, and the initial state
 {$\ket{\uparrow\uparrow\uparrow\uparrow\cdots}$}.
}
\label{manytime}
\end{figure*}

In this section, we show  that useful QVLs can be derived and result in tighter inequalities than the previous ones for many-body systems.
As seen below, our results provide meaningful convergent bounds even for large systems, in contrast with a naive application of previous speed limits, which lead to divergent bounds due to, e.g., the divergence of $\Delta H$.
While this fact has been pointed out before~\cite{PhysRevX.9.011034,PhysRevResearch.3.023074,PhysRevX.11.011035,PRXQuantum.3.020319}, previous results are not satisfactory in the following sense:
Refs.~\cite{PhysRevX.9.011034,PhysRevX.11.011035} relied on an unproven conjecture that is applied to only limited situations where quantum systems are controlled by the change of external parameters; Ref.~\cite{PhysRevX.11.011035} mentioned the problem but did not explicitly consider many-body situations;
 {results in \cite{PhysRevResearch.3.023074} are based on the metric of the states, which cannot fully take account of observables' properties (see  Sec.~\ref{distinct} for the detail);}
results in Refs.~\cite{PRXQuantum.3.020319,PhysRevLett.130.010402} can only be applied to a limited type of observables~\footnote{Indeed, observables should be diagonal with respect to the basis of the vertices of the graph, which describes the system.}. In stark contrast, we rigorously show  {a set of QVLs concerning local observables for general many-body dynamics. As a notable example, we show}
that the speed of a general local observable is bounded by the local energy fluctuation (for the case of the unitary dynamics), which does not diverge even in the thermodynamic limit and provides a tighter bound.
Moreover, we also clarify how we can improve the bound  {using the correlation} when the ``bath" of the system is finite.

We also note that, while several other approaches exist to bound the speed of the many-body dynamics, such as the Lieb-Robinson bound and its applications~\cite{lieb1972finite,nachtergaele2006lieb,PhysRevLett.97.050401,lashkari2013towards,PhysRevLett.115.256803,kuwahara2016floquet,abanin2017rigorous,PhysRevB.96.060301,PhysRevLett.124.210606,PhysRevA.101.052122}, our bounds have the advantage in that they have the information-theoretical meaning.
Furthermore, for unitary dynamics, our bounds include  {the quantum fluctuation of the observable and} the energy fluctuation of part of the Hamiltonians, in accordance with the original spirit   {of the Mandelstam-Tamm bound for observables}. 

\subsection{Quantum velocity limit for decomposed dynamics}

For the above purpose,
we first illustrate a general inequality as a variant of the QVL discussed in Eq.~\eqref{scalar_bound}.
Let  $\mc{L}$ be decomposed as $\mc{L}=\mc{L}_1+\mc{L}_2$, where 
$\mc{{L}}_2[\hat{\rho}]$  does not change the expectation value of $K$ projected observables $\hat{A}_k -\sum_\mu\braket{\hat{A}_k,\hat{\Lambda}_\mu}\hat{\Lambda}_\mu$ of our interest.
That is, defining the SLD $\hat{{L}}_{1,2}$ for $\mc{L}_{1,2}$, we assume 
\aln{
\braketL{\hat{A}_k-\sum_\mu\braket{\hat{A}_k,\hat{\Lambda}_\mu}\hat{\Lambda}_\mu,\hat{L}_2}=0
}
for all $1\leq k\leq K$.
Under this condition, we can show  (see Appendix~\ref{manproof} for proof)
\aln{\label{many0}
\vec{B}^\mathsf{T}D^{-1}\vec{B}\leq \mc{F}_{11}-\frac{\mc{F}_{12}^2}{\mc{F}_{22}}=\frac{\mr{Det}[\mc{F}]}{\mc{F}_{22}}.
}
Here,
\aln{\label{genfis}
\mc{F}_{zz'}
&=\braketL{\hat{L}_z-\sum_{\mu=1}^M\braket{\hat{L}_z,\hat{\Lambda}_\mu}\hat{\Lambda}_\mu,\hat{L}_{z'}-\sum_{\mu=1}^M\braket{\hat{L}_{z'},\hat{\Lambda}_\mu}\hat{\Lambda}_\mu} \nonumber\\
&=\mc{I}_{zz'}
-\sum_{\mu=1}^M\braket{\hat{L}_z,\hat{\Lambda}_\mu}\braket{\hat{\Lambda}_\mu,\hat{L}_{z'}}
}
is the modified quantum Fisher information matrix with 
$
\mc{I}_{zz'}=\braket{\hat{L}_z,\hat{L}_{z'}}
$
being the standard quantum Fisher information matrix (see also Appendix~\ref{sec:multiproof}).
As discussed in Appendix~\ref{manproof}, the bound in the right hand-side of~\eqref{many0} can be replaced simply with $\mc{I}$:
\aln{\label{qvldec}
\vec{B}^\mathsf{T}D^{-1}\vec{B}\leq \mc{I}_{11}-\frac{\mc{I}_{12}^2}{\mc{I}_{22}}=\frac{\mr{Det}[\mc{I}]}{\mc{I}_{22}}.
}

As a primary example, 
let us consider the Hamiltonian dynamics $\hat{H}=\hat{H}_1+\hat{H}_2$ and a pure state.
In this case, we have $\mc{I}_{11}=4\Delta H_1^2$, $\mc{I}_{22}=4\Delta H_2^2$ and $ {\mc{I}_{12}}=4\mr{cov}(\hat{H}_1,\hat{H}_2)$.
In particular, for  $M=1$ with $\hat{\Lambda}_1=\mbb{\hat{I}}$, we obtain the following QVL:
\aln{\label{mbboundv}
\vec{B}^\mathsf{T}C^{-1}\vec{B}\leq 4\Delta H_1^2(1-\phi_{H_1H_2}^2),
}
provided that $\braket{[\hat{A}_k,\hat{H}_2]}=0$ for all $k$.
Furthermore, for a single observable $\hat{A}$ ($K=1$), we have
\aln{\label{mbbounds}
\lrv{\npartial{\hat{A}}}\leq \mc{B}_\mr{MB}:=2\Delta A\Delta H_1\sqrt{1-\phi_{H_1H_2}^2},
}
when $\braket{[\hat{A},\hat{H}_2]}=0$.

\subsection{Bounds for many-body dynamics}
As an important application, we consider quantum many-body spin systems on lattice sites $\Omega$ and dynamics caused by local interactions.
Let us focus on a set of observables $\hat{A}_k$ that act on a subsystem $S\subset \Omega$. The dynamics is then decomposed as $\mc{L}=\mc{L}_S+\mc{L}_I+\mc{L}_{\Omega\backslash S}$,
where $\mc{L}_S\:(\mc{L}_{\Omega\backslash S})$ acts nontrivially only  on $S\:(\Omega\backslash S)$ and $\mc{L}_I$ represents the interaction between $S$ and $\Omega\backslash S$ (see Fig.~\ref{manytime}(a)).
Note that we can also regard $\Omega\backslash S$ as the ``bath" for the subsystem.

If we take $M=0$, for which $D_{kl}=\braket{\hat{A}_k,\hat{A}_l}$, or $M=1$ with $\hat{\Lambda}_1=\mbb{\hat{I}}$, for which $D=C$, we can see that inequality~\eqref{many0} holds with, e.g.,  setting $\mc{L}_1=\mc{L}_S+\mc{L}_I$ and $\mc{L}_2=\mc{L}_{\Omega\backslash S}$ since $\braket{\hat{A}_k,\hat{L}_2}=\mr{cov}(\hat{A}_k,\hat{L}_2)=0$.
For the small subsystem $S$, the bound in \eqref{many0} [or \eqref{mbbounds}] becomes much tighter than Eq.~\eqref{scalar_bound} [or $\mc{B}_\mr{MT}$] because $\mc{I}_{11}\ll I_Q$ in general.
Furthermore, the second term in the right-hand side of~\eqref{many0} [or \eqref{mbboundv} and \eqref{mbbounds}] also indicates a nontrivial consequence that the correlation between the subsystem and the rest suppresses the speed limit.
 {In the following, let us show some concrete examples.
}

 {
As a first example, we consider a speed limit for unitary dynamics whose Hamiltonian is given by $\hat{H}=\hat{H}_S+\hat{H}_I+\hat{H}_{B}=\hat{H}_{SI}+\hat{H}_{B}$ (note that $\mc{L}_{\Omega\backslash S}=-i[\hat{H}_B,\cdots]$).
In this case, we have 
\aln{\label{HSIHB}
\lrv{\npartial{\hat{A}}}\leq2\Delta A\Delta H_{SI} \sqrt{1-\phi_{H_{SI}H_{B}}^2},
}
which is convergent even for $\Delta H\ra\infty$ in the thermodynamic limit, since $\Delta H_{SI}$ is convergent.
Note that $|\phi_{H_{SI}H_{B}}|$  can be larger as we decrease the system size if the correlation length between the subsystem and the rest is finite.
%Nonetheless, this term is essential, especially for finite-size systems, improving the upper bound.
This means that we can further tighten the bound compared with the naive bound $2\Delta A\Delta H_{SI}$ using the correlation factor $\sqrt{1-\phi_{H_{SI}H_{B}}^2}$, especially when the bath of the system is finite.
Such a situation can occur in actual  experiments using artificial quantum systems, e.g.,   trapped ions.
Furthermore, this correction ensures that the bound in \eqref{HSIHB} becomes always better than the conventional bound in Eq.~\eqref{MTpure}, as detailed in next subsection.
}

 {
As a second example, consider an observable $\hat{A}$ for which $[\hat{H}_{I},\hat{A}]=0$.
In this case, we can take $\hat{H}_1=\hat{H}_S$ and $\hat{H}_2=\hat{H}_{IB}=\hat{H}_I+\hat{H}_B$,  obtaining 
\aln{\label{secondmanbound}
\lrv{\frac{d\braket{\hat{A}}}{dt}}\leq 2\Delta A\Delta H_{S}\sqrt{1-\phi_{H_{S}H_{IB}}^2},
}
where the fluctuation of $\hat{H}_S$, not $\hat{H}_{SI}$, appears.
}

 {As a third example, we take a specific spin-1/2 many-body system on $L$ lattice sites to demonstrate the bound \eqref{mbbounds}.
We consider unitary dynamics whose Hamiltonian is given by 
\aln{\label{Ham}
\hat{H}=\sum_{j=1}^{L-1}J\hat{\sigma}_j^z\hat{\sigma}_{j+1}^z+\sum_{j=1}^{L}g\hat{\sigma}_j^x+h\hat{\sigma}_j^z.
}
for a pure state.
We focus on the local magnetization at the first site, $\hat{A}=\hat{\sigma}_1^z$.
In this case, we can set $\hat{H}_1=g\hat{\sigma}_1^x$ and $\hat{H}_2=\hat{H}-\hat{H}_1$, since $[\hat{A},\hat{H}_2]=0$ (this choice is slightly different from the decompositions into $\hat{H}_{SI}\:(\hat{H}_{S})$ and $\hat{H}_{B}\:(\hat{H}_{IB})$ discussed in the previous paragraphs).
Figure~\ref{manytime}(b,c) shows the bound $\mc{B}_\mr{MB}$ in Eq.~\eqref{mbbounds}, the slightly loose bound that neglects the correlation, i.e., $\mc{B}_\mr{MB}'=2\Delta A\Delta H_1$, and the Mandelstam-Tamm bound $\mc{B}_\mr{MT}$ in Eq.~\eqref{MTpure}.
Figure~\ref{manytime}(b,c) shows that $\mc{B}_\mr{MB}$ and $\mc{B}_\mr{MB}'$ become much better than $\mc{B}_\mr{MT}$, which tends to be loose for larger $L$.
We also find that $\mc{B}_\mr{MB}$ becomes better than $\mc{B}_\mr{MB}'$ especially for relatively small $L$, indicating that the correlation factor can capture finite-size corrections for the speed limit.
}

 {As a final example, let us discuss the tradeoff relation for anti-commuting observables in many-body systems.
We consider a set of observables $\hat{A}_1,\cdots,\hat{A}_K$ acting on $S$, which are assumed to satisfy $\{\hat{A}_k,\hat{A}_l\}=0$ for $k\neq l$.
Then, taking $M=0$ and assuming initial pure states in isolated systems in ~\eqref{qvldec}, we obtain 
\aln{\label{manytrade}
\sum_{k=1}^K\frac{1}{\braket{\hat{A}_k^2}}\lrv{\npartial{\hat{A}_k}}^2
\leq 4\Delta H_{SI}^2(1-\phi_{H_{SI}H_B}^2),
}
which is a many-body counterpart for \eqref{tradegenanti}. For example, if we take mutually anti-commuting Pauli strings $\{\hat{\Sigma}_k\}$ on a local subsystem (such as $\{\hat{\sigma}_i^x,\hat{\sigma}_i^y,\hat{\sigma}_i^z\}$ with $i\in S$), we have 
\aln{\label{manytradep}
\sum_{k=1}^K\lrv{\npartial{\hat{\Sigma}_q}}^2\leq  4\Delta H_{SI}^2(1-\phi_{H_{SI}H_B}^2)
}
as a many-body counterpart for  \eqref{paulisum}.
This means that the energy scale that governs the tradeoff relation among anti-commuting local observables is the local energy fluctuation $\Delta H_{SI}$, not the entire energy fluctuation $\Delta H$.
}

 {
\subsection{Distinction compared with Ref.~\cite{PhysRevResearch.3.023074}}\label{distinct}
In this subsection, we discuss the distinct features of our results compared with results in Ref.~\cite{PhysRevResearch.3.023074}.
In Ref.~\cite{PhysRevResearch.3.023074}, they considered the reduced density matrix $\hat{\rho}_S(t)$ of $\hat{\rho}(t)$ for the local subsystem $S$ and showed
\aln{
\arccos(\sqrt{F_S(t_1,t_2)})\leq \int_{t_1}^{t_2}{\Delta H_{SI}}dt,
}
where $
F_S(t_1,t_2)=\lrs{\mr{Tr}\lrm{\sqrt{\sqrt{\hat{\rho}_S(t_1)}\hat{\rho}_S(t_2)\sqrt{\hat{\rho}_S(t_1)}}}}^2
$
is the mixed-state fidelity for the reduced density matrix.
}

\begin{figure}
\begin{center}
\includegraphics[width=\linewidth]{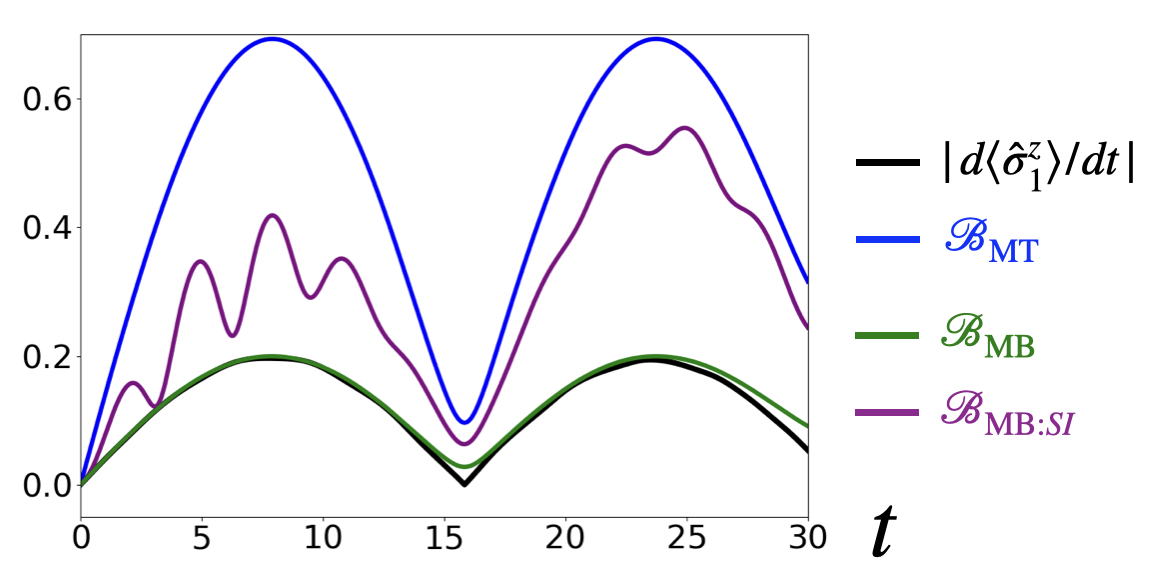}
\end{center}
\caption{ {
Time dependence of the speed of  an observable and speed limits for a spin system in Eq.~\eqref{Ham} with $L=12$.
We show the speed of $\braket{\hat{\sigma}_1^z}$ (black), the Mandelstam-Tamm bound $\mc{B}_\mr{MT}$ (blue), the bound   $\mc{B}_\mr{MB}$ based on $\Delta H_1=|g|\Delta\sigma_1^x$ (green), and the bound   $\mc{B}_{\mr{MB}:SI}$ based on $\Delta H_{SI}$ in \eqref{HSIHB}  (purple), where $\hat{H}_{SI}=g\hat{\sigma}_1^x+h\hat{\sigma}_1^z+J\hat{\sigma}_1^z\hat{\sigma}_2^z$. 
The bound $\mc{B}_\mr{MB}$ provides much tighter bounds compared with $\mc{B}_\mr{MT}$ and $\mc{B}_{\mr{MB}:SI}$.
We use $J=1, g=0.1$, $h=1$, and the initial state
{$\ket{\uparrow\downarrow\downarrow\downarrow\cdots}$}}}.
\label{encompare}
\end{figure}

 {
However, our results have some notable advantages compared with the results in  Ref.~\cite{PhysRevResearch.3.023074}.
First, we note that 
the bound in \eqref{HSIHB}, where $\Delta H_{SI}$ appears as in Ref.~\cite{PhysRevResearch.3.023074}, is just a specific example of \eqref{mbbounds}.
In particular, if an observable $\hat{A}$ of our interest commutes with $\hat{H}_2$ for isolated systems, \eqref{mbbounds} holds.
Then, instead of \eqref{HSIHB}, 
we can have nontrivial bounds using different energy fluctuations $\Delta H_1$ (such as $\Delta H_S$) from \eqref{mbbounds} if there is some commutativity structure between the observable and the partial Hamiltonian.
This is already demonstrated in the second and third examples (see inequality \eqref{secondmanbound} and Fig.~\ref{manytime}(b,c), respectively).
Furthermore, in the setting of the third example, we find that $\mathcal{B}_\mathrm{MB}$ based on $\Delta H_1=|g|\Delta \sigma_1^x$ can in fact  be much tighter than the bound \eqref{HSIHB} based on $\Delta H_{SI}$ (see Fig.~\ref{encompare}).
Note that the energy scale $\Delta H_1$ is obtained only after the consideration of the commutativity property of the observable and cannot be inferred from the discussion in Ref.~\cite{PhysRevResearch.3.023074}, which divides the subsystem and the rest from the outset and necessarily includes $\Delta H_{SI}$.
}

  {
 Second, Ref.~\cite{PhysRevResearch.3.023074} does not discuss observable-based inequalities. 
 In contrast, our approach offers nontrivial consequences by focusing on observable-based inequalities.
For a notable consequence, our inequalities show that the correlation between $\hat{H}_{SI}$ (or, more generally, $\hat{H}_1$) and $\hat{H}_B$ ($\hat{H}_2$) improves the bound for the first time.
This improvement cannot be directly understood from the results in Ref.~\cite{PhysRevResearch.3.023074}.
Furthermore, we find that the correction plays a qualitatively significant role in the following sense:
while $\Delta H_{SI}$ can be larger than $\Delta H$ in general (e.g., when the state is the eigenstate of $\hat{H}$, but not $\hat{H}_{SI}$), $\Delta H_{SI}\sqrt{1-\phi_{H_{SI}H_B}^2}$ cannot exceed $\Delta{H}$.
Indeed, we have
\aln{
&\Delta H^2-\Delta H_{SI}^2(1-\phi_{H_{SI}H_B}^2)\nonumber\\
&=\lrs{\Delta H_B+\frac{\mr{cov}(\hat{H}_{SI},\hat{H}_B)}{\Delta H_B}}^2\geq 0.
}
This means that our inequality in \eqref{HSIHB} [or, more generally, \eqref{mbbounds}] is always better than the conventional Mandelstam-Tamm inequality, while the result in Ref.~\cite{PhysRevResearch.3.023074} is not.
}

 {
For another important consequence, we elucidate general QVLs for multiple observables that are useful even in many-body systems (see \eqref{qvldec}).
This is demonstrated as the fourth example in the previous subsection, i.e., the tradeoff relation for the anti-commuting observables (e.g., certain Pauli strings) given in \eqref{manytrade} and \eqref{manytradep}.
These bounds cannot be immediately obtained from the metric-based method in Ref.~\cite{PhysRevResearch.3.023074} since we should account for relative relations among different observables.
}

\section{velocity limit based on local conservation law of probability}\label{SecVII}
While we have discussed QVLs based on the Fisher information so far, we can obtain a distinct type of velocity limits based on the local conservation law of  probability for multiple observables. 
These velocity limits, as for the recently found speed limits~\cite{PRXQuantum.3.020319} for each single observable, are especially advantageous in discussing macroscopic transitions.
As in the case for the bounds in Eqs.~\eqref{matrix_bound} and \eqref{scalar_bound}, our multiple-observable velocity limits can provide better bounds than the speed limit obtained previously~\cite{PRXQuantum.3.020319,PhysRevLett.130.010402}.
%Our results also provide some insights into the optimal transport theory~\cite{villani2009optimal},

\subsection{Review on the speed limit for single observables}
We first review the speed limit for a single observable (see Eq.~\eqref{fund}) based on Ref.~\cite{PRXQuantum.3.020319} with a slight modification of avoiding  the explicit introduction of the graph structure.
Thanks to this modification, the results in this manuscript can be applied to  a broader class of macroscopic systems, which were difficult previously~\cite{PRXQuantum.3.020319,PhysRevLett.130.010402} (see the final paragraph of this subsection).

To begin with, we consider a discrete system whose state space is described by the basis set $\{i\}$ and a probability distribution $\{p_i\}$ on it.
We assume that  the local conservation law of probability is satisfied, which leads to the continuity equation of probability,
\aln{
\fracd{p_i}{t}=-\sum_{j(\neq i)}J_{ji},
}
where $J_{ji}=-J_{ij}$ is the probability current from $i$ to $j$.
Note that this equation and the following formalism apply to various systems.
For example, for a classical stochastic system whose time evolution is given by $dp_i/dt=\sum_jR_{ij}p_j$ with a transition rate matrix $R$, $J_{ij}=R_{ij}p_j-R_{ji}p_i$. 
For a unitary quantum dynamics with $d\hat{\rho}/dt=-i[\hat{H},\hat{\rho}]$, 
we can take some fixed basis set  $\{\ket{i}\bra{i}\}$ and define $p_i=\braket{i|\hat{\rho}|i}$.
Then, we find $J_{ij}=-iH_{ij}\rho_{ji}+\mr{c.c}$, where $H_{ij}=\braket{i|\hat{H}|j}$ and $\rho_{ji}=\braket{j|\hat{\rho}|i}$.
We can also consider open quantum systems, as discussed in Ref.~\cite{PRXQuantum.3.020319}.

Let us consider an observable written as a function of $\{i\}$. 
We first focus on quantum systems and a single observable given by $\hat{A}=\sum_ia_i\ket{i}\bra{i}$ (for example, when $\ket{i}$ represents the Fock basis, we can take, e.g., the sum of the particle positions and the on-site interactions as $\hat{A}$). 
Then, we have
$
\npartial{\hat{A}}=-\sum_{i\neq j}a_iJ_{ji}=-\frac{1}{2}\sum_{i\neq j}(a_i-a_j)J_{ji},
$
where $J_{ij}=-J_{ji}$ is used.
Now, we introduce $r_{ij}\geq 0$, 
%which satisfies $r_{ij}=0$ if $(a_i-a_j)J_{ji}=0$ and $r_{ij}\neq 0$ otherwise. We note that $r_{ii}=0$.
which satisfies $r_{ij}>0$ if $J_{ij}\neq 0$ (equivalently, $J_{ij}=0$ if $r_{ij}=0$).
Then
\aln{
\npartial{\hat{A}}=-\frac{1}{2}\sum_{i\neq j,r_{ij}>0}r_{ij}(a_j-a_i)\frac{J_{ij}}{r_{ji}}=\braket{\nabla A,\mbf{u}}_r,
}
where
\aln{
\braket{\mbf{Y},\mbf{Z}}_r=\frac{1}{2}\sum_{i\neq j,r_{ij}>0}r_{ij}Y_{ij}Z_{ij},
}
$
(\nabla A)_{ij}=a_i-a_j,
$
and 
$
(\mbf{u})_{ij}=\frac{J_{ij}}{r_{ij}}.
$

Now, We can use the Cauchy-Schwarz inequality to find
\aln{\label{fund}
\lrv{\npartial{\hat{A}}} \leq \sqrt{\braket{\nabla A,\nabla A}_r}\sqrt{U},
}
where we define
\aln{
U=\braket{\mbf{u},\mbf{u}}_r.
}
This inequality leads to important consequences.
First, the factor $\braket{\nabla A,\nabla A}_r$ can be small even when $\|\hat{A}\|_\infty$ or $\Delta A$ can be large (see the example below).
Second, $U$ is often bounded by a physically relevant quantity.
For example, for unitary dynamics, by taking $r_{ij}=|H_{ij}\rho_{ji}|$, we find~\cite{PRXQuantum.3.020319}
\aln{\label{vbound}
U\leq 2\sum_{i\neq j}r_{ij}-\frac{2E_\mr{trans}^2}{\sum_{i\neq j}r_{ij}}\leq 2C_H-\frac{2E_\mr{trans}^2}{C_H},
}
where 
\aln{
E_\mr{trans}:=\sum_{i\neq j}H_{ij}=\braket{\hat{H}}-\sum_iH_{ii}
}
is the transition part of the energy and 
\aln{\label{CH}
C_H:=\max_i\sum_{j(\neq i)}|H_{ij}|
}
is the strength of the transition, which is easily known from the Hamiltonian.

Instead, if we consider a classical stochastic system, $U$ may be bounded by the entropy production rate $\dot{\Sigma}$ (note that we consider a classical observable $A=\{a_i\}$ instead of $\hat{A}$ in this case).
For example, if the system is attached to a single heat bath satisfying the detailed balance condition, we take $r_{ij}=R_{ij}p_j+R_{ji}p_i$ and obtain~\cite{PhysRevLett.121.070601}
\aln{
U\leq \frac{\dot{\Sigma}}{2},
}
where $\dot{\Sigma}:=\sum_{i\neq j}R_{ij}p_j\ln\frac{R_{ij}p_j}{R_{ji}p_i}$.
We note that this thermodynamic inequality is recently found to be related to the optimal transport problem~\cite{villani2009optimal}, where the entropy production is bounded using the square of the order-2 Wasserstein distance~\cite{PhysRevResearch.3.043093,PhysRevResearch.5.013017}.

For a simple example, let us consider far-from-equilibrium transport of a quantum particle in one dimension, whose Hamiltonian is given by
\aln{\label{singlep}
\hat{H}=\hat{V}_\mr{pot}+\sum_iJ_h\hat{b}_{i+1}^\dag\hat{b}_i+\mr{h.c.},
}
 where $\hat{b}_i$ is the annihilation operator of the particle at site $i$ and $\hat{V}_\mr{pot}$ determines an arbitrary on-site potential.
 Note that the basis $\ket{i}$ is taken as the basis for the single-particle position.
We focus on the position operator given by
$
\hat{x}=\sum_i i\ket{i}\bra{i}.
$
For an infinitely large system, $\Delta x$ diverges unboundedly in time (because of, e.g., diffusion), which makes the inequality $|d\braket{\hat{x}}/dt|\leq 2\Delta x\Delta H$ meaningless (note that $|d\braket{\hat{x}}/dt|$ is convergent). 

In contrast, if we apply \eqref{fund}, we find a convergent bound. 
If we take $r_{ij}=|H_{ij}\rho_{ji}|=J_h(\delta_{i+1,j}+\delta_{i-1,j})|\rho_{ji}|$, we find
$
\braket{\nabla A,\nabla A}_r=J_h\sum_i|\rho_{i,i+1}|\leq J_h,
$
where we have used $|\rho_{i,i+1}|\leq \sqrt{p_ip_{i+1}}\leq (p_i+p_{i+1})/2$.
We also have $C_H=2J_h$ in Eq.~\eqref{CH}.
Then, using~\eqref{vbound}, we obtain
\aln{\label{spsp}
\lrv{\npartial{\hat{x}}}\leq \sqrt{J_hU}\leq \sqrt{4J_h^2-E_\mr{trans}^2},
}
which provides a convergent bound even for a macroscopic system.
As discussed in Ref.~\cite{PRXQuantum.3.020319}, we can derive a similar speed limit useful for macroscopic transition even in (possibly interacting) many-particle systems.

Let us argue that the speed limit in \eqref{fund}  can be applied to  a broader class of macroscopic systems, which were difficult previously~\cite{PRXQuantum.3.020319,PhysRevLett.130.010402}.
This is because we can avoid the explicit introduction of the graph structure, which was done in  Refs.~\cite{PRXQuantum.3.020319,PhysRevLett.130.010402}.
In Refs.~\cite{PRXQuantum.3.020319,PhysRevLett.130.010402}, the speed limit for an observable is described by, e.g., the graph analogue of the Lipshitz constant of it ($\max_{(i,j)\in \mathcal{E}}|a_i-a_j|$, where $\mathcal{E}$ denotes the edge of the graph).
This factor is significantly changed if we alter the graph structure.
However, such change is physically unfavorable for a weakly perturbed system.
For example, QSLs for particles in Refs.~\cite{PRXQuantum.3.020319,PhysRevLett.130.010402} should be loosened even if we include very small long-range hoppings since they dramatically change the graph structure (especially $\mathcal{E}$).
However, our bound in Eq.~\eqref{fund} does not have such a problem, i.e., such  small long-range hoppings do not greatly change the bound.
This is because we use $\braket{\nabla A,\nabla A}_r$ instead of the Lipshitz constant of $A$; the fact that perturbation is weak is encoded through $r$.
For example, let us consider a single-particle quantum system with small long-range hopping amplitudes, where
 $|H_{ij}|$ becomes nonzero but small for $|i-j|\gg 1$.
Then, $r_{ij}=|H_{ij}\rho_{ji}|$ becomes automatically small for $|i-j|\gg 1$, which does not alter the right-hand side of \eqref{fund} much.

Finally, we mention that the above speed limit can be discussed in a continuous system.
Let us assume that the space coordinate is given by $\mbf{x}$, and that we can define a probability distribution $P(\mbf{x},t)$ on it, which satisfies the continuity equation
\aln{
\fracpd{P}{t}=-\nabla\cdot\mbf{J}
}
for a probability current $\mbf{J}(\mbf{x},t)$.
We consider an observable $A(\mbf{x})$ whose expectation value is given by 
\aln{
\braket{A(t)}=\int d\mbf{x}P(\mbf{x},t)A(\mbf{x}),
}
Assuming that $P\rightarrow 0$ for $|\mbf{x}|\ra\infty$, we obtain~\cite{PRXQuantum.3.020319}
\aln{
\npartial{{A}}=-\braket{\nabla A,\mbf{u}}_r^c.
}
Here,
\aln{
\braket{\mbf{Y},\mbf{Z}}_r^c:=\int_{r(\mbf{x})>0} d\mbf{x}r(\mbf{x})\mbf{Y}(\mbf{x})\cdot\mbf{Z}(\mbf{x})
}
and $\mbf{u}(\mbf{x})=\mbf{J}(\mbf{x})/r(\mbf{x})$,
where we assume $r(\mbf{x})>0$ if $\mbf{J}(\mbf{x})\neq 0$.
Then, the Cauchy-Schwarz inequality leads to 
\aln{\label{fundcont}
\lrv{\npartial{{A}}}\leq \sqrt{\braket{\nabla A,\nabla A}_r^c}\sqrt{U_c},
}
where $U_c=\braket{\mbf{u},\mbf{u}}_r^c$, in analogy with Eq.~\eqref{fund}.
Because $\braket{\nabla A,\nabla A}_r^c$ provides a small value compared with $\Delta A$, Eq.~\eqref{fundcont} is useful for macroscopic transitions in continuous systems as in Eq.~\eqref{fund}.

Importantly, $\braket{\mbf{u},\mbf{u}}_r^c$ is often bounded by some physical quantities.
For example, for the nonlinear Schrodinger equation that can describe, e.g., the mean-field dynamics of a Bose gas~\cite{RevModPhys.71.463},
we find $\braket{\mbf{u},\mbf{u}}_r^c\leq 2E_\mr{kin}$ by taking $r=P(\mbf{x})$~\cite{PRXQuantum.3.020319}. Here, $E_\mr{kin}=\int d\mbf{x}P|\nabla\theta|^2\hbar^2/(2m^2)$
is the kinetic energy of the Bose gas per particle, where the (normalized) wave function is given by $\psi(\mbf{x})=\sqrt{P(\mbf{x})}e^{i\theta(\mbf{x})}$ with a quantum phase $\theta(\mbf{x})$.
Another example is the thermodynamic Fokker-Planck equation~\cite{PhysRevE.97.062101,dechant2018current}, where we find $\braket{\mbf{u},\mbf{u}}_r^c\leq \mu T\dot{\Sigma}$ by taking $r=P(\mbf{x})$. 
Here, $\mu$ is the mobility of the particle, $T$ is the temperature, and $\dot{\Sigma}=\frac{\langle\mathbf{J}^2/P^2\rangle}{\mu T}$ is the entropy production rate for the Fokker-Planck system.

\subsection{Velocity limit for multiple observables}
We now discuss our new velocity limit for multiple observables based on the local conservation law of probability.
Similar to the derivation in Eqs.~\eqref{matrix_bound} and~\eqref{scalar_bound}, we obtain the matrix inequality (see Appendix~\ref{vel_loc_app})
\aln{
\vec{B}\vec{B}^\mathsf{T}\preceq U\mc{D}
}
and the  {equivalent} scalar inequality
\aln{\label{trans_bound}
\vec{B}^\mathsf{T}\mc{D}^{-1}\vec{B}\leq U
}
for a set of observables $\{\hat{A}_k\}$ given by $\hat{A}_k=\sum_i(a_k)_i\ket{i}\bra{i}$ for all $k$.
Here, we define
 a $K\times K$ matrix $\mc{D}$ whose components are given by
\aln{\label{dkl}
\mc{D}_{kl}=\braket{\nabla A_k,\nabla A_l}_r-\sum_\mu\braket{\nabla A_k,\nabla\Lambda_\mu }_r\braket{\nabla \Lambda_\mu,\nabla A_l}_r,
}where $\hat{\Lambda}_\mu$ are invariant observables that are assumed to have the form  $\hat{\Lambda}_\nu=\sum_i(\lambda_\nu)_i\ket{i}\bra{i}$ and satisfy the orthonormalized condition $\braket{\nabla\Lambda_\mu, \nabla\Lambda_{\mu'}}_r=\delta_{\mu\mu'}$.
We have also assumed that $\mc{D}$ has an inverse.
For $K=1$ and $M=0$, we recover Eq.~\eqref{fund}.

The velocity limit based on the probability current has many nontrivial consequences not obtained by the velocity limit based on the Fisher information.
Below, we detail examples of an asymmetric lower and upper bound and a tradeoff relation for  observables whose gradients are uncorrelated.

We first discuss the asymmetric lower and upper bound.
Choosing $K=2$ in \eqref{trans_bound}, we have
\aln{
\frac{\mc{D}_{22}\lrv{\npartial{\hat{A}_1}}^2-2\mc{D}_{12}\npartial{\hat{A}_1}\npartial{\hat{A}_2}+\mc{D}_{11}\lrv{\npartial{\hat{A}_2}}^2}{\mc{D}_{11}\mc{D}_{22}-|\mc{D}_{12}|^2}\leq U.
}

\begin{figure*}
\begin{center}
\includegraphics[width=\linewidth]{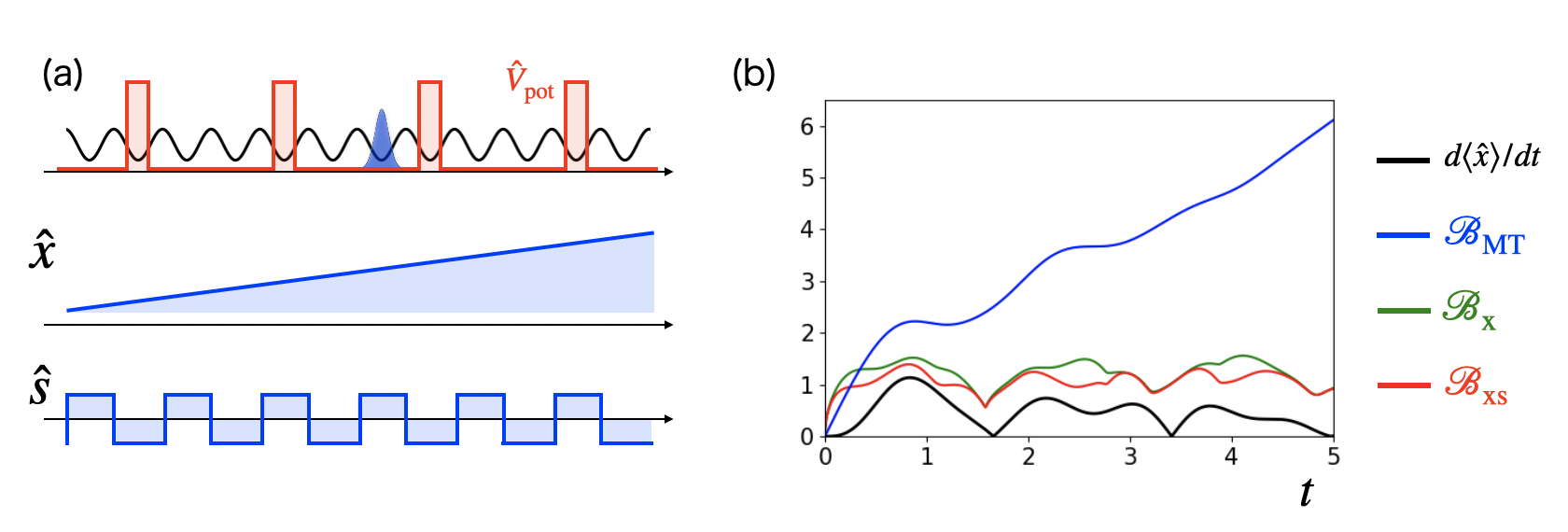}
\end{center}
\caption{
(a) Quantum transport of a single particle described by the Hamiltonian in Eq.~\eqref{singlep}. The potential $\hat{V}_\mr{pot}$ is chosen as $\hat{V}_\mr{pot}=\sum_iV_\mr{pot}^i\ket{i}\bra{i}$, where $V_\mr{pot}^i=5.0$  for $i\equiv 2\:(\mr{mod} 3)$ and $V_\mr{pot}^i=0$ otherwise.
The initial state is $\ket{\psi(0)}=\ket{i=100}$.
We observe the position of the particle $\hat{x}$ and the imbalance between odd and even sites, $\hat{s}$, whose gradients are uncorrelated with each other for an extended system (we use $L=200$ in the numerical simulation).
(b) The speed of $\hat{x}$ (black), the bound obtained from 
Eq.~\eqref{dpdp}, i.e., $\mc{B}_\mr{xs}=\sqrt{J_hU-|d\braket{\hat{s}}/dt|^2/4}$ (red), the bound
$ \mc{B}_\mr{x}=\sqrt{J_hU}$ in Eq.~\eqref{spsp} (green), and the bound $\mc{B}_\mr{MT}$ in Eq.~\eqref{MTpure} (blue).
We find that $\mc{B}_\mr{xs}$ provides a better bound than $\mc{B}_\mr{x}$, meaning that the knowledge of the speed of the imbalance can tighten the bound for the speed of the position.
We also find that 
$\mc{B}_\mr{MT}$ becomes divergent for large $t$.
We choose $J_h=1$.
}
\label{transport}
\end{figure*}

To gain physical insight, we here focus on the transport dynamics of a single particle on an extended one-dimensional lattice, while generalization to many-particle systems and higher dimensions is possible (as in the single-observable case).
We first consider the unitary dynamics whose Hamiltonian is given in Eq.~\eqref{singlep}.
We focus on two observables $\hat{A}_{1,2}=\sum_i(a_{1,2})_i\ket{i}\bra{i}$. 
After some calculation, we have an asymmetric lower and upper bound for the velocity of $\hat{A}_1$ in terms of that of $\hat{A}_2$ (see Appendix~\ref{twoc_app}),
\begin{comment}
\aln{
\frac{\mc{C}_{22}|\partial_t\braket{\hat{A}_1}|^2-2\mc{C}_{12}\partial_t\braket{\hat{A}_1}\partial_t\braket{\hat{A}_2}+\mc{C}_{11}|\partial_t\braket{\hat{A}_2}|^2}{\mc{C}_{11}\mc{C}_{22}-|\mc{C}_{12}|^2}\leq  4J_h^2-E_\mr{T}^2,
}
where $E_\mr{T}=\braket{\hat{H}-\hat{V}_\mr{pot}}$ is the energy for the transition part
and $\mc{C}_{kl}=\braket{\partial_i\hat{A}_k,\partial_i\hat{A}_l}=\mr{Tr}[\{\partial_i\hat{A}_k,\partial_i\hat{A}_l\}\hat{\rho}]/2$.
\end{comment}
\begin{align}\label{twocurrent}
\tilde{\chi} \mc{V}_2
& -\sqrt{(1-\tilde{\chi}^2)(J_hU-\mc{V}_2^2)}
\leq \mc{V}_1\nonumber\\
&\leq {\chi \mc{V}_2 +\sqrt{(1-\tilde{\chi}^2)(J_hU-\mc{V}_2^2)}},
\end{align}
where $U$ can be replaced with $4J_h-2E_\mr{trans}^2/J_h$ (see Eq.~\eqref{vbound}).
Here,  $\mc{V}_k:=(d{\hat{A}_k}/dt)/\sqrt{\mc{C}_{kk}}$ and $\tilde{\chi}:=\mc{C}_{12}/\sqrt{\mc{C}_{11}\mc{C}_{22}}\leq 1$ with
\aln{\label{corsimple}
&\mc{C}_{kl}\nonumber\\
&=\sum_i\frac{p_i}{2}\{(\nabla A_k)_{i,i+1}(\nabla A_{k'})_{i,i+1}+(\nabla A_k)_{i,i-1}(\nabla A_{k'})_{i,i-1}\}.
}
This inequality provides a tighter inequality than that in Ref.~\cite{PRXQuantum.3.020319} for a single observable in light of the knowledge of another observable.

Instead of the unitary time evolution, we can discuss classical stochastic dynamics for single particle transport. If we assume the transition rate matrix as $R_{ij}=R(\delta_{i,j+1}+\delta_{i,j-1})\:(i\neq j)$, we have 
\begin{align}
\tilde{\chi} \mc{V}_2
& -\sqrt{(1-\tilde{\chi}^2)(2RU-\mc{V}_2^2)}
\leq \mc{V}_1\nonumber\\
&\leq {\chi \mc{V}_2 +\sqrt{(1-\tilde{\chi}^2)(2RU-\mc{V}_2^2)}},
\end{align}
where $U$ can be replaced with $\dot{\Sigma}/2$.

Note that \eqref{corsimple} is often easily calculated.
For example, for $\hat{A}_1=\hat{x}=\sum_ii\ket{i}\bra
{i}$ and
$\hat{A}_2=\hat{x}^2=\sum_ii^2\ket{i}\bra
{i}$, we find
$\mc{C}_{12}=\mc{C}_{21}=2\braket{\hat{x}}$, $\mc{C}_{11}=1$, and $\mc{C}_{22}=4\braket{\hat{x}^2}+1$.

As a special case for the above argument, 
we can find the additivity principle for two observables whose gradients are uncorrelated.
For example, if we consider the position of the particle $\hat{A}_1=\hat{x}=\sum_ii\ket{i}\bra{i}$ and the even-odd imbalance of the density $\hat{A}_2=\hat{s}=\sum_i(4\lfloor i/2\rfloor -2i+1)\ket{i}\bra{i}$, we have $\mc{C}_{11}=1,\mc{C}_{12}=\mc{C}_{21}=0, \mc{C}_{22}=4$, and $\tilde{\chi}=0$.
Then, substituting them into~\eqref{twocurrent} shows that 
\aln{\label{dpdp}
\lrv{\npartial{\hat{x}}}^2+\frac{1}{4}\lrv{\npartial{\hat{s}}}^2\leq J_hU\leq 4J_h^2-E_\mr{trans}^2
}
for unitary dynamics, which is stronger than Eq.~\eqref{spsp}.
Similarly, the corresponding bound for classical stochastic systems reads
\aln{
\lrv{\npartial{\hat{x}}}^2+\frac{1}{4}\lrv{\npartial{\hat{s}}}^2\leq 2RU\leq R\dot{\Sigma}.
}

Figure~\ref{transport} verifies the bound obtained from 
Eq.~\eqref{dpdp}, i.e., $|d\braket{\hat{x}}/dt|\leq \mc{B}_\mr{xs}=\sqrt{J_hU-|d\braket{\hat{s}}/dt|^2/4}$ for a single-particle quantum system in Eq.~\eqref{singlep}.
We also compare it with the bounds 
$ \mc{B}_\mr{x}=\sqrt{J_hU}$ in Eq.~\eqref{spsp} and $\mc{B}_\mr{MT}$ in Eq.~\eqref{MTpure}.
We find that $\mc{B}_\mr{xs}$ provides a better bound than $\mc{B}_\mr{x}$, meaning that the knowledge of the speed of the imbalance can tighten the bound for the speed of the position.
We also find that 
$\mc{B}_\mr{MT}$ becomes divergent for large $t$.

We can obtain a similar velocity limit for multiple observables in continuous systems.
For this purpose, we introduce the velocity vector 
$\vec{B}_c=\{B_k\}_{k=1}^K$ with  $B_k=d\braket{A_k}/dt$ from $A_k(\mbf{x})$ and the invariant observables $\{\Lambda_\mu(\mbf{x})\}_{\mu=1}^M$ satisfying $\braket{\nabla\Lambda_\mu,\nabla\Lambda_\nu}_r^c=\delta_{\mu\nu}$.
We then have $\vec{B}_c\vec{B}_c^\mathsf{T}\preceq U_c\mc{D}_c$ and
\aln{
\vec{B}_c\mc{D}_c^{-1}\vec{B}_c^\mathsf{T}\leq U_c,
}
 where $\mc{D}_c$ is obtained by the replacement of $\braket{\cdots,\cdots}_r$ in 
Eq.~\eqref{dkl} with $\braket{\cdots,\cdots}_r^c$.
Again, we remind that $U_c$ is bounded by physical quantities, e.g., the kinetic energy for  the nonlinear Schr\"{o}dinger equation and the entropy production rate for the Fokker-Planck equation.

\begin{comment}
Finally, we briefly mention the possible application of the velocity limit in terms of (I) and (IV) introduced in the introduction.
(I) The speed limit under invariant observables systems will  be relevant in systems with additional local conservation laws. 
For example, we can consider a system with dipolar-moment conservation~\cite{}, for which $\nabla\hat{\Lambda}$ becomes nontrivial.
(IV) For many-body systems, 
\end{comment}

\begin{comment}
\section{Other extensions}\label{SecVII'}
Here, we show that the concept of velocity limits is not limited to the inequalities discussed in the previous sections.
We argue that the velocity limits can also be found for continuous systems.
We also show that similar bounds are obtained even for fluctuations of multiple observables.

\subsection{Velocity limits in the continuous space}
Let us denote the coordinate of the continuous space by $\mbf{x}$.
Defining the 
\end{comment}

%\subsection{Velocity limits for fluctuations}

\section{Conclusion and outlook}\label{SecVIII}
In this paper, we have introduced the notion of the quantum velocity limits for multiple observables for the first time.
We show that the velocity limits provide tighter bounds due to the knowledge of other observables than the conventional speed limits for a single observable. 
As a first type of the quantum velocity limits, we have derived the universal information-theoretical bound for the velocity vector of observables (\eqref{matrix_bound} and \eqref{scalar_bound}).
Remarkably, this velocity limit has various consequences, including the speed limits tightened by conserved quantities~\eqref{inv_speed}, the nontrivial lower bound for the velocity of an observable~\eqref{uplo}, the tradeoff relation between uncorrelated observables~\eqref{uncor} and its relation to anti-commutivity, and the convergent  {velocity limits for observables in general many-body systems~\eqref{qvldec}.}
We also show the distinct type of the multiple-observable velocity limit based on the local conservation law of probability~\eqref{trans_bound}.
These velocity limits serve as a hitherto unknown concept toward a universal theory of far-from-equilibrium quantum dynamics of multiple observables.

While we have discussed our velocity limits for the expectation values of observables, 
it is intriguing to extend the concept to other quantities.
Recently,  speed limits are investigated for, e.g., the correlations~\cite{pandey2022speed,hasegawa2023thermodynamic,carabba2022quantum,hornedal2023geometric},   {fluctuations~\cite{hamazaki2023limits}}, 
entropies~\cite{PhysRevResearch.2.013161,PhysRevE.103.032105,mohan2022quantumic,PRXQuantum.3.020319,campaioli2022resource,PhysRevA.106.012403,pandey2023fundamental}, and operator complexity~\cite{hornedal2022ultimate,hornedal2023geometric}.
Even for those quantities, the knowledge of other observables, such as conserved quantities, may be used to tighten the bounds.
Furthermore, it is also interesting to investigate how unique quantum properties affect the dynamics of those quantities, like the non-equilibrium tradeoff relation for nontrivially anti-commuting observables found in this work.

Since practical quantum velocity limits are obtained for many-body systems, we may be able to apply them to evaluate the timescale of quantum many-body dynamics~\cite{PhysRevLett.111.140401,goldstein2015extremely,reimann2016typical,de2018equilibration,PhysRevB.99.174313,PhysRevX.7.031027}.
This is relevant for the problem of thermalization of isolated quantum  systems~\cite{eisert2015quantum}, which offers a foundation of quantum statistical mechanics.
Interestingly, our strategy of considering local observables  in locally interacting systems, not the entire state, is analogous to the standard method in discussing this problem.
We leave it a future problem to discuss the detailed relation between our bounds and the timescale of thermalization.

Though we have mainly discussed examples in unitary quantum systems, the velocity limits can be universally applied to open quantum systems, classical stochastic systems, and even nonlinear systems, such as population dynamics.
It is left as a future issue to examine such systems on the basis of our bounds.
It is also essential to discuss how our velocity limits can be used to evaluate the controllability of quantum systems under the knowledge of, e.g., conserved quantities and Hamiltonian structures.
This may be accomplished by combining our results with  optimization techniques developed in other fields~\cite{PhysRevX.12.011039}.

Finally, as seen in Sec.~\ref{SecII}, the information-theoretical quantum velocity limits can also be regarded as a generalized version of the quantum Cram\'er-Rao bound, the fundamental bound in quantum information theory.
In contrast with the conventional quantum Cram\'er-Rao bound, our bound accounts for multiple observables of a system, such as conserved quantities.
Then, our rigorous inequality will have profound implications even outside the context of non-equilibrium dynamics, e.g., quantum metrology under the knowledge of other observables.
Indeed, a similar motivation has recently been appreciated in multi-parameter quantum metrology~\cite{PhysRevX.10.031023}.
Thus,  our formalism may be beneficial for such an application, too.

\begin{acknowledgments}
We are grateful to Francesco Albarelli for letting us know relevant references in the field of quantum metrology.
 {We also thank the anonymous referee for informing us of the alternative proof of Eq.~\eqref{scalar_bound} given in Appendix~\ref{AppA2}.}
The numerical calculations were carried out with the help of QUSPIN~\cite{SciPostPhys.2.1.003,weinberg2019quspin}.
This work was supported by JST ERATO Grant Number JPMJER2302, Japan.
%R.H. was supported by JST ERATO-FS Grant Number JPMJER2204, Japan.

\end{acknowledgments}

\appendix

\section{Details of Eqs.~\eqref{matrix_bound} and \eqref{scalar_bound} in the main text}\label{sec:multiproof}
\subsection{Proof of Eqs.~\eqref{matrix_bound} and \eqref{scalar_bound}  {and their generalization}}
We first give a proof of Eqs.~\eqref{matrix_bound} and \eqref{scalar_bound} in the main text.
To keep the derivation as general as possible, we consider the case with multiple observables and multi-parameters simultaneously.
More precisely, we focus on a set of $K$ observables $\hat{A}_1,\cdots,\hat{A}_K$ and a set of $Z$ parameters $y_1,\cdots, y_Z$. We take $Z=1$ and $y_1=t$ at the end to obtain the formula in the main text.
We define 
\aln{\label{bkz}
B_{kz}:=\partial_{y_z}\braket{\hat{A}_k}=\braket{\hat{A}_k,\hat{L}_z},
}
where $\hat{L}_z$ is the symmetric logarithmic derivative (SLD) for the parameter $y_z$.

We assume that the set of $M$ invariant observables $\hat{\Lambda}_\mu$ exists and satisfies  {$\partial_{y_z}\braket{\hat{\Lambda}_\mu}-\braket{\partial_{y_z}{\hat{\Lambda}_\mu}}=0$} for all $z$.
As discussed in the main text, we can assume the orthonormalization condition $\braket{\hat{\Lambda}_\mu,\hat{\Lambda}_\nu}=\delta_{\mu\nu}$.
Then,  ${B}_{kz}$ is given by
\aln{
B_{kz}=\braketL{\hat{A}_k-\sum_\mu f_{k\mu}\hat{\Lambda}_{\mu},\hat{L}_z}
}
for arbitrary $f_{k\mu}$, which we assume real.

Now, we introduce two real vectors $(a_1,\cdots,a_K)$ and $(b_1,\cdots,b_Z)$ and consider
\aln{
\sum_{kz}a_kb_zB_{kz}=\braketL{\sum_ka_k\lrs{\hat{A}_k-\sum_\mu f_{k\mu}\hat{\Lambda}_{\mu}},\sum_zb_z\hat{L}_z}.
}
Using the Cauchy-Schwarz inequality, we have
\aln{\label{df}
\sum_{kzk'z'}a_ka_{k'}b_zb_{z'}B_{kz}B_{k'z'}
\leq 
\sum_{kzk'z'}a_ka_{k'}b_zb_{z'}D^f_{kk'}\mc{I}_{zz'},
}
where
\aln{
D^f_{kk'}:=\braketL{\hat{A}_k-\sum_\mu f_{k\mu}\hat{\Lambda}_{\mu},\hat{A}_{k'}-\sum_\mu f_{k'\mu}\hat{\Lambda}_{\mu}}
}
and 
\aln{
\mc{I}_{zz'}:=\braket{\hat{L}_z,\hat{L}_{z'}}
}
is the SLD quantum Fisher information matrix.

We can choose $f_{k\mu}$ as $\braket{\hat{A}_k,\hat{\Lambda}_\mu}$ to optimize the right-hand side of inequality~\eqref{df}, for which $D^f$ becomes $D$ introduced in the main text.
In fact, we can prove the matrix inequality $D\preceq D^f$ for all $f_{k\mu}$ since
\aln{
\sum_{kk'}a_ka_k'(D^f_{kk'}-D_{kk'})=\sum_\mu\lrv{\sum_ka_k(\braket{\hat{A}_k,\hat{\Lambda}_\mu}-f_{k\mu})}^2\geq 0.
}
We thus obtain 
\aln{\label{dfopt}
\sum_{kzk'z'}a_ka_{k'}b_zb_{z'}B_{kz}B_{k'z'}
\leq 
\sum_{kzk'z'}a_ka_{k'}b_zb_{z'}D_{kk'}\mc{I}_{zz'}.
}

 {
Now, the condition \eqref{dfopt} can be recast as 
\aln{
\min_{\vec{a},\vec{b}}\lrm{
(\vec{a}^\mathsf{T}D\vec{a})(\vec{b}^\mathsf{T}\mathcal{I}\vec{b})-
(\vec{a}^\mathsf{T}B\vec{b})^2}\geq 0,
}
where $B=\{B_{kz}\}$.
By introducing an auxiliary real variable $\lambda$, we can rewrite the above equation as
\aln{
\min_{\lambda,\vec{a},\vec{b}}\lrm{\lambda^2+
2\lambda(\vec{a}^\mathsf{T}B\vec{b})+
(\vec{a}^\mathsf{T}D\vec{a})(\vec{b}^\mathsf{T}\mathcal{I}\vec{b})}\geq 0.
}
For a fixed $\lambda$ and $\vec{b}$, we can find that $\vec{a}=-\frac{\lambda D^{-1}B\vec{b}}{\vec{b}^\mathsf{T}\mathcal{I}\vec{b}}$ minimizes the left-hand side, obtaining a novel matrix inequality
\aln{\label{generalDI}
B^\mathsf{T}D^{-1}{B}\preceq \mathcal{I}.
}
Instead, we find that $\vec{b}=-\frac{\lambda \mathcal{I}^{-1}B^\mathsf{T}\vec{a}}{\vec{a}^\mathsf{T}D\vec{a}}$ minimizes the left-hand side for a fixed $\lambda$ and $\vec{a}$, obtaining
\aln{\label{generalID}
B\mathcal{I}^{-1}{B}^\mathsf{T}\preceq D,
}
where we have assumed that $\mc{I}$ has an inverse.
Note that inequalities \eqref{generalDI} and \eqref{generalID} are equivalent because they are the Schur complements for a large matrix
\aln{
\begin{pmatrix}
D & {B} \\
{B}^\mathsf{T} & \mc{I} \\
\end{pmatrix}.
}
}

 {
Inequalities~\eqref{generalDI} and \eqref{generalID} give the inequalities generalizing our results discussed in the main text.
We here take $Z=1$, with which we have $\mc{I}_{11}=I_Q$.
Then, \eqref{generalID} reduces to the matrix inequality in Eq.~\eqref{matrix_bound} in the main text,
\aln{
\vec{B}\vec{B}^\mathsf{T}\preceq DI_Q.
}
As an equivalent inequality, \eqref{generalDI} reduces to the scalar inequality  in Eq.~\eqref{scalar_bound} in the main text,
\aln{\label{scalareq}
\mathcal{K}(\{A_k\};\{\Lambda_\mu\}):=\vec{B}^\mathsf{T}D^{-1}\vec{B}\leq I_Q.
}
}

 {
Note that, if we instead consider $K=1$ and $Z\neq 1$, we have a matrix inequality
\aln{
\vec{\mathsf{B}}\vec{\mathsf{B}}^\mathsf{T}\preceq \mc{I}\lrs{\braket{\hat{A}^2}-\sum_\mu\braket{\hat{A},\hat{\Lambda}_\mu}^2}
}
and an equivalent scalar inequality
\aln{
\vec{\mathsf{B}}^\mathsf{T}\mc{I}^{-1}\vec{\mathsf{B}}\leq \lrs{\braket{\hat{A}^2}-\sum_\mu\braket{\hat{A},\hat{\Lambda}_\mu}^2},
}
where $\vec{\mathsf{B}}=(B_{11},\cdots,B_{1Z})^\mathsf{T}$.
}

 {
\subsection{Alternative proof of \eqref{scalar_bound} for $M=1$ with $\hat{\Lambda}_1=\hat{\mathbb{I}}$}\label{AppA2}
In this subsection, we show that \eqref{scalar_bound} for $M=1$ with $\hat{\Lambda}_1=\hat{\mathbb{I}}$ is also obtained by combining the well-known inequality $\lrv{\frac{d\braket{\hat{A}}}{dt}}^2\leq \Delta A^2{I_Q}$ and the ideas introduced in our manuscript, i.e., introduction of velocity vector for multiple observables and suitable optimization.
To see this, we take $\hat{A}$ as a linear combination of multiple observables of our interest as  $\hat{A}=\sum_kz_k\hat{A}_k$.
We  then have 
\aln{
\lrv{\frac{d\braket{\hat{A}}}{dt}}^2=(\vec{z}^\mathsf{T}\vec{B})^2
}
and
\aln{
\Delta A^2=\vec{z}^\mathsf{T}C\vec{z}.
}
Now, we notice
\aln{
\max_{\vec{z}}\frac{\lrv{\frac{d\braket{\hat{A}}}{dt}}^2}{\Delta A^2}
&=\max_{\vec{z}}\frac{((C^{1/2}\vec{z})^\mathsf{T}C^{-1/2}\vec{B})^2}{\vec{z}^\mathsf{T}C\vec{z}}=\vec{B}^\mathsf{T}C^{-1}\vec{B},
}
where we have used the Cauchy-Schwarz inequality, whose equality condition is satisfied for $\vec{z}\propto C^{-1}\vec{B}$.
This implies $\vec{B}^\mathsf{T}C^{-1}\vec{B}\leq I_Q$, as desired.
We stress that while this derivation seems to originate from the single-observable speed limit, we additionally need to introduce velocity vector and solve suitable optimization problem for multiple variables (associated with multiple observables). 
Therefore, our QVLs for multiple observables cannot be regarded as a mere consequence of QSLs for a single observable, 
offering novel and useful inequalities.
We also stress that the ideas of introducing velocity vectors and solving optimization problem are first employed to understand quantum dynamics in our manuscript.
}

\section{Proof of the bound for a finite time interval}\label{appfintime}
We here show Eq.~\eqref{finitetime} in the main text.
For this purpose, we take an arbitrary real vector $\{a_k\}$ and consider 
\aln{
\sum_ka_k\mathfrak{B}_k=\int_0^T\sum_ka_k{B}_k(t)dt.
}
We then have
\aln{
\sum_{kk'}a_ka_{k'}\mathfrak{B}_k\mathfrak{B}_{k'}&=\lrv{\sum_ka_k\mathfrak{B}_k}^2\nonumber\\
&\leq T\int_0^T\lrs{\sum_ka_k{B}_k(t)}^2dt\nonumber\\
&\leq T\int_0^T \sum_{kk'}a_ka_{k'}D(t)_{kk'}I_Qdt\nonumber\\
&\leq T^2\sum_{kk'}a_ka_{k'} \av{D_{kk'}I_Q}.
}
Here, we have used the Cauchy-Schwarz inequality from the first to the second line and the instantaneous velocity limit from the second to the third line.
Since this inequality holds for all $\{a_k\}$, we have
\aln{
\mathfrak{B}\mathfrak{B}^\mathsf{T}\preceq T^2\av{DI_Q}.
}

\section{Proof of the equality condition for a single spin-1/2 system}\label{equality_anyspin1}
Here, we discuss the proof of the equality conditions discussed in Sec.~\ref{singlespin}.
We first consider the case with $\hat{H}=g\hat{\sigma}^x$ and $\hat{A}=c_I\hat{\mbb{I}}+\sum_{\alpha=x,y,z}c_\alpha\hat{\sigma}^\alpha$.
A straightforward calculation leads to 
\aln{
\frac{d\braket{\hat{A}}}{dt}&=2g(-c_y\braket{\hat{\sigma}^z}+c_z\braket{\hat{\sigma}^y})\nonumber\\
\braket{\hat{A}^2}-\sum_{\mu=1,2}\braket{\hat{A},\hat{\Lambda}_\mu}^2&
=c_y^2+c_z^2-\frac{(c_y\braket{\hat{\sigma}^y}+c_z\braket{\hat{\sigma}^z})^2}{1-\braket{\hat{\sigma}^x}^2}\nonumber\\
\Delta H^2&=g^2(1-\braket{\hat{\sigma}^x}^2),
}
where  $\hat{\Lambda}_1=\hat{\mbb{I}}$ and 
\aln{
\hat{\Lambda}_2=\frac{\hat{H}-\braket{\hat{H}}}{\sqrt{\braket{(\hat{H}-\braket{\hat{H}})^2}}}.
}
Thus, the equality 
\aln{\label{eco}
\lrv{\frac{d\braket{\hat{A}}}{dt}}=2\Delta H\sqrt{\braket{\hat{A}^2}-\sum_{\mu=1,2}\braket{\hat{A},\hat{\Lambda}_\mu}^2}
}
holds true, where we have used $\braket{\hat{\sigma}^x}^2+\braket{\hat{\sigma}^y}^2+\braket{\hat{\sigma}^z}^2=1$.

Now, we consider a more general Hamiltonian $\hat{H}$.
For two-level systems, the Hamiltonian can always be represented as
\aln{
\hat{H}=g\hat{V}^\dag\hat{\sigma}^x\hat{V}+h\hat{\mbb{I}},
}
where $\hat{V}$ is a untiary operator.
Then, for the state $\ket{\psi}$, we have
\aln{
\frac{d\braket{\hat{A}}}{dt}&=\braket{i[\hat{H},\hat{A}]}=\braket{\psi'|i[g\hat{\sigma}^x,\hat{A}']|\psi'},
}
where $\ket{\psi'}=\hat{V}\ket{\psi}$ and $\hat{A}'=\hat{V}\hat{A}\hat{V}^\dag$.
Using the above discussion, we have
\aln{
\lrv{\frac{d\braket{\hat{A}}}{dt}}=2\Delta H'\sqrt{\braket{\hat{A}'^2}'-\sum_{\mu=1,2}\braket{\hat{A}',\hat{\lambda}_\mu}'^2},
}
where $\braket{\cdots}'=\braket{\psi'|\cdots|\psi'}$ and
\aln{
\Delta H'^2=g^2(1-\braket{\hat{\sigma}^x}'^2),
}
$\lambda_1=\hat{\mbb{I}}$, and
\aln{
\hat{\lambda}_2=\frac{\hat{\sigma}^x-\braket{\hat{\sigma}^x}'}{\sqrt{\braket{(\hat{\sigma}^x-\braket{\hat{\sigma}^x}')^2}}}.
}
Now, straightforward calculation leads to $\Delta H'=\Delta H$, $\braket{\hat{A}'^2}'=\braket{\hat{A}^2}$, and
$\braket{\hat{A}',\hat{\lambda}_\mu}'=\braket{\hat{A},\hat{\Lambda}_\mu}$.
Then, we find the equality condition in the form of Eq.~\eqref{eco} holds even in this general case.

\section{Other bounds based on invariant observables}\label{otherinv}
Here, we discuss two additional applications of our  bounds based on invariant observables in Eq.~\eqref{inv_speed} in the main text, assuming the unitary time evolution $\frac{d\hat{\rho}(t)}{dt}=-i[\hat{H}(t),\hat{\rho}(t)]$.
The first one is to consider the (instantaneous) energy eigenstates of $\hat{H}(t)$, $\ket{E_\alpha}\:(\alpha=1,2,\cdots)$.
We can take
\aln{
\hat{\Lambda}_\mu=\frac{\ket{E_\mu}\bra{E_\mu}}{\sqrt{\braket{E_\mu|\hat{\rho}|E_\mu}}},
}
which satisfy the orthonormalization condition.
We then have
\aln{
\lrv{\frac{d\braket{\hat{A}}}{dt}}\leq \sqrt{\braket{\hat{A}^2}-\sum_\mu\frac{\braket{E_\mu|\{\hat{\rho},\hat{A}\}|E_\mu}^2}{4\braket{E_\mu|\hat{\rho}|E_\mu}}}\sqrt{I_Q}.
}

Another example is to take the power of $\hat{\rho}\:(=\hat{\rho}(t))$
assuming the mixed state, since $d\braket{\hat{\rho}^n}/dt=0$ due to the conservation of purity.
We here consider the simplest case
$\hat{\Lambda}_1=\mbb{\hat{I}}$ and
$\hat{\Lambda}_2=({\hat{\rho}}-\braket{{\hat{\rho}}})/\sqrt{\braket{({\hat{\rho}}-\braket{{\hat{\rho}}})^2}}$.
In this case, we find
\aln{
\lrv{\frac{d\braket{\hat{A}}}{dt}}\leq \Delta A\sqrt{I_Q}\sqrt{1-\phi_{A\rho}^2}.
}

\begin{comment}
Finally, if we consider the diagonalize the general mixed state as $\hat{\rho}=\sum_mp_m\ket{p_m}\bra{p_m}$, 
we can take
\aln{
\hat{\Lambda}_\mu=\frac{\ket{p_\mu}\bra{p_\mu}}{\sqrt{p_\mu}},
}
which again satisfies the orthonormalization condition.
With this choice, we have
\aln{
|\partial_t\braket{\hat{A}}|&\leq \sqrt{ \braket{\hat{A}^2}-\sum_\mu p_\mu\braket{p_\mu|\hat{A}|p_\mu}^2}\sqrt{I_Q}\nonumber\\
&=\sqrt{\sum_\mu p_\mu\Delta A_\mu^2}\sqrt{I_Q},
}
where 
\aln{
\Delta A_\mu^2=\braket{p_\mu|\hat{A}^2|p_\mu}-\braket{p_\mu|\hat{A}|p_\mu}^2.
}
Note that this bound is tighter than the conventional bound since $\sum_\mu p_\mu\Delta A_\mu^2\leq \Delta A^2$.
\end{comment}

\section{Proof of Eq.~\eqref{pnoeq}}\label{pnoeqproof}
We express the Hamiltonian as 
\aln{
\hat{H}=h_I\hat{\mbb{I}}+h_x\hat{\mbb{\sigma}}^x+h_y\hat{\mbb{\sigma}}^y+h_z\hat{\mbb{\sigma}}^z.
}
We have
\aln{
&\Delta H^2=h_x^2(1-\braket{\hat{\mbb{\sigma}}^x}^2)
+h_y^2(1-\braket{\hat{\mbb{\sigma}}^y}^2)
+h_z^2(1-\braket{\hat{\mbb{\sigma}}^z}^2)\nonumber\\
&-2h_xh_y\braket{\hat{\mbb{\sigma}}^x}\braket{\hat{\mbb{\sigma}}^y}-2h_xh_z\braket{\hat{\mbb{\sigma}}^x}\braket{\hat{\mbb{\sigma}}^z}
-2h_zh_y\braket{\hat{\mbb{\sigma}}^z}\braket{\hat{\mbb{\sigma}}^y}
}
and
\aln{
\lrv{\npartial{\hat{\sigma}^x}}^2
&=4(h_y\braket{\hat{\sigma}^z}-h_z\braket{\hat{\sigma}^y})^2\nonumber\\
\lrv{\npartial{\hat{\sigma}^y}}^2
&=4(h_x\braket{\hat{\sigma}^z}-h_z\braket{\hat{\sigma}^x})^2\nonumber\\
\lrv{\npartial{\hat{\sigma}^z}}^2
&=4(h_y\braket{\hat{\sigma}^x}-h_x\braket{\hat{\sigma}^y})^2.
}
Using $\braket{\hat{\sigma}^x}^2+\braket{\hat{\sigma}^y}^2+\braket{\hat{\sigma}^z}^2=1$, we obtain Eq.~\eqref{pnoeq}.

\section{Proof of Eq.~\eqref{many0}}\label{manproof}
To prove Eq.~\eqref{many0}, we first note that
\aln{
B_k&=\braketL{\hat{A}_k-\sum_\mu\braket{\hat{A}_k,\hat{\Lambda}_\mu}\hat{\Lambda}_\mu,\hat{L}}\nonumber\\
&=\braketL{\hat{A}_k-\sum_\mu\braket{\hat{A}_k,\hat{\Lambda}_\mu}\hat{\Lambda}_\mu,\hat{L}_1}\nonumber\\
&=\braketL{\hat{A}_k-\sum_\mu\braket{\hat{A}_k,\hat{\Lambda}_\mu}\hat{\Lambda}_\mu,\hat{L}_1-\sum_\mu\braket{\hat{L}_1,\hat{\Lambda}_\mu}\hat{\Lambda}_\mu}
}
for $1\leq k\leq K$.

We next consider $(K+1)$-dimensional vector $\vec{B}'$, which is given by
\aln{
\vec{B}'=\lrs{\vec{B},\braketL{\hat{L}_2-\sum_\mu\braket{\hat{L}_2,\hat{\Lambda}_\mu}\hat{\Lambda}_\mu,\hat{L}_1-\sum_\mu\braket{\hat{L}_1,\hat{\Lambda}_\mu}\hat{\Lambda}_\mu}}^{\mathsf{T}}.
}
In a manner similar to Appendix~\ref{sec:multiproof}, we have 
\aln{
({\vec{B}'})^\mathsf{T}(D')^{-1}\vec{B}'\leq \mc{F}_{11},
}
where $\mc{F}_{zz'}$ is given in Eq.~\eqref{genfis} and 
$D'$ is block-diagonalized as
\aln{\label{dprime}
D'=
\begin{pmatrix}
D & \vec{0} \\
\vec{0}^\mathsf{T} & \mc{F}_{22} \\
\end{pmatrix},
}
where we have used the relation
\aln{
0&=\braketL{\hat{A}_k-\sum_\mu\braket{\hat{A}_k,\hat{\Lambda}_\mu}\hat{\Lambda}_\mu,\hat{L}_2}\nonumber\\
&=\braketL{\hat{A}_k-\sum_\mu\braket{\hat{A}_k,\hat{\Lambda}_\mu}\hat{\Lambda}_\mu,\hat{L}_2-\sum_\mu\braket{\hat{L}_2,\hat{\Lambda}_\mu}\hat{\Lambda}_\mu}.
}
Then we have
\aln{
(D')^{-1}=
\begin{pmatrix}
D^{-1} & \vec{0} \\
\vec{0}^\mathsf{T} & \mc{F}_{22}^{-1} \\
\end{pmatrix}.
}
From Eq.~\eqref{dprime}, we immediately obtain Eq.~\eqref{many0}:
\aln{
\vec{B}^\mathsf{T}D^{-1}\vec{B}\leq \mc{F}_{11}-\frac{\mc{F}_{12}^2}{\mc{F}_{22}}=\frac{\mr{Det}[\mc{F}]}{\mc{F}_{22}}.
}

While we have subtracted $\sum_\mu\braket{\hat{L}_k,\hat{\Lambda}_\mu}\hat{\Lambda}_\mu$ from $\hat{L}_k$ above, a similar discussion can be made without this subtraction.
In this case, we obtain
\aln{
\vec{B}^\mathsf{T}D^{-1}\vec{B}\leq \mc{I}_{11}-\frac{\mc{I}_{12}^2}{\mc{I}_{22}}=\frac{\mr{Det}[\mc{I}]}{\mc{I}_{22}}.
}
instead.

\section{Velocity limit for multiple observables based on the local conservation law of probability}\label{vel_loc_app}
Here, we discuss the detail of the velocity limit for multiple observables based on the local conservation law of probability.
%We first briefly review the derivation of the speed limit for a single observable. We then discuss the speed limit for multiple observables and show Eq.~(@@@) in the main text.

\subsection{Derivation of the case for multiple observables}
%We generalize the current-based speed limit discussed above to multiple observables.
While we can consider both classical and quantum systems, we here use the quantum description to keep generality.
We assume that (invariant) observables are diagonal in the $i$-basis, i.e.,
\aln{
\hat{A}_k=\sum_i(a_k)_i\ket{i}\bra{i}
}
and
\aln{
\hat{\Lambda}_\nu=\sum_i(\lambda_\nu)_i\ket{i}\bra{i}.
}
We also assume the orthonormalization condition for $\nabla\Lambda_\nu$, instead of $\hat{\Lambda}_\nu$. 
That is,
\aln{
\braket{\nabla\Lambda_\mu,\nabla\Lambda_\nu}_r=\delta_{\mu\nu}.
}

The proof of the multiple-observable bound is similar to that in Appendix~\ref{sec:multiproof}.
Let us consider the general  case with multi-parameters $y_1,\cdots,y_Z$ as in Appendix~\ref{sec:multiproof} and assume the continuity equation 
\aln{
\partial_{y_z}p_i=-\sum_{j(\neq i)}J_{ji}^z,
}
We first notice that $B_{kz}$ in Eq.~\eqref{bkz}  is given by
\aln{
B_{kz}=\braket{\nabla A_k,\mbf{u}_z}_r
}
with $(\mbf{u}_z)_{ij}=J_{ij}^z/r_{ij}$.
As we assume that invariant operators satisfy  {$\partial_{y_z}\braket{\hat{\Lambda}_\mu}-\braket{\partial_{y_z}{\hat{\Lambda}_\mu}}=0$} for all $z$ and $\mu$, we have $\braket{\nabla \Lambda_\mu,\mbf{u}_z}_r=0$.
Then we have
\aln{
\sum_{kz}a_kb_zB_{kz}=\braketL{\sum_ka_k\lrs{\nabla A_k-\sum_\mu f'_{k\mu}\nabla\Lambda_\mu},\sum_zb_z\mbf{u}_z}_r.
}
for  real vectors $(a_1,\cdots,a_K)$, $(b_1,\cdots,b_Z)$, and $\{f_{k\mu}'\}$.
As in Appendix~\ref{sec:multiproof},
we find that $f'_{k\mu}=\braket{\nabla A_k,\nabla\Lambda_\mu}_r$ provides a tight bound compared with the other choices, so we assume this choice in the following.

Using the Cauchy-Schwarz inequality, we have
\aln{\label{csin}
\sum_{kzk'z'}a_ka_{k'}b_zb_{z'}B_{kz}B_{k'z'}
\leq 
\sum_{kzk'z'}a_ka_{k'}b_zb_{z'}\mc{D}_{kk'}\mc{U}_{zz'},
}
where
\aln{
\mc{D}_{kk'}:=\braket{\nabla A_k,\nabla A_{k'}}_r-\sum_\mu
\braket{\nabla A_k,\nabla\Lambda_\mu}_r\braket{\nabla\Lambda_\mu,\nabla A_{k'}}_r
}
and 
\aln{
\mc{U}_{zz'}:=\braket{\mbf{u}_z,\mbf{u}_{z'}}_r.
}

 {
Just as in Appendix~\ref{sec:multiproof}, we can show two equivalent matrix inequalities
\aln{
B^\mathsf{T}\mc{D}^{-1}B\preceq \mc{U}
}
and
\aln{
B\mc{U}^{-1}B^\mathsf{T}\preceq \mc{D}.
}
}

If we take $Z=1$, \eqref{csin} indicates
\aln{
\vec{B}\vec{B}^\mathsf{T}\preceq \mc{D}U
}
and
\aln{
\vec{B}^\mathsf{T}\mc{D}^{-1}\vec{B}\leq U,
}
which we have presented in the main text.
If we instead consider $K=1$ and $Z\neq 1$, we have a matrix inequality
\aln{
\vec{\mathsf{B}}\vec{\mathsf{B}}^\mathsf{T}\preceq \mc{U}\lrs{\braket{\nabla A,\nabla A}_r-\sum_\mu\braket{\nabla A,\nabla {\Lambda}_\mu}_r^2}
}
and a scalar inequality
\aln{
\vec{\mathsf{B}}^\mathsf{T}\mc{U}^{-1}\vec{\mathsf{B}}\leq \lrs{\braket{\nabla A,\nabla A}_r-\sum_\mu\braket{\nabla A,\nabla {\Lambda}_\mu}_r^2},
}
where we have assumed that $\mc{U}$ has an inverse.

\subsection{Derivation of Eq.~\eqref{twocurrent} in the main text}\label{twoc_app}
We derive Eq.~\eqref{twocurrent} in the main text, assuming the single-particle quantum system whose Hamiltonian is given in Eq.~\eqref{singlep}.
We take $M=0$, i.e., neglect the term concerning $\nabla\Lambda_\mu$.

We first show that, if $r_{ij}\leq r'_{ij}$ for all $i$ and $j$, we have a matrix inequality $\mc{D}\preceq \mc{D}'$, where 
$\mc{D}_{kk'}:=\braket{\nabla A_k,\nabla A_{k'}}_{r}$ and $\mc{D}'_{kk'}:=\braket{\nabla A_k,\nabla A_{k'}}_{r'}$.
Indeed, $\mc{D}'-\mc{D}$ has matrix elements given by 
\aln{
\mc{D}'_{kk'}-\mc{D}_{kk'}=\braket{\nabla A_k,\nabla A_{k'}}_{r'-r},
}
which is positive semidefinite.
Thus, $\mc{D}\preceq \mc{D}'$ holds,
and then
\aln{
\vec{B}^\mathsf{T}\mc{D'}^{-1}\vec{B}\leq \vec{B}^\mathsf{T}\mc{D}^{-1}\vec{B}\leq U.
}

We now take 
\aln{
r_{ij}=|H_{ij}\rho_{ji}|\leq r'_{ij}=|H_{ij}|\frac{p_i+p_j}{2}
}
with $H_{ij}=J_h(\delta_{i,j+1}+\delta_{i,j-1})$.
In this case,
\aln{
&{\mc{D}'}_{kk'}:=\nonumber\\
&J_h\sum_{i}\frac{p_i}{2}\lrs{(\nabla A_k)_{i,i+1}(\nabla A_{k'})_{i,i+1}+(\nabla A_k)_{i,i-1}(\nabla A_{k'})_{i,i-1}},
}
where we neglect the boundary contribution, assuming that $p_i$ becomes vanishing for $i\ra\pm\infty$.

For $K=2$, we especially have
\aln{\label{two_obsc}
\frac{\mc{D}'_{22}|\npartial{\hat{A}_1}|^2-2\mc{D}'_{12}\npartial{\hat{A}_1}\npartial{\hat{A}_2}+\mc{D}'_{11}|\npartial{\hat{A}_2}|^2}{\mc{D}'_{11}\mc{D}'_{22}-|\mc{D}'_{12}|^2}\leq U,
}
where $U\leq 2C_H-2E_\mr{trans}^2/C_H$ (see \eqref{vbound}).
Noting that $C_H=2J_h$ and introducing $\mc{C}=\mc{D}'/J_h$,
we have
\aln{
\frac{\mc{C}_{22}|\npartial{\hat{A}_1}|^2-2\mc{C}_{12}\npartial{\hat{A}_1}\npartial{\hat{A}_2}+\mc{C}_{11}|\npartial{\hat{A}_2}|^2}{\mc{C}_{11}\mc{C}_{22}-|\mc{C}_{12}|^2}\leq 4J_hU.
}
Finally, straightforward calculation leads to Eq.~\eqref{twocurrent} in the main text:
\aln{
|\mc{V}_1-\tilde{\chi}\mc{V}_2|&\leq \sqrt{(1-\tilde{\chi}^2)(4J_hU-\mc{V}_2^2)}\nonumber\\
&\leq \sqrt{(1-\tilde{\chi}^2)(4J_h^2-E_\mr{trans}^2-\mc{V}_2^2)},
}
where $\mc{V}_k:=\npartial{\hat{A}_k}/\sqrt{\mc{C}_{kk}}$, and $\tilde{\chi}:=\mc{C}_{12}/\sqrt{\mc{C}_{11}\mc{C}_{22}}\leq 1$.

As an example, let us consider the position operator $\hat{A}_1=\hat{x}=\sum_ii\ket{i}\bra{i}$ and the imbalance operator for odd-even density, $\hat{s}=\sum_i(4\lfloor i/2\rfloor -2i+1)\ket{i}\bra{i}$.
We then have 
\aln{
(\nabla x)_{i,i+1}=1,\quad (\nabla s)_{i,i+1}=-2(-1)^i,
}
and $\mc{C}_{11}=1, \mc{C}_{12}=\mc{C}_{21}=0$, and $\mc{C}_{22}=4$.
Thus, $\tilde{\chi}=0$ and we have
\aln{
|\partial_t\braket{\hat{x}}|^2+\frac{|\partial_t\braket{\hat{s}}|^2}{4}\leq 4J_h^2-E_\mr{trans}^2.
}

\bibliography{derh2.bib}

\end{document}